\PassOptionsToPackage{prologue,dvipsnames}{xcolor}
\documentclass[acmsmall]{acmart}
\usepackage{todonotes}
\usepackage{tikz}
\usetikzlibrary{calc, arrows.meta, shapes, arrows, positioning, trees, ext.paths.ortho}
\newcommand\tikzmark[2]{
    \tikz[remember picture,baseline=(#1.base),inner sep=0pt] \node (#1) {#2};}

\usepackage[clockwise]{rotating}
\usepackage{makecell}
\usepackage{wrapfig}

\usepackage{stackengine}
\usepackage{amsmath}
\usepackage[frozencache,cachedir=.]{minted}
\usepackage[most,minted]{tcolorbox}
\usepackage{upquote}
\usepackage{textcomp}
\usepackage{tabularx}
\usepackage{algorithm}
\usepackage{algpseudocode}
\usepackage{graphicx}
\usepackage{pgf-pie}
\usepackage{pgfplots}
\usepackage{multirow}
\usepackage{bbding}
\usepackage[font=bf]{caption}
\usepackage{svg}
\usepackage{subcaption}
\usepackage{xcolor}
\usepackage{pbox}
\usepackage[normalem]{ulem}
\usepackage{tabularray}
\usepackage{mdframed}
\usepackage{adjustbox}

\tcbset{listing engine=listings}
\surroundwithmdframed[leftline=false, rightline=false,innerleftmargin=18pt, innerrightmargin=0pt]{minted}
\setminted[php]{linenos,
startinline, 
breaklines=true, 
fontsize=\footnotesize, 
numbersep=1em}

\makeatletter
\newcommand{\currentfontsize}{\fontsize{\f@size}{\f@baselineskip}\selectfont}
\makeatother

\setmintedinline[php]{style=bw}

\newcolumntype{Y}{>{\centering\arraybackslash}X}
\newcolumntype{S}{>{\centering\arraybackslash\hsize=.5\hsize}X}
\newcommand{\artemis}{\textsc{Artemis}}

\pgfplotsset{compat=1.18}
\lstdefinestyle{mystyle}{
    basicstyle=\footnotesize\ttfamily,
    breaklines=true,
    breakindent=0pt
}
\lstdefinelanguage{prompt}{
    keywords = {User,Assistant,System,ToolDef,ToolCall,ToolCallResult},
}

\newcommand{\ballnumber}[1]{\tikz[baseline=(myanchor.base)] \node[circle,fill=.,inner sep=1pt] (myanchor) {\color{-.}\footnotesize #1};}

\newcommand\red[1]{}
\newcommand\blue[1]{\textcolor{black}{#1}}
\newcommand\rev[1]{\textcolor{black}{#1}}

\newcommand\spacer[1]{\rule[-#1]{0pt}{#1}}

\usepackage{pifont}

\newcommand{\xmark}{\text{\ding{55}}}

\definecolor{safetystring}{RGB}{243,243,243}
\definecolor{internationalorange}{rgb}{1.0, 0.31, 0.0}
\definecolor{hotmagenta}{rgb}{1.0, 0.11, 0.81}
\definecolor{greenpigment}{rgb}{0.0, 0.65, 0.31}
\definecolor{darkspringgreen}{rgb}{0.09, 0.45, 0.27}
\definecolor{darkpastelgreen}{rgb}{0.01, 0.75, 0.24}
\definecolor{chartreuse}{rgb}{0.5, 1.0, 0.0}
\definecolor{babyblueeyes}{rgb}{0.63, 0.79, 0.95}
\colorlet{lightblue}{babyblueeyes!40}
\definecolor{bleudefrance}{rgb}{0.19, 0.55, 0.91}
\definecolor{aogreen}{rgb}{0.0, 0.5, 0.0}
\definecolor{brightpink}{rgb}{1.0, 0.0, 0.5}
\definecolor{brilliantrose}{rgb}{1.0, 0.33, 0.64}
\definecolor{darkorange}{rgb}{1.0, 0.55, 0.0}
\definecolor{cobalt}{rgb}{0.0, 0.28, 0.67}
\definecolor{usercolor}{RGB}{231,239,246}
\definecolor{assistantcolor}{RGB}{254,247, 224}
\definecolor{call}{RGB}{0,128,255}
\colorlet{taintedsrc}{darkorange!10}
\colorlet{taintedtgt}{babyblueeyes!40}
\colorlet{taintedarrow}{darkorange}
\colorlet{constraint}{babyblueeyes!40}
\colorlet{constrainttext}{red}
\colorlet{sinks}{green!20}
\colorlet{dataflow}{red}
\colorlet{formula}{black}

\newcommand{\redarrowdashed}{\raisebox{2pt}{\tikz{\draw[line width=0.25mm, -{Latex}, dashed, red](0,0) -- (5mm,0);}}}

\newcommand{\implicitcallarrow}{\raisebox{2pt}{\tikz{\draw[line width=0.25mm,densely dotted, -{Latex}, call](0,0) -- (5mm,0);}}}
\newcommand{\magicarrow}{\raisebox{2pt}{\tikz{\draw[line width=0.25mm,  densely dotted, -{Latex},dataflow](0,0) -- (5mm,0);}}}
\newcommand{\proparrow}{\raisebox{2pt}{\tikz{\draw[line width=0.25mm, -{Latex}, dataflow](0,0) -- (5mm,0);}}}
\newcommand{\prodashedparrow}{\raisebox{2pt}{\tikz{\draw[line width=0.25mm, dashed, -{Latex}, dataflow](0,0) -- (5mm,0);}}}
\newcommand{\blockarrow}{\raisebox{2pt}{\tikz{\draw[line width=0.25mm, -{Latex}](0,0) -- (5mm,0);}}}

\tikzstyle{label}=[draw=none, align=left, font=\scriptsize\ttfamily]

\usepackage{accents}
\newcommand{\halfapprox}{\clipbox{0em 0em 0em 0.22em}{$\approx$}}

\usepackage{soul}
\usepackage{url}

\NewDocumentCommand{\rot}{O{45} O{1em} m}{\makebox[#2][l]{\rotatebox{#1}{#3}}}%

\setcopyright{cc}
\setcctype{by}
\acmDOI{10.1145/3720488}
\acmYear{2025}
\acmJournal{PACMPL}
\acmVolume{9}
\acmNumber{OOPSLA1}
\acmArticle{128}
\acmMonth{4}
\received{2024-10-16}
\received[accepted]{2025-02-18}

\begin{document}

\title[\textsc{Artemis}: Toward Accurate Detection of Server-Side Request Forgeries]{\textsc{Artemis}: Tow\underline{ar}d Accura\underline{te} Detection of Server-Side Request Forgeries through LL\underline{M}-Ass\underline{is}ted Inter-procedural Path-Sensitive Taint Analysis}

\author{Yuchen Ji}
\orcid{0000-0002-4828-5338}
\affiliation{%
  \institution{ShanghaiTech University}
  \city{Shanghai}
  \country{China}
}
\email{jiych2022@shanghaitech.edu.cn}

\author{Ting Dai}
\orcid{0000-0003-0257-2304}
\affiliation{%
  \institution{IBM Research}
  \city{Yorktown Height}
  \country{USA}
}
\email{ting.dai@ibm.com}

\author{Zhichao Zhou}
\orcid{0000-0002-4543-262X}
\affiliation{%
  \institution{ShanghaiTech University}
  \city{Shanghai}
  \country{China}
}
\email{zhouzhch@shanghaitech.edu.cn}

\author{Yutian Tang}
\orcid{0000-0001-5677-4564}
\affiliation{%
  \institution{University of Glasgow}
  \city{Glasgow}
  \country{United Kingdom}
}
\email{yutian.tang@glasgow.ac.uk}

\author{Jingzhu He}
\orcid{0009-0005-9448-5022}
\affiliation{%
  \institution{ShanghaiTech University}
  \city{Shanghai}
  \country{China}
}
\email{hejzh1@shanghaitech.edu.cn}
\authornote{Jingzhu He is the corresponding author.}



\begin{abstract}
Server-side request forgery (SSRF) vulnerabilities are inevitable in PHP web applications. Existing static tools in detecting vulnerabilities in PHP web applications neither contain SSRF-related features to enhance detection accuracy nor consider PHP's dynamic type features.
In this paper, we present \textsc{Artemis}\footnote{Artemis is the Greek goddess of the hunt who defeated Apate, the goddess of forgery.}, 
a static taint analysis tool for detecting SSRF vulnerabilities in PHP web applications. 
First, \textsc{Artemis} extracts both PHP built-in and third-party functions as candidate source and sink functions. 
Second, \textsc{Artemis} constructs both explicit and implicit call graphs to infer functions' relationships.
Third, \textsc{Artemis} performs taint analysis based on a set of rules that prevent over-tainting and pauses when SSRF exploitation is impossible.
Fourth, \textsc{Artemis} analyzes the compatibility of path conditions to prune false positives.
We have implemented a prototype of \textsc{Artemis} and evaluated it on 250 PHP web applications.
\textsc{Artemis} reports 207 true vulnerable paths (106 true SSRFs) with 15 false positives. 
Of the 106 detected SSRFs, 35 are newly found and reported to developers, with 24 confirmed and assigned CVE IDs.
\end{abstract}

\maketitle

\section{Introduction}
\label{sec:intro}
\begin{wrapfigure}{r}{0.4\textwidth}
    \centering
    \includegraphics[width=\linewidth]{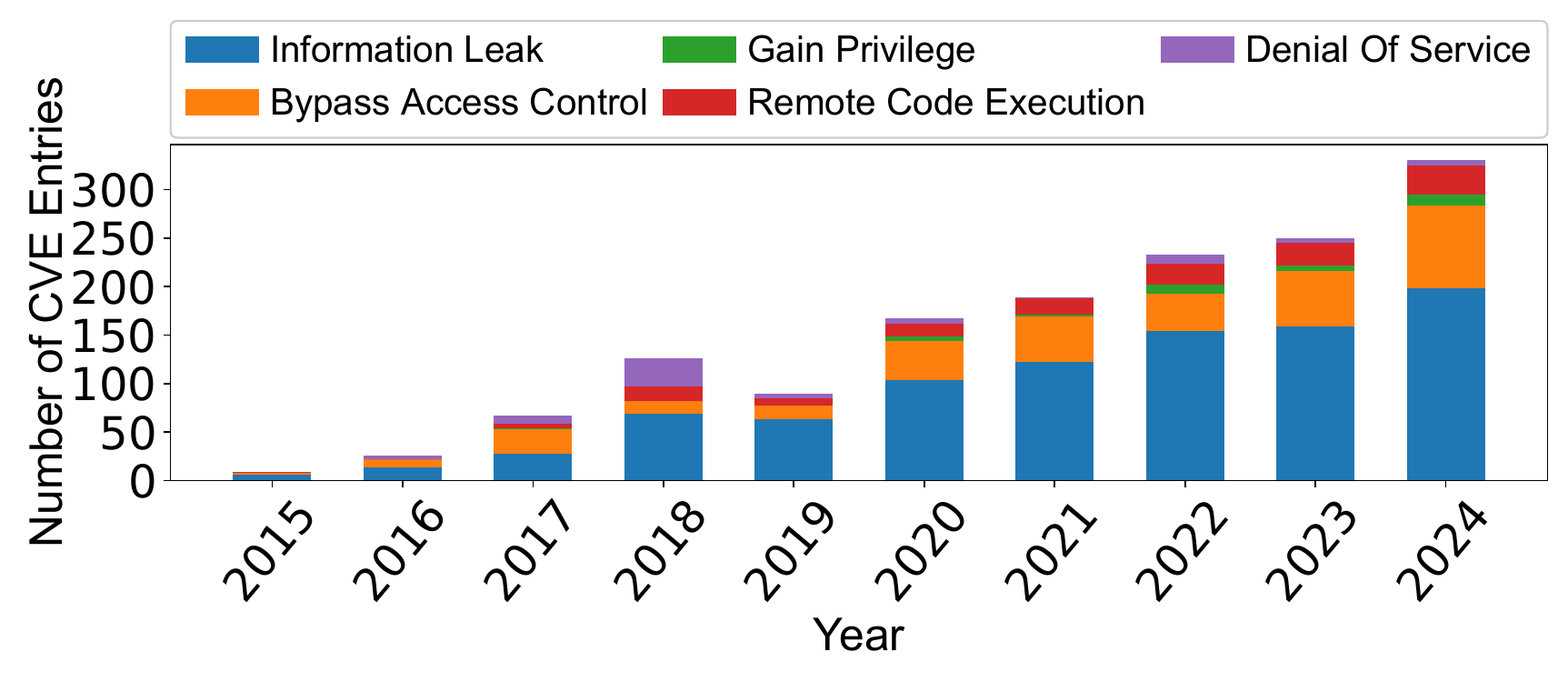}
    \caption{The statistics of SSRFs in recent 10 years with impacts.}
    \label{fig:ssrf-trend}
\end{wrapfigure}

PHP is currently the dominant programming language for building web applications \cite{php-survey}.
PHP web applications allow developers to use server-side requests to interact with third-party applications~\cite{ssr, server-side-browser}. 
Attackers often manipulate user input to send forged server-side requests, leading to the exploitation of server-side request forgery (SSRF) vulnerabilities. 
Attackers pretend like the server sends the request, bypassing access controls such as firewalls that prevent direct access to specific URLs~\cite{ssrf-cwe}. %
Exploitation of SSRF vulnerabilities can lead to severe consequences for applications, including denial of service (DoS), leakage of sensitive data, and remote code execution~\cite{ssr}. 
In 2019, the exploitation of an SSRF vulnerability in Capital One's service resulted in the leakage of credit card information for over 100 million consumers~\cite{1-capital-one}.
Existing research has explored the detection and prevention of various vulnerabilities in web applications, including cross-site scripting (XSS)~\cite{xss-3,xss-9,xss-11,xss-13,xss-16,xss-20}, SQL injection~\cite{sql-7, sql-27}, DoS~\cite{dos-12, dos-21}, cross-site request forgery (CSRF)~\cite{csrf-2, csrf-8, csrf-14, csrf-15}, prototype injection~\cite{proto-23, proto-28, proto-29, proto-30}, business logic flow vulnerabilities~\cite{logic-1, logic-5, logic-17, logic-18, logic-19, logic-22, logic-24,logic-31}, and recurring vulnerabilities~\cite{skyport, recurscan}.
However, limited research focuses on SSRF detection, even though reported SSRF cases are rapidly increasing and the impacts are severe, as illustrated in Figure~\ref{fig:ssrf-trend}. %
Since 2021, SSRFs have been ranked as the top 10 vulnerabilities by OWASP, based on the occurrence, impacts, incident rates, and number of CVEs~\cite{1-top10}. 

Static taint analysis tools~\cite{tchecker,rips, 1-phan-plugin,phpjoern,psalm} are widely used by developers to detect vulnerabilities in PHP web applications. The key idea is to track the flow of sources containing untrusted user input and check whether the tainted data reaches specified sink functions that send server-side requests.
However, using existing static taint analysis tools to detect SSRFs in PHP results in high false positives and false negatives due to four key challenges.

First, existing tools only consider PHP's built-in sources and sinks. In our preliminary study, we find that 61\% (153 in 250) applications use third-party APIs to handle user input or send server-side requests. Existing tools produce false negatives without considering third-party sources and sinks. %
Manually extracting these sources and sinks is time-consuming and does not scale.%

Second, \rev{while the implicit call graph construction has been researched recently in statically typed languages~\cite{java-cg}, existing PHP static analysis tools often fail to resolve implicit call targets in their call graphs. For example, magic methods~\cite{magic}, automatically called for undefined object methods, and dynamic constructs like variable classes and methods~\cite{var-meth}, are ignored.} 
Failing to include implicit method calls when constructing call graphs leads to mis-detection of SSRFs. 
For example, \textsc{Rips}~\cite{rips} overlooks method calls because \textsc{Rips} generates call graphs by matching function signatures, disregarding object-oriented methods. 
\textsc{Phpjoern}~\cite{phpjoern} constructs call graphs by identifying unique method names. When a method name is not unique, i.e., several classes have methods with the same name, \textsc{Phpjoern} disregards the call.
\textsc{TChecker}~\cite{tchecker}, \textsc{Phan}~\cite{1-phan-plugin} and \textsc{psalm}~\cite{psalm} all generate call graphs using type inference and account for object-oriented methods. However, if the variable type cannot be statically inferred due to features such as reflection, any method call on that variable is ignored. Additionally, magic methods are ignored.

Third, existing tools contain both over-tainting and under-tainting rules. 
Over-tainting rules overlook data flow paths and string sanitizations, which may result in non-vulnerable code being flagged as tainted. Under-tainting rules exclude certain data structures, code blocks, and indirect paths, which can lead to missed vulnerabilities.
For example, \textsc{Rips} recklessly marks the return value of a function as tainted if any argument is tainted, regardless of whether the argument actually affects the return value through data flows. 
\textsc{TChecker} treats functions that are not connected in the call graph as dead code and omits their analysis. However, bypassed functions can be invoked by reflection in third-party libraries and may contain vulnerabilities. 
\blue{\textsc{Phan} and \textsc{Psalm}} taint a string if any of its components are tainted, even if the string is not a URL.%

Fourth, existing tools do not consider path conditions \rev{of SSRF exploitation}, leading to false positives. \blue{
Specifically, two types of path conditions prevent attacker-controlled URLs from reaching the sink functions. First, the sinks may be on an impossible branch. Second, path conditions may forbid attacker-controlled URLs.}
Although several existing tools, such as \textsc{Rips}, \textsc{TChecker}, and \textsc{Phan}, attempt to model specific sanitization rules for sanitizer functions, these tools do not model path conditions effectively because they solely rely on name matching to recognize sanitizers. %

\subsection{A Motivating Example}
\newlength\myboxwidth

\setlength{\myboxwidth}{\dimexpr\columnwidth-2\fboxsep}
\begin{wrapfigure}{R}{0.51\textwidth}
    \centering
    \renewcommand\theFancyVerbLine{%
\ifnum\value{FancyVerbLine}=9 
  \tikzmark{call1}{\arabic{FancyVerbLine}}%
\else\ifnum\value{FancyVerbLine}=14
  \tikzmark{func1}{\arabic{FancyVerbLine}}%
\else
\arabic{FancyVerbLine}%
\fi\fi
}

\begin{minted}[fontsize=\tiny,breakanywhere, escapeinside=<>, 
highlightlines={5}, 
highlightcolor=constraint]{php}
<\tikzmark{t1}{\colorbox{gray!10}{\$url}}> =<\tikzmark{s1}{\colorbox{taintedsrc}{\$\_POST[\textquotesingle{}url\textquotesingle{}]}}>; //Source fetches user input
...
function getRss(<\tikzmark{t2}{\colorbox{gray!10}{\$url}}>) 
{
  if (!empty(<\tikzmark{t3}{\colorbox{gray!10}{\$url}}>) && strstr(<\tikzmark{t4}{\colorbox{gray!10}{\$url}}>, 'blogspot')) {
    ...
    $rss = new SimpleCluvPie();
    ...
    new $rss->file_class(<\tikzmark{t5}{\colorbox{gray!10}{\$url}}>, $rss->timeout, 5, null, $rss->useragent, $rss->force_fsockopen );
  }
}

class SimpleCluvPie_File {
  public function __construct(<\tikzmark{t6}{\colorbox{gray!10}{\$url}}>, $timeout, $redirects, $headers, $useragent, $force_fsockopen) {
    $fp = curl_init();
   <\colorbox{sinks}{curl\_setopt}>($fp,CURLOPT_URL,<\tikzmark{t7}{\colorbox{gray!10}{\$url}}>);//Sink sends request
  }
}
\end{minted}
\begin{tikzpicture}[remember picture, overlay,
arr/.style = {draw=#1, rounded corners=1.3mm, -{Latex}}]

\draw[draw=call, rounded corners=1.3mm, thick, -{Latex}, densely dotted]  (call1.west) -- ($(call1.west)-(2.2mm,1mm)$) -- ($(func1.west)-(1.4mm,-1.5mm)$) --  (func1.west);

\path (s1.north) edge[ dataflow, -{Latex}, bend right=10] (t1.north);
\path (t1.south) edge[ dataflow, -{Latex}] (t2.north);

\path (t2.south) edge[ dataflow, -{Latex}] (t3.north);
\path (t3.north) edge[ dataflow, dashed,-{Latex},bend left=10] node[text=constrainttext,font=\footnotesize\bfseries,xshift=4mm,yshift=1mm] {not empty}(t4.north);
\draw[draw=dataflow, dashed, -{Latex}] (t4.south) -- node[text=constrainttext,font=\footnotesize\bfseries,xshift=12mm,yshift=0mm] {contains blogspot} (t5.north);
\path (t5.south) edge[ dataflow, -{Latex}] (t6.north);
\path (t6.south) edge[ dataflow, -{Latex}] (t7.north);

\end{tikzpicture}
\caption{A new SSRF in CommentLuv v3.0.4 whose taint propagation path from \colorbox{taintedsrc}{source} to \colorbox{sinks}{sink} contains \colorbox{constraint}{path conditions} and implicit call flows. {\protect\implicitcallarrow} represents implicit call flows. 
{\protect\proparrow} represents data flows.
{\protect\prodashedparrow} represents (controlled) data flows with path conditions.}
    \label{fig:motivating}
    \vspace{-5pt}
\end{wrapfigure}

We present a previously unknown SSRF vulnerability in Figure~\ref{fig:motivating} to illustrate how it is exploited and how to detect it using static taint analysis. The application takes a user input at line \#1. The input URL reaches the request-sending function on line \#16 if it is non-empty and contains the string \texttt{blogspot}. An attacker could exploit the vulnerability by crafting a URL like \texttt{http://blogspot.evil.com}, where \texttt{evil.com} is a domain controlled by the attacker and resolves to \texttt{127.0.0.1}. %
Thus, this URL bypasses the string check and causes the server to send requests to internal addresses. Exploitation of this SSRF causes sensitive data leakage. %

To detect this SSRF vulnerability using taint analysis, we taint the built-in source \mintinline{php}{$_POST} that takes a user input at line \#1 and the built-in sink \mintinline{php}{curl_setopt} that sends the server-side request at line \#16. We aim to find whether there exists a feasible path from the tainted source to sink, like the path indicated by the red solid lines in Figure~\ref{fig:motivating}. 

However, existing taint analysis tools fail to detect this vulnerability due to two reasons. First, the implicit call flow between line \#9 and line \#14 is obscured by the dynamically assigned class name \mintinline{php}{$rss->file_class}, making it difficult for static analysis tools to build a complete call graph. As a result, the taint flow terminates at line \#9. 
Second, existing tools fail to consider the path conditions within an if branch at line \#5. The path condition performs string check on the \mintinline{php}{$url}. Only when \mintinline{php}{$url} contains the sub-string \texttt{blogspot}, there exists a feasible path from line \#5 to line \#9, making the sink function reachable.

\textsc{Artemis} overcomes the above challenges to detect this SSRF. First, by examining all \mintinline{php}{__construct} methods within the application (\rev{keyword \texttt{new} represents a call to the constructor method}), \textsc{Artemis} resolves the implicit call from line \#9 to line \#14, identifying the correct class (\mintinline{php}{SimpleCluvPie_File}) based on the constructor's argument count \rev{because the constructor on line \#14 is the only constructor that accepts six parameters.} 
Second, \textsc{Artemis} extracts the path condition and analyzes the string check imposed by \mintinline{php}{strstr}. \textsc{Artemis} confirms that attacker-controlled URLs containing \texttt{blogspot} reach the sink function, making exploitation feasible. 

\subsection{Contributions}
In this work, we present \textsc{Artemis}, a holistic taint analysis tool to detect SSRFs in PHP web applications. \textsc{Artemis} consists of four integrated modules: 1) \emph{source and sink identification}; 2) {\emph{statically inferred call graph construction}}; 3) {\emph{rule-based taint analysis}}; and 4) {\emph{false positive pruning}}. 
First, \textsc{Artemis} extracts both PHP built-in and \rev{third-party} source and sink functions. For functions \rev{in third party libraries}, \textsc{Artemis} extracts all the functions' signatures with PHPDoc comments~\cite{phpdoc}. \textsc{Artemis} then leverages a large language model (LLM) to further determine the candidate sources and sinks. 
Second, \textsc{Artemis} constructs the call graphs considering PHP's dynamic typed features. Specifically, \textsc{Artemis} builds the implicit caller-callee relationships statically when the callee is implemented using magic methods or the callee has a variable class or method name. 
Third, \textsc{Artemis} performs taint analysis to report candidate paths from sources to sinks based on augmented propagation rules to prevent over-tainting and under-tainting. \textsc{Artemis} also constructs implicit data flows to improve detection coverage. Moreover, \textsc{Artemis} applies unique safety string assurance rules to eliminate the tainted paths along which SSRF exploitation is impossible.
Lastly, \textsc{Artemis} prunes false positives by conducting path condition analysis to eliminate infeasible paths with incompatible conditions. Specifically, \textsc{Artemis} identifies always unsatisfied conditions and URL rejection conditions from the extracted path conditions. 
During result validation, we use a multi-turn LLM conversation to create concrete SSRF exploits for \textsc{Artemis}'s reported vulnerable paths. Automatically generating exploits with LLM significantly reduces the manual effort required to confirm taint analysis results. 
\rev{\textsc{Artemis} makes the following contributions:} 
\begin{itemize} 
\item \rev{\textbf{Third-party sources and sinks extraction.} We leverage an LLM-assisted method to augment the set of SSRF sources and sinks by incorporating third-party library APIs besides PHP built-in functions. We show that 50\% SSRF vulnerabilities detected by \textsc{Artemis} are introduced by third-party libraries.} 

\item \rev{\textbf{Implicit call graph construction schemes.} We observe that 13\% calls targets are implicit including magic methods, variable class names of callees, and variable method names of callees in widely used PHP applications. 
We develop a set of implicit call graph construction schemes through slight over-approximation. The results show that \textsc{Artemis} detects 17 additional vulnerabilities without introducing any new false positives when incorporating implicit call graph construction.}

\item \rev{\textbf{Augmented taint analysis rules.} We augment taint propagation rules to prevent over-tainting and under-tainting. We incorporate PHP array semantics and develop array-specific taint rules. We develop PHP-specific implicit data flow reconstruction rules. Moreover, we analyze the strings to eliminate the tainted paths along which SSRF exploitation is impossible. The results show that the augmented taint rules increase the detection coverage by 4.5\% and reduce the false positives by 90.2\% for reported paths.}

\item \rev{\textbf{SSRF-specific false positive pruning.} We develop simple but effective SSRF-specific schemes to eliminate infeasible paths with incompatible constraints purely statically. Specifically, we extract path conditions of tainted paths and prune paths with always unsatisfied and URL rejection conditions to reduce false positives. Our false positive pruning schemes reduce the number of reported false positive paths from 35 to 15. }

\end{itemize}
\rev{We have implemented a prototype of \textsc{Artemis} and evaluated it on 250 PHP web applications. Our results show that \textsc{Artemis}  reports 207 true vulnerable paths (106 true SSRFs) and 15 false positives, significantly outperforming existing tools. Among the 106 detected SSRFs, 35 are new, and 24 of them have been confirmed by developers with assigned CVE IDs.}

The rest of the paper is organized as follows. 
Section~\ref{sec:related} discusses related work. Section~\ref{sec:design} describes the design details of the \textsc{Artemis} system.  Section~\ref{sec:evaluation} presents the experimental evaluation. Finally, the paper concludes in Section~\ref{sec:conclusion}.

\section{Related work}
\label{sec:related}
\textbf{Taint-based Vulnerability Detection.}
Existing work has designed generic vulnerability detection tools based on taint analysis. \textsc{Rips}~\cite{rips} performed taint analysis and built call graphs by matching function signatures and specified over 900 propagation rules for PHP built-in functions. \blue{However, it ignored object-oriented methods when building call graphs.} 
\blue{\textsc{Phpjoern}~\cite{phpjoern} improved call graph construction by analyzing object-oriented methods. However, when functions in different classes had identical names, \textsc{Phpjoern} could not distinguish them and omitted these functions.} 
\blue{\textsc{TChecker}~\cite{tchecker} is built on \textsc{Phpjoern} by enhancing call graph construction for object-oriented methods through type inference. \textsc{TChecker} also added propagation rules for class field variables during taint analysis.} \blue{However, if a variable's type cannot be statically inferred, \textsc{TChecker} ignores its method calls.}
\textsc{WAP}~\cite{wap} utilized supervised machine learning methods to prune false positive patterns on the vulnerable paths identified by taint analysis, \blue{which requires manual effort to collect vulnerable code samples for training.} \blue{Compared to prior work, \textsc{Artemis} recognizes third-party sources and sinks, constructs implicit call graphs, reduces both over-tainting and under-tainting during taint analysis and prunes false positives caused by path conditions. As a result, \textsc{Artemis} detects more SSRFs with significantly fewer false positives and does not rely on manual effort such as collecting vulnerable code samples.} 
Existing work has also introduced taint-based detection tools for specific types of vulnerability. 
\textsc{Splendor}~\cite{splendor} analyzed database operations in propagation paths using a heuristic token matching strategy to detect second-order XSS vulnerabilities in PHP applications. 
\textsc{Torpedo}~\cite{torpedo} investigated whether retrieved string values from the database could lead to DoS vulnerabilities \blue{using string analysis}. \blue{In contrast to these tools, \textsc{Artemis} focuses on SSRF.}

\textbf{Fuzzing-based Vulnerability Detection}. Existing work has also explored vulnerability detection approaches based on fuzzing.  
For example, \textsc{Ufuzzer}~\cite{ufuzzer} targeted file-upload vulnerabilities by fuzzing inputs to reach sink functions along the propagation paths identified by symbolic execution tools. 
\textsc{fuse}~\cite{fuse} crafted mutation strategies focused on modifying standard upload requests to detect file-upload vulnerabilities in applications built in various languages. 
\textsc{Navex}~\cite{navex} performed taint analysis to identify vulnerable paths and then fuzzed input to detect vulnerabilities.
\textsc{witcher}~\cite{witcher} increased code coverage during fuzzing by modifying user inputs to generate SQL special characters and shell commands, aiming to detect SQL injections and code injections that could lead to remote code execution in applications built with various languages.
\textsc{Atropos}~\cite{atropos} used a feedback-driven approach by modifying the PHP interpreter to generate logs that guide fuzzing mutations to detect potential vulnerabilities such as XSS and object injection in PHP web applications. 
\blue{Pellegrino et al.~\cite{ssr} examined the security implications of server-side requests (SSR) and developed a black-box fuzzing tool named \textsc{Guenther} to detect SSR misuses, including SSRF. However, \textsc{Guenther} requires manual input of the URL and parameters for fuzzing, which involves significant manual effort to crawl and analyze.  
The recently introduced \textsc{SSRFuzz}~\cite{ssrfuzz} detects SSRF in PHP applications by combining dynamic taint analysis with black-box fuzzing. The authors first examined every function in the PHP manual to identify all potential sinks. Then, \textsc{SSRFuzz} crawled the application, dynamically tracking taints and logging HTTP requests and parameters when tainted input reached sinks. Finally, \textsc{SSRFuzz} applied black-box fuzzing with SSRF-specific mutations and monitoring rules to identify vulnerabilities.
Compared to \textsc{SSRFuzz}, \textsc{Artemis} identifies both PHP built-in and third-party functions as sources and sinks using LLM and performs static taint analysis. The exploit is then constructed automatically using LLM from the results of static taint analysis, without the need for crawling and fuzzing. %
}

\section{System Design}
\label{sec:design}
\begin{figure*}[tbp]
    \centering
    \includegraphics[width=\textwidth]{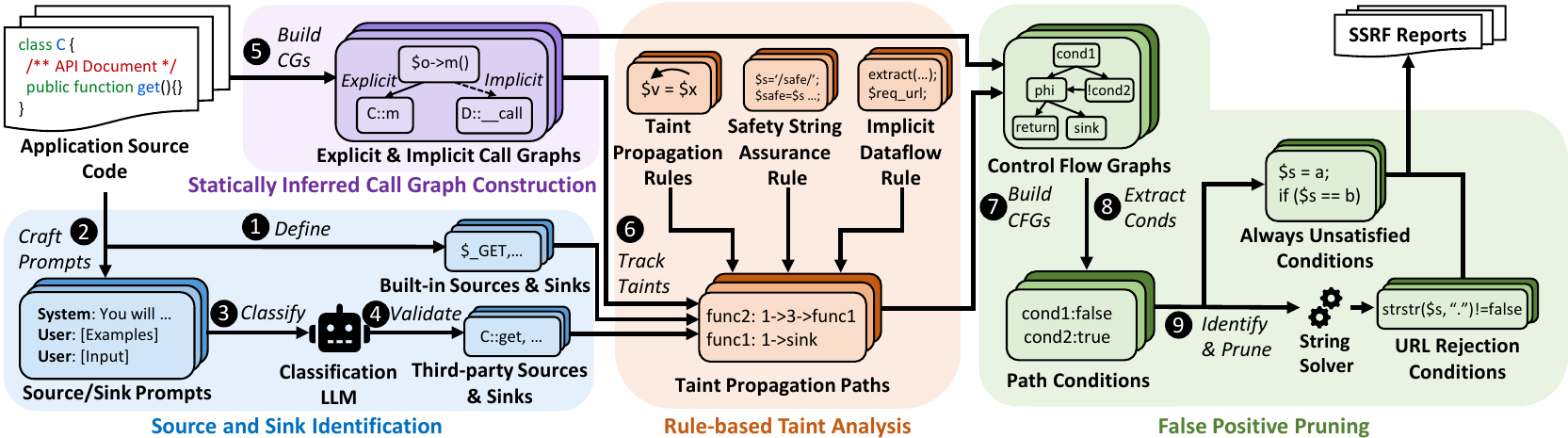}
    \caption{The overall architecture of \artemis. %
    }
    \vspace{-10pt}
    \label{fig:arch}
\end{figure*}

In this section, we present the design details of \textsc{Artemis}. Figure~\ref{fig:arch} shows the overall architecture of \textsc{Artemis}. 
First, \textsc{Artemis} takes the source code of the application as inputs and extracts both PHP built-in and third-party functions as candidate sources and sinks. 
Second, \textsc{Artemis} constructs both explicit and implicit call graphs statically from the source code. 
Third, \textsc{Artemis} performs a taint analysis based on constructed call graphs, augmented propagation rules, implicit data flow rules, and safety string assurance rules to locate tainted paths from candidate sources to sinks. 
Lastly, \textsc{Artemis} prunes false positives that are caused by path insensitivity.

\subsection{Source and Sink Identification}
\label{subsec:source-sink}
Source and sink functions are crucial for SSRF detection, where source functions take the user inputs and sink functions send server-side requests. \textsc{Artemis} identifies them from both PHP standard libraries and third-party libraries.

\textbf{Built-in Sources and Sinks.}
We consider 5 built-in superglobals~\cite{superglobal} as sources, including \mintinline{php}{$_GET}, \mintinline{php}{$_POST}, \mintinline{php}{$_REQUEST}, \mintinline{php}{$_COOKIE}, and \mintinline{php}{$_SERVER}, because they commonly retrieve user inputs from incoming requests in PHP language. %
We follow the practice of previous efforts~\cite{ssrfuzz} to extract built-in SSRF sinks that can handle both network-based URLs (e.g., \texttt{http://} and \texttt{ftp://}) and file-accessing URLs (e.g., \texttt{file://} and \texttt{phar://}) to send server-side requests. By combining results from previous efforts and existing tools~\cite{ssrfuzz, psalm, 1-phan}, 86 built-in sinks are collected. The 86 built-in sinks include \blue{11} network request sending functions such as \mintinline{php}{curl_init()} from \texttt{cURL} and \mintinline{php}{fsockopen()} from the \texttt{socket} library and \blue{75} remote file accessing functions such as \mintinline{php}{file_get_contents()}. %

\textbf{Third-party Sources and Sinks.}
Third-party library functions are commonly used as sources and sinks in modern PHP applications. These libraries simplify development by encapsulating built-in sources with additional layers of input encoding, validation, and sanitization~\cite{1-request-1, 1-request-2}. Similarly, they wrap built-in sinks with features like argument validation, response parsing, and error handling mechanisms such as timeouts and retries~\cite{1-sink-1, 1-sink-2}. 
We perform an offline analysis to identify source and sink functions from third-party libraries. This allows us to cache the results for quick retrieval during analysis, eliminating the need for repeated reanalysis.

We extract candidate source and sink functions from third-party libraries, with a focus on public functions with PHPDoc comments~\cite{phpdoc}, as functions lacking PHPDocs are less likely intended for external use.
Next, we prune those source candidates with return\footnote{The return types can be extracted from the \texttt{@return} tag in functions' corresponding PHPDocs.} types of \texttt{void}, \texttt{int}, \texttt{float}, or \texttt{bool}, as these types of user input cannot effectively manipulate URLs for sending server-side requests to user-specified but restricted destinations. %

Furthermore, we refine the source and sink candidates by utilizing few-shot learning with GPT-4o~\cite{gpt4}. We develop a custom prompt template~\ref{subsec:identify-prompt} that includes function names and PHPDocs, along with one positive and one negative example. We apply the template to each source and sink candidate with its corresponding function name and PHPDoc and send the prompt to GPT-4o to classify each candidate function as a source, sink, or neither.%

To ensure accuracy, we manually verify the third-party source and sinks generated by GPT-4o and ensure that no false positives were made at this stage. 
\rev{GPT-4o reports 48 sources and 42 sinks. After manual verification, we confirm 42 sources and 40 sinks as our third-party source/sink set.}

\subsection{Statically Inferred Call Graph Construction}
\label{sec:cg}
To statically construct call graphs in dynamically typed PHP applications, \textsc{Artemis} identifies both explicit call targets with literal string class and method names, as well as implicit call targets involving magic methods, variable class names, and variable method names. 
We design call target connection strategies to cover all types of PHP method invocations while incorporating acceptable estimations to address the challenges of statically inferring dynamic type features.
In designing our strategies, we conduct a preliminary statistical analysis of 442,129 method call sites from 55 PHP applications with reported SSRF CVEs. We apply a slight overestimation in challenging cases with lower frequencies to ensure that these strategies do not generate substantial false positives but enhance the efficiency of our analysis. In Section~\ref{subsec:tp-comp}%
, we demonstrate that ignoring implicit call targets, despite their infrequent appearance, results in non-negligible false negatives. %

\textbf{Method Invocations.} PHP methods can be invoked in seven forms: 
\ballnumber{1} \mintinline{php}{C::m}, 
\ballnumber{2} \mintinline{php}{$c::$m},
\ballnumber{3} \mintinline{php}{$v->m}, 
\ballnumber{4} \mintinline{php}{$v->$m},
\ballnumber{5} \mintinline{php}{new C()},
\ballnumber{6} \mintinline{php}{new $c()},
and \ballnumber{7} \mintinline{php}{call_user_func} and \mintinline{php}{call_user_func_array}.
The first two forms apply to static methods, the third and fourth apply to instance methods, the fifth and sixth apply to constructor calls, which are equivalent to \mintinline{php}{$v->__construct()}, and the seventh is equivalent to one of the first four. 
\texttt{C} denotes a literal class name in static method calls.
\texttt{\$c} denotes a variable class name in static method calls.
\texttt{\$v} denotes the receiving variable in instance method call. The class name of \texttt{\$c} or \texttt{\$v} is literal \texttt{O}, \rev{denoted by \mintinline{php}{$c:O} or \mintinline{php}{$v:O}} when it can be inferred through type inference~\cite{1-phan, tchecker}.
\rev{If type inference cannot infer any concrete types, the class name is considered variable, represented by \mintinline{php}{$v:any}.}
\texttt{m} denotes a literal method name. 
\texttt{\$m} denotes a variable method name.

\abovedisplayskip=2pt
\belowdisplayskip=2pt
\textbf{Explicit Call Target Connection.}
\textsc{Artemis} identifies explicit call targets by matching their signatures in both PHP built-in and target application code bases. 
For function calls, \textsc{Artemis} considers their signatures as their literal names with parameter count. 
For method calls, \textsc{Artemis} considers their signatures as their class name and method literal name with parameter count and types. 
For a static method call in the form of \mintinline{php}{C::m(arg)}, its signature is \mintinline{php}{C::m(1)}, where \texttt{1} is the parameter count.
For a method call in the form of \mintinline{php}{$o->m(a1,a2)} where \mintinline{php}{$o = new O}, its signature is \mintinline{php}{O::m(2)}. PHP does not support method overloading, meaning that it is not possible to define two methods of the same name in the same class. As a result, explicit method calls can be resolved uniquely in one class. \rev{When \mintinline{php}{$o} is inferred to have multiple types due to dynamic typing~\cite{php-typing} or reflection~\cite{php-reflect}, each candidate class of \mintinline{php}{$o} is checked to determine explicit call targets.} 
To address method inheritance~\cite{php-inherit} from a parent class to a subclass, \textsc{Artemis} conducts additional matching if the current method's signature does not match anything. It iteratively replaces the class name in the signature with its parent class name until a match is found or reaches the root class without identifying a valid method.
\rev{The formal rules to connect explicit calls are denoted as:}
\[
\scriptsize
\color{formula}
\begin{aligned}
    &\verb|$v->m(k)|, \verb|$v:O|,\exists \verb|C::m(k)|,\verb|C|\supseteq \verb|O| \implies \verb|C::m| \\
    &\verb|O::m(k)|, \exists \verb|C::m(k)|,\verb|C|\supseteq \verb|O| \implies \verb|C::m|
\end{aligned}
\]
\begin{figure}[t]
    \centering
\begin{subfigure}[t]{0.49\textwidth}
    \renewcommand\theFancyVerbLine{%
\ifnum\value{FancyVerbLine}=1 
  \tikzmark{f1}{\arabic{FancyVerbLine}}%
\else\ifnum\value{FancyVerbLine}=7
  \tikzmark{f2}{\arabic{FancyVerbLine}}%
\else\ifnum\value{FancyVerbLine}=9
  \tikzmark{f3}{\arabic{FancyVerbLine}}%
\else\ifnum\value{FancyVerbLine}=14
  \tikzmark{f4}{\arabic{FancyVerbLine}}%
\else
\arabic{FancyVerbLine}%
\fi\fi\fi\fi
}
\begin{minted}[fontsize=\tiny,escapeinside=||]{php}
ClientWrapper::|\tikzmark{s1}{\colorbox{green!10}{get}}|($_GET['uri']);

class ClientWrapper {
 /** @var Client */
 private static $client;

 public static function |\colorbox{green!10}{\_\_callStatic}|(|\tikzmark{t1}{\colorbox{orange!10}{\$m}}|, $o){
  $url = $o[0];
  self::$client::{|\tikzmark{t4}{\colorbox{orange!10}{\$m}}|}($url, []);
 }
}

class Client {
 public static function |\tikzmark{t5}{\colorbox{orange!10}{get}}|($url, $args){}
 ...
 public static function request($m, $uri, $ops) {}
}
\end{minted}
\begin{tikzpicture}[overlay,remember picture]
    \path (s1.south) edge[-{Latex}, densely dotted, dataflow] (t1.north);
    \path (t1.west) edge[-{Latex}, dataflow] (t4.north);
    \path (t4.south) edge[-{Latex}, densely dotted, dataflow] (t5.north);
    
    \draw[draw=call, rounded corners=1.3mm, densely dotted, thick, -{Latex}]  (f1.west) -- ($(f1.west)-(1.4mm,1mm)$) -- ($(f2.west)-(1.4mm,-1.5mm)$) --  (f2.west); 
    \draw[draw=call, rounded corners=1.3mm, densely dotted, thick, -{Latex}]  (f3.west) -- ($(f3.west)-(2.2mm,1mm)$) -- ($(f4.west)-(1mm,-1.5mm)$) --  (f4.west); 
\end{tikzpicture}
    \caption{CVE-2023-40969 with \colorbox{green!10}{magic method call} and \colorbox{orange!10}{variable method name}.
    }
    \label{fig:magic}
\end{subfigure}
\begin{subfigure}[t]{0.49\textwidth}
\renewcommand\theFancyVerbLine{%
\ifnum\value{FancyVerbLine}=6 
  \tikzmark{f1}{\arabic{FancyVerbLine}}%
\else\ifnum\value{FancyVerbLine}=11
  \tikzmark{f2}{\arabic{FancyVerbLine}}%
\else
\arabic{FancyVerbLine}%
\fi\fi
}
\begin{minted}[fontsize=\tiny,linenos,
startinline, 
breaklines=true, 
numbersep=1em,
escapeinside=||
]{php}
class AJXP_Controller {
  public static function find($action, $httpVars, $fileVars) {
  $callback = getCallback($actionName);
 |\colorbox{yellow!10}{\$plugin}| = $callback->getPlugin();
 |\colorbox{orange!10}{\$methodName}| = $callback->get("methodName");
  return |\colorbox{yellow!10}{\$plugin}|->|\colorbox{orange!10}{\$methodName}|($action, $httpVars, $fileVars);
 } 
}

class HttpDownloader extends AJXP_Plugin {
 public function switchAction($action, $httpVars, $fileVars) {
  ...
 }
}
\end{minted}
\begin{tikzpicture}[overlay,remember picture]
    \draw[draw=call, rounded corners=1.3mm, densely dotted, thick, -{Latex}]  (f1.west) -- ($(f1.west)-(2mm,1mm)$) -- ($(f2.west)-(1.1mm,-1.5mm)$) --  (f2.west); 
\end{tikzpicture}
\caption{CVE-2019-15033 with \colorbox{yellow!10}{variable class name} and \colorbox{orange!10}{variable method name}.
}
\label{fig:dynamic-call}
\end{subfigure}
    \caption{Examples of implicit call targets. {\protect\proparrow}, {\protect\magicarrow}, and {\protect\implicitcallarrow} represent the explicit/implicit data flows and implicit call flows.}
    \vspace{-5pt}
    \label{fig:enter-label}
\end{figure}

\textbf{Implicit Call Target Connection.}
When a method call target's method name and class name are known \rev{(e.g., \mintinline{php}{O::m(2)}) but its definition in the corresponding class hierarchy is missing, \textsc{Artemis} considers it as a magic call and} searches for magic methods, including \mintinline{php}{__call()} and \mintinline{php}{__callStatic()}, within the inheritance chain.
\rev{The formal rules to connect magic calls are denoted as:}
\[
\scriptsize
\color{formula}
\begin{aligned}
    &\verb|$v->m(k)|, \verb|$v:O|, \nexists \verb|C::m(k)|,\exists \verb|C::__call|,\verb|C|\supseteq \verb|O| \implies \verb|C::__call| \\
    &\verb|O::m(k)|, \nexists \verb|C::m(k)|,\exists \verb|C::__callStatic|,\verb|C|\supseteq \verb|O| \implies \verb|C::__callStatic|
\end{aligned}
\]
The definition of a magic method in a class can be treated as the definition of any undefined method in the same class with a minor twist.
Specifically, a magic method takes two parameters---the first is a string representing the invoked method name, and the second is an array containing the actual arguments passed to the invoked method.
For example, as shown in Figure~\ref{fig:magic},
the signature of call at line \#1 is \mintinline{php}{ClientWrapper::get(1)}\rev{, which does not match existing method signatures in \mintinline{php}{ClientWrapper}. However, a magic method \mintinline{php}{__callStatic()} exists, therefore}
\textsc{Artemis} builds an implicit call flow from line \#1 to line \#7. %

When a call target has a literal method name but a variable class name, \rev{(e.g., \mintinline{php}{$v::m(k)})} \textsc{Artemis} first searches through all built-in and application classes to find the class methods whose signatures match the literal method name.
If no match is found, but magic methods exist in classes, \textsc{Artemis} considers the magic methods (i.e., \mintinline{php}{__call()} or \mintinline{php}{__callStatic()}) in this class as candidate call sites, regardless of their parameters.
\rev{The formal rules to connect callees with literal method names but variable class names are denoted as:}
\[
\scriptsize
\color{formula}
\begin{aligned}
    &\verb|$v->m(k)|, \verb|$v:any|,\exists \verb|C::m(k)|,\verb|D::m(k)|,... \implies \verb|C::m|,\verb|D::m|,...\\
    &\verb|$c::m(k)|,\verb|$c:any|,\exists \verb|C::m(k)|,\verb|D::m(k)|,... \implies \verb|C::m|,\verb|D::m|,...
\end{aligned}
\]
In the motivating example in Figure~\ref{fig:motivating}, the implicit call target is a class constructor \blue{(\mintinline{php}{__construct()})} call at line \#9 that has a variable class name \mintinline{php}{$rss->file_class}. 
By searching all classes in the CommentLuv application, \textsc{Artemis} identifies that the constructor function of the \mintinline{php}{SimpleCluvPie_File} class is the only constructor method with the same number of parameters as the call site, thereby revealing the implicit call flow. %

When a call target has a literal class name but a variable method name \rev{(e.g., \mintinline{php}{C::$m(k)})}, \textsc{Artemis} searches all the methods in the corresponding class and classes in the class hierarchy. All methods that have the same number of parameters as the call site \rev{(e.g., \mintinline{php}{C:f(k)}, \mintinline{php}{B::g(k)} where \texttt{B} is parent class of \texttt{C})} are considered as candidate call targets. %
Moreover, all magic methods in the target classes are considered candidate call sites without parameter checking.
\rev{The formal rules to connect callees with literal class names but variable method names are denoted as:}
\[
\scriptsize
\color{formula}
\begin{aligned}
    &\verb|$v->$m(k)|, \verb|$v:O|, \exists \verb|C::m(k)|,\verb|C::n(k)|,...,\verb|C|\supseteq \verb|O| \implies \verb|C::m|,\verb|C::n|,...\\
    &\verb|O::$m(k)|, \exists \verb|C::m(k)|,\verb|C::n(k)|,...,\verb|C|\supseteq \verb|O| \implies \verb|C::m|,\verb|C::n|,...
\end{aligned}
\]
For example, in Figure~\ref{fig:magic}, the call target at line \#14 has a literal class name \mintinline{php}{Client}
and a variable method name \mintinline{php}{$method}. \textsc{Artemis} can not only identify the call site as \mintinline{php}{Client::get} from the aforementioned magic call flow construction but also search for all methods in the \mintinline{php}{Client} class and pinpoint \mintinline{php}{get} as the call site, as it has the same number of arguments as the call target at line \#14. \blue{In our experiments, 132 (0.03\%) of the call sites are classified under this category. Thus, even though our method overestimates the \textit{parameter count}, it does not result in a notable increase in false positives in detection while enhancing coverage.}

When a call target has a variable class name and a variable method name, \rev{(e.g., \mintinline{php}{$c::$m}),} \textsc{Artemis} searches all methods, including magic call methods, in all built-in and application classes to find those with the same \textit{number of parameters} and \textit{compatible types} as the call target, with magic methods not requiring parameter checks. \rev{Types of formal parameters $t_f$ and actual parameters $t_a$ match if a type in $t_a$ can be cast to a type in $t_f$. If a type cannot be inferred, the type \texttt{mixed} is used, which is compatible with all types.} 
\rev{The formal rules to connect callees with variable class names and variable method names are denoted as:}
\[
\scriptsize
\color{formula}
\begin{aligned}
    &\verb|$v->$m|(\texttt{a1:}t_{a_1},...\texttt{an}:t_{a_n}), \verb|$v:any|, \exists \verb|C::m|(\texttt{A1:}t_{f_1},...,\texttt{An:}:t_{f_n}),  t_{a_1}\subseteq t_{f_1}, \dots , t_{a_n} \subseteq t_{f_n} \implies \verb|C::m| \\
    &\verb|$c::$m|(\texttt{a1:}t_{a_1},...\texttt{an}:t_{a_n}), \verb|$c:any|, \exists \verb|C::m|(\texttt{A1:}t_{f_1},...,\texttt{An:}:t_{f_n}),  t_{a_1}\subseteq t_{f_1}, \dots , t_{a_n} \subseteq t_{f_n} \implies \verb|C::m| \\
\end{aligned}
\]
For instance, in the example shown in Figure~\ref{fig:dynamic-call}, the call at line \#6 involves both an unknown class name and an unknown method name. Since the method at line \#11 takes three parameters, and the parameter types match\footnote{\rev{The types are inferred by type inference where \mintinline{php}{$action} is string while \mintinline{php}{$httpVars} and \mintinline{php}{$httpVars} are arrays.}}, we consider it a match and treat it as a potential call target. \blue{In our experiments, 88 call sites (0.02\%) fall into this category. Therefore, our conservative call resolution, which over-approximates potential call targets by matching parameter count and type, not only ensures completeness but also keeps false positives at a manageable level.}

\rev{Variadic methods \cite{variadic}, which accept varying numbers of arguments, are not included to simplify our design. In our experiment, each application averages only two such methods, and manual validation showed that they are not used in SSRFs.} %

\subsection{Rule-based Taint Analysis}
\label{sec:taintanalysis}
\begin{wrapfigure}{R}{0.45\textwidth}
\centering
\begin{tblr}{
    colspec = {ll},
    rows={m,rowsep=0pt,font=\scriptsize,mode=math},
    row{even} = {bg=gray!10},
    colsep=2pt,
}
v:=v_1 & \text{/*variable assignment*/}\\
v:=\ominus v_1 & \text {/*unary op assignment*/} \\
v:=v_1 \oplus v_2 & \text {/*binary op assignment*/} \\
v:=(T)v_1 & \text {/*type cast*/} \\
a:=[\left({k_1}\Rightarrow\right)^*{v_1},\cdots] & \text{/*array initialization*/} \\
a[k^*]:=v_1 & \text {/*array element assignment*/} \\
v:=a[k] & \text {/*assign. from array element*/} \\
\texttt{foreach}(a \ \texttt{as} \ \left(k\Rightarrow\right)^*v)\{s\} & \text{/*foreach loop */} \\
v:=\phi(v_t,v_f) & \text{/*branch value merging*/} \\
\texttt{call}\left(m, v_a \rightarrow v_f\right) & \text{/*v}_a\text{/v}_f \text{ actual/formal arg.*/} \\
v:=\texttt{return}\left(m, v_r\right) & \text{/*method } m \text{ returns var. } v_r\text{*/} \\
v:=\texttt{yield}\left(m, v_r\right) & \text{/*}m \text{ returns } v_r \text{ in a generator*/} \\
v:=\texttt{new }  T\left( v_a \rightarrow v_f\right) & \text{/*v}_a\text{/v}_f\text{ constructor call*/} \\
\end{tblr}
\caption{Abstracted PHP language statement syntax in SSA form. * represents an optional element. %
}
\label{fig:phpsyntax}
\end{wrapfigure}

\textsc{Artemis} performs rule-based taint analysis to track the propagation of tainted data within the constructed call graphs. This analysis traces how attacker-controlled input moves through various call sites and control flows until it reaches sink functions, leading the server-side requests to unintended destinations.
We design a set of taint propagation rules to handle generic operations in PHP syntax. To ensure high detection accuracy, our design focuses on mitigating the issues of over-tainting and under-tainting prevalent in existing tools. 
Furthermore, we design taint clearance rules to terminate the tracking of tainted data when it is no longer relevant or has been neutralized.

\textbf{Language Abstraction.}
We perform an abstraction\footnote{The full language abstration description can be found in Appendix~\ref{subsec:language-abstraction}.} of the PHP language syntax geared toward taint analysis~\footnote{The syntax abstraction along with the taint propagation is performed on PHP v7.4.}, specifically investigating the propagation of tainted sources to sink operations, which may lead to requests being directed to unintended destinations via various control and data flows. In this abstraction, a PHP program is conceptualized as a series of statements in the static single assignment (SSA) forms illustrated in Figure~\ref{fig:phpsyntax}.

\textbf{Variable States.}
For a variable with scalar types, such as \texttt{string}, \textsc{Artemis} considers such variable $v$ as tainted, i.e., $v:\tau$, when $v$ directly or indirectly derives its value from a tainted source, inheriting the tainted status. Otherwise, the variable $v$ is safe, i.e., $v:\mu$, because it is not influenced by any tainted source in any way.
For complex variable types, such as \texttt{array},
\textsc{Artemis} considers an array variable $a$ as tainted, if and only if all elements inside of $a$ are tainted, denoted by $a:\tau \equiv \forall v \in a, v:\tau$. %
The variable $a$ is safe if and only if all the elements inside $a$ are safe, denoted by $a:\mu \equiv \forall v \in a, v:\mu $. Otherwise, $a$ is partially tainted, i.e., $a:\accentset{\halfapprox}{\tau}$.

To represent taint states, \textsc{Artemis} uses a dedicated data structure \texttt{T} that comprises a triplet of components: $\verb|self|, \verb|arr|_s, \verb|arr|_r$. 
\texttt{self} represents the tainted or safe state of the variable itself. 
$\texttt{arr}_s$ and $\texttt{arr}_r$ are exclusively applicable for array variables.
$\texttt{arr}_s$ tracks the states of the array elements with statically known keys. 
$\texttt{arr}_r$ tracks the states of the array elements whose keys can only be inferred during runtime execution.

\blue{To prevent over-tainting, we use two separate taint structures for each variable. $\texttt{T}_f$ is for local file URLs and $\texttt{T}_r$ is for request URLs. Unless stated otherwise, the rules apply to both taint structures.}

\begin{table*}[htbp]
  \renewcommand{\ballnumber}[1]{\tikz[baseline=(myanchor.base)] \node[circle,fill=.,inner sep=1pt,,scale=0.7, transform shape] (myanchor) {\color{-.}\scriptsize #1};}
    \newcommand{\ballnumbertabsmall}[1]{\tikz[baseline=(myanchor.base)] \node[circle,fill=.,inner sep=0.3pt,scale=0.7, transform shape] (myanchor) {\color{-.}\scriptsize #1};}
    
  \caption{Generalized and array-specific taint propagation rules in both SSA and AST forms for different PHP syntax with source code examples. The tainted state is propagated from the \colorbox{taintedsrc}{$src$} variable(s) to the \colorbox{taintedtgt}{$tgt$} variable(s).%
  }
  \label{tab:propagation-var}

\begin{minipage}[t]{0.28\linewidth}
  \vspace{0pt}
  \begin{tblr}[]{
    hline{1,3} = {solid},
    colspec={|X[6]X[5]|},
    rows={m,rowsep=0pt,font=\tiny},
    cell{1}{1} = {m,bg=gray!20,fg=black},
    cell{1}{2} = {r=2}{m},
    colsep=3pt
}
    {\ballnumber{1}\textbf{ Variable Assignment:}\\$\{v_1:\tau\}, v:=v_1 \models \{v:\tau,v_1:\tau\}$} 
  & \begin{adjustbox}{width=0.4\textwidth,keepaspectratio}
\begin{tikzpicture}[
    every node/.style={draw=none, align=left,font=\tiny\ttfamily},
    level distance=6mm,
    edge from parent/.style={draw},
    sibling distance=10mm
]
    \node {ASSIGN}
        child {node[fill=taintedtgt] {\shortstack[l]{VAR\\name:$v$}} edge from parent node[label,left] {var} }
        child {node[fill=taintedsrc] {\shortstack[l]{VAR\\name:$v_1$}} edge from parent node[label,right] {expr} };
\end{tikzpicture}
\end{adjustbox}
  \\\begin{minipage}{0.5\columnwidth}
\begin{minted}{php}
|\spacer{0mm}|
|\tikzmark{a}{\colorbox{taintedtgt}{\$d}}| =|\tikzmark{b}{\colorbox{taintedsrc}{\$\_POST}}|;
\end{minted}
        \begin{tikzpicture}[overlay,remember picture]
            \path (b.north) edge[bend right=10, taintedarrow, -{Latex}] (a.north);
        \end{tikzpicture}
         \vspace{0.5pt}
    \end{minipage} 
  & \\
\end{tblr}
\end{minipage}
\begin{minipage}[t]{0.29\linewidth}
  \vspace{0pt}
  \begin{tblr}[]{
    hline{1,3} = {solid},
    colspec={|X[12]X[11]|},
    rows={m,rowsep=0pt,font=\tiny},
    cell{1}{1} = {m,bg=gray!20,fg=black},
    cell{1}{2} = {r=2}{m},
    colsep=3pt
}
    {\ballnumber{2} \textbf{Unary Operation:}\\$\{v_1:\tau\}, v:=\ominus v_1 \models \{v:\tau,v_1:\tau\}$} 
  & \begin{adjustbox}{width=0.4\textwidth,keepaspectratio}
\begin{tikzpicture}[
   every node/.style={draw=none, align=left,font=\tiny\ttfamily},
   level distance=6mm,
    edge from parent/.style={draw},
    sibling distance=10mm
]
    \node {ASSIGN}
        child {node[fill=taintedtgt] {\shortstack[l]{VAR\\name:$v$}} edge from parent node[label,left] {var}}
        child {node {UNARY\_OP}
            child {node[fill=taintedsrc] {\shortstack[l]{VAR\\name:$v_1$}}edge from parent node[label,right] {expr}} edge from parent node[label,right] {expr}
        };
\end{tikzpicture}
\end{adjustbox}
  \\\begin{minipage}{0.5\columnwidth}
\begin{minted}{php}
|\spacer{0mm}|
|\tikzmark{t}{\colorbox{taintedtgt}{\$c}}| = clone|\tikzmark{s}{\colorbox{taintedsrc}{\$r}}|;
\end{minted}
        \begin{tikzpicture}[overlay,remember picture]
            \path (s.north) edge[ bend right=10,taintedarrow, -{Latex}] (t.north);
        \end{tikzpicture}
        \vspace{0.1pt}
    \end{minipage} 
  & \\
\end{tblr}
\end{minipage}
\begin{minipage}[t]{0.38\linewidth}
  \vspace{0pt}
\begin{tblr}[]{
    hline{1,3} = {solid},
    colspec={|X[8]X[7]|},
    rows={m,rowsep=0pt,font=\tiny},
    cell{1}{1} = {m,bg=gray!20,fg=black},
    cell{1}{2} = {r=2}{m},
    colsep=3pt
}
    {\ballnumber{3} \textbf{Binary Operation:}\\$\{v_1\vee v_2:\tau\}, v:=v_1\oplus v_2 \models \{v:\tau,v_1\vee v_2:\tau\}$} 
  & \begin{adjustbox}{width=0.4\textwidth,keepaspectratio}
\begin{tikzpicture}[
    every node/.style={draw=none, align=left,font=\tiny\ttfamily},
    level distance=6mm,
    edge from parent/.style={draw},
    sibling distance=11mm,
    level 1/.style={sibling distance=12mm},
    level 2/.style={sibling distance=11mm},
]
\node {ASSIGN}
    child {node[fill=taintedtgt] {\shortstack[l]{VAR\\name:$v$}} edge from parent node[label,left] {\texttt{var}}}
    child {node {BINARY\_OP}
        child {node[fill=taintedsrc] {\shortstack[l]{VAR\\name:$v_1$}} edge from parent node[label,left] {\texttt{left}}}
        child {node[fill=taintedsrc] {\shortstack[l]{VAR\\name:$v_2$}} edge from parent node[label,right] {\texttt{right}}} edge from parent node[label,right] {\texttt{expr}}
    };
\end{tikzpicture}
\end{adjustbox} %
  \\\begin{minipage}{0.6\columnwidth}
\begin{minted}{php}
|\spacer{0mm}|
|\tikzmark{t}{\colorbox{taintedtgt}{\$r}}| =|\tikzmark{s1}{\colorbox{taintedsrc}{\$req}}| .|\tikzmark{s2}{\colorbox{taintedsrc}{\$url}}|;
\end{minted}
        \begin{tikzpicture}[overlay,remember picture]
            \path (s2.north) edge[ bend right=10,taintedarrow,-{Latex}] (t.north);
            \path (s1.south) edge[ bend left=10,taintedarrow,-{Latex}] (t.south);
        \end{tikzpicture}
        \vspace{7pt}
     \end{minipage}
  & \\
\end{tblr}
\end{minipage}

\begin{minipage}[t]{0.455\linewidth}
  \vspace{0pt}
  \begin{tblr}[]{
    hline{1,3} = {solid},
    colspec={|X[3]X[2]|},
    rows={m,rowsep=0pt,font=\tiny},
    cell{1}{1} = {m,bg=gray!20,fg=black},
    cell{1}{2} = {r=2}{m},
    colsep=3pt
}
    {\ballnumber{4} \textbf{Type Cast:}\\$\{v_1:\tau\}, v:=(T) v_1 \models  \{v:\tau,v_1:\tau\}$, when $T$ is object, string, or array.} 
  & \begin{adjustbox}{width=0.35\textwidth,keepaspectratio}
\begin{tikzpicture}[
    every node/.style={draw=none, align=left,font=\tiny\ttfamily},
    level distance=6mm,
    edge from parent/.style={draw},
    sibling distance=11mm,
    level 1/.style={sibling distance=12mm},
    level 2/.style={sibling distance=10mm},
]
\node {ASSIGN}
    child {node[fill=taintedtgt] {\shortstack[l]{VAR\\name:$v$}} edge from parent node[label,left] {var}}
    child {node {CAST\_T}
        child {node {\shortstack[l]{type:$T$}} edge from parent node[label,left] {flag} }
        child {node[fill=taintedsrc] {\shortstack[l]{VAR\\name:$v_1$}} edge from parent node[label,right] {expr} } edge from parent node[label,right] {expr}
    };
\end{tikzpicture}
\end{adjustbox}
  \\\begin{minipage}{0.6\columnwidth}
\begin{minted}{php}
|\spacer{0mm}|
|\tikzmark{t}{\colorbox{taintedtgt}{\$urls}}| = (array)|\tikzmark{s}{\colorbox{taintedsrc}{\$\_POST[\textquotesingle{}q\textquotesingle{}]}}|;
\end{minted}
        \begin{tikzpicture}[overlay,remember picture]
            \path (s.north) edge[bend right=10,taintedarrow, -{Latex}] (t.north);
        \end{tikzpicture}
        \vspace{4pt}
    \end{minipage}
  & \\
\end{tblr}
\end{minipage}
\begin{minipage}[t]{0.5\linewidth}
  \vspace{0pt}
\begin{tblr}[]{
    hline{1,3} = {solid},
    colspec={|X[3]X[5]|},
    rows={m,rowsep=0pt,font=\tiny},
    cell{1}{1} = {m,bg=gray!20,fg=black},
    cell{1}{2} = {r=2}{m},
    colsep=3pt
}
     {\ballnumber{5} \textbf{Method Argument:}\\$\{v_a:\tau\}, \texttt{call}(m, v_a \to v_f) \models \{v_f:\tau,v_a:\tau\}$} 
   & \begin{adjustbox}{width=0.5\textwidth,keepaspectratio}
\begin{minipage}[t]{0.3\columnwidth}
\begin{tikzpicture}[
    every node/.style={draw=none, align=left,font=\tiny\ttfamily},
    level distance=6mm,
    edge from parent/.style={draw=none},
    sibling distance=9mm,
]
\node (f){FUNC\_DECL}
    child {node (m){\shortstack[l]{$m$}}}
    child {node (l){PARAM\_LIST}
        child {node[fill=taintedtgt] (p) {\shortstack[l]{PARAM\\name:$v_f$}}}
    }
    child {node (s){...}};
\draw[solid] ($(f.south)-(7mm,0mm)$) -- node[xshift=-2.2mm,yshift=0.3mm] {\scriptsize{name}} ($(m.north)-(0mm,0mm)$);
\draw[solid] ($(f.south)-(-7mm,0mm)$) -- node[xshift=-3.5mm,yshift=0.5mm] {\scriptsize{stmts}} ($(s.north)-(0mm,0mm)$);
\draw[solid] ($(f.south)-(0mm,0mm)$) -- node[xshift=-3.5mm,yshift=0mm] {\scriptsize{params}} ($(l.north)-(0mm,0mm)$);
\draw[solid] ($(l.south)-(0mm,0mm)$) -- ($(p.north)-(0mm,0mm)$);
\end{tikzpicture}
\end{minipage}
\hspace{0.05\textwidth}
\begin{minipage}[t]{0.25\columnwidth}
\begin{tikzpicture}[
    every node/.style={draw=none, rectangle, align=left,font=\tiny\ttfamily},
    level distance=6mm,
    edge from parent/.style={draw},
    sibling distance=10mm
]
\node {CALL}
    child {node {\shortstack[l]{NAME\\name:$m$}} edge from parent node[label,left] {expr}}
    child {node {ARG\_LIST}
        child {node[fill=taintedsrc] {\shortstack[l]{VAR\\name:$v_a$}}} edge from parent node[label,right] {args}
    };
\end{tikzpicture}
\end{minipage}
\end{adjustbox}
   \\\begin{minipage}{0.5\columnwidth}
   \vspace{0.1pt}
\begin{minted}{php}
$c = get(|\tikzmark{s}{\colorbox{taintedsrc}{\$\_POST[\textquotesingle{}link\textquotesingle{}]}}|);
function get(|\tikzmark{t}{\colorbox{taintedtgt}{\$url}}| ) {...}
\end{minted}
        \begin{tikzpicture}[overlay,remember picture]
            \path ([xshift=4mm]s.south) edge[taintedarrow, -{Latex}] (t.east);
        \end{tikzpicture}
     \end{minipage}
   & \\
\end{tblr}
\end{minipage}

\begin{minipage}[t]{0.495\linewidth}
  \vspace{0pt}
\begin{tblr}[]{
    hline{1,3} = {solid},
    colspec={|X[11]X[14]|},
    rows={m,rowsep=0pt,font=\tiny},
    cell{1}{1} = {m,bg=gray!20,fg=black},
    cell{1}{2} = {r=2}{m},
    colsep=3pt
}
   {\ballnumber{6} \textbf{Method Return/Yield:}\\$\{v_r:\tau\},$ $v:=\texttt{return/}$ $\texttt{yield}(m, v_r) \models \{v:\tau,v_r:\tau\}$} 
   & \begin{adjustbox}{width=0.45\textwidth,keepaspectratio}
\begin{minipage}{0.21\columnwidth}
\vspace{0pt}
    \begin{tikzpicture}[
    every node/.style={draw=none, rectangle, align=left,font=\tiny\ttfamily},
    level distance=6mm,
    edge from parent/.style={draw},
    sibling distance=9mm
]
\node {ASSIGN}
    child {node[fill=taintedtgt] {\shortstack[l]{VAR\\name:$v$}} edge from parent node[label,left] {var}}
    child {node {CALL}
        child {node {\shortstack[l]{NAME\\name:$m$}} edge from parent node[label,right] {expr}} edge from parent node[label,right] {expr}
    };
\end{tikzpicture}
\end{minipage}
\hspace{0.05\textwidth}
\begin{minipage}{0.27\columnwidth}
\vspace{0pt}
\begin{tikzpicture}[
    every node/.style={draw=none, align=left,font=\tiny\ttfamily},
    level distance=6mm,
    edge from parent/.style={draw},
    sibling distance=11mm,
    level 1/.style={level distance=5mm,sibling distance=8mm},
    level 2/.style={level distance=4mm,sibling distance=11mm},
    level 3/.style={level distance=6mm},
]
\node (f){FUNC\_DECL}
    child {node (m){\shortstack[l]{$m$}} edge from parent node[label,left] {name}}
    child {node (s){STMT\_LIST}
        child {node {...}}
        child {node {RETURN/YIELD}
        child {node[fill=taintedsrc] {\shortstack[l]{VAR\\name:$v_r$}} edge from parent node[label,right] {expr}
    }} edge from parent node[label,right] {stmts}
    };
\end{tikzpicture}
\end{minipage}
\end{adjustbox}
   \\\begin{minipage}{0.5\columnwidth}
\begin{minted}{php}
 function get(){
   return|\tikzmark{s}{\colorbox{taintedsrc}{\$\_GET[\textquotesingle{}name\textquotesingle{}]}}|;
 }
|\tikzmark{t}{\colorbox{taintedtgt}{\$filename}}| = get();
\end{minted}
        \begin{tikzpicture}[overlay,remember picture]
            \path (s.south) edge[taintedarrow, taintedarrow, -{Latex}] ([yshift=1mm]t.east);
        \end{tikzpicture}
     \end{minipage}
   & \\
\end{tblr}
\end{minipage}
\begin{minipage}[t]{0.46\linewidth}
  \vspace{0pt}
\begin{tblr}[]{
    hline{1,Z} = {solid},
    colspec={|X[5]X[5]|},
    rows={m,rowsep=0pt,font=\tiny},
    cell{1}{1} = {m,bg=gray!20,fg=black},
    cell{1}{2} = {r=2}{m},
    colsep=3pt
}
   {\ballnumber{8} \textbf{Foreach Condition:}\\ 
      $\{(a:\tau) \lor (a:\accentset{\halfapprox}{\tau},\phi\notin S)\}, \texttt{foreach}(a \texttt{ as } k\Rightarrow v)\{S\} \models \{v:\tau, a:[k:\tau,...]\}$
      }
   & \begin{adjustbox}{width=0.45\textwidth,keepaspectratio}
\begin{tikzpicture}[
    every node/.style={draw=none, rectangle, align=left,font=\tiny\ttfamily},
    level distance=6mm,
    edge from parent/.style={draw=none},
    sibling distance=10mm
]
\node (root) {\shortstack[l]{FOREACH}}
    child {node[fill=taintedsrc] (expr) {\shortstack[l]{VAR\\name:$a$}} }
    child {node (key) {\shortstack[l]{VAR\\name:$k$}}}
    child {node[fill=taintedtgt] (value) {\shortstack[l]{VAR\\name:$v$}}}
    child {node (stmts){\shortstack[l]{...}}};
\draw[solid] ($(root.west)-(0mm,1mm)$) -- node[xshift=-4mm,yshift=0mm] {\scriptsize{expr}} ($(expr.north)-(0mm,0mm)$);
\draw[solid] ($(root.south)-(1mm,0mm)$) -- node[xshift=-3mm,yshift=0mm] {\scriptsize{key}} ($(key.north)-(-3mm,0mm)$);
\draw[solid] ($(root.south)-(-1mm,0mm)$) -- node[xshift=3.5mm,yshift=0mm] {\scriptsize{value}} ($(value.north)-(3mm,0mm)$);
\draw[solid] ($(root.east)-(0mm,1mm)$) -- node[xshift=3mm,yshift=1mm] {\scriptsize{stmts}} ($(stmts.north)-(0mm,0mm)$);
\end{tikzpicture}
\end{adjustbox}
   \\\begin{minipage}{0.5\columnwidth}
\begin{minted}{php}
|\spacer{1mm}|
 foreach(|\tikzmark{s}{\colorbox{taintedsrc}{\$u}}| as|\tikzmark{t1}{\$k}|=>|\tikzmark{t2}{\colorbox{taintedtgt}{\$v}}|){}
\end{minted}
        \begin{tikzpicture}[overlay,remember picture]
            \path (s.south) edge[ bend right=10, taintedarrow, -{Latex}] (t2.south);
        \end{tikzpicture}
        \vspace{9.1pt}
    \end{minipage}
   & \\
\end{tblr}
\end{minipage}

\begin{minipage}{0.96\linewidth}
    \begin{tblr}[]{
    hline{1,3} = {solid},
    vline{2,4} = {dotted},
    colspec={|X[2]X[1]X[2]X[1]|},
    rows={m,rowsep=0pt,font=\tiny},
    cell{1}{1} = {c=2}{m,bg=gray!20,fg=black},
    cell{1}{3,4} = {r=2}{m},
    colsep=3pt
}
    {\ballnumber{7} \textbf{Phi Assignment:}\\$\{v_t:\tau_1,v_f:\tau_2\}, v:=\phi(v_t,v_f) \models \{v:\tau_1\vee\tau_2,v_t:\tau_1,v_f:\tau_2\}$} 
   & & \begin{adjustbox}{width=0.3\textwidth,keepaspectratio}
\begin{tikzpicture}[
    every node/.style={draw=none, align=left,font=\tiny\ttfamily},
    level distance=6mm,
    sibling distance=10mm,
    edge from parent/.style={draw},
    level 1/.style={level distance=6mm,sibling distance=21mm},
    level 2/.style={sibling distance=10mm},
]
\node {IF}
    child {node {IF\_ELEM}
      child {node {...}  edge from parent node[label,left] {cond}}
      child {node {STMT\_LIST}  
        child {node [fill=taintedsrc] {ASSIGN} 
            child {node [fill=taintedtgt] {\shortstack[l]{VAR\\name:$v$}} edge from parent node[label,left] {var}}
            child {node [fill=taintedsrc] {\shortstack[l]{VAR\\name:$v_t$}} edge from parent node[label,right] {expr}}
        }
      edge from parent node[label,right] {stmts}
      }
    }
    child {node {IF\_ELEM}
      child {node {null} edge from parent node[label,left] {cond}}
      child {node {STMT\_LIST}  
        child {node [fill=taintedsrc] {ASSIGN} 
            child {node [fill=taintedtgt] {\shortstack[l]{VAR\\name:$v$}} edge from parent node[label,left] {var}}
            child {node [fill=taintedsrc] {\shortstack[l]{VAR\\name:$v_f$}} edge from parent node[label,right] {expr}}
        }
      edge from parent node[label,right] {stmts}
      }
    };
\end{tikzpicture}
\end{adjustbox} & \begin{adjustbox}{width=0.15\textwidth,keepaspectratio}
\begin{tikzpicture}[
    every node/.style={draw=none, align=left,font=\tiny\ttfamily},
    level distance=6mm,
    sibling distance=10mm,
    edge from parent/.style={draw},
    level 1/.style={level distance=6mm,sibling distance=21mm},
    level 2/.style={sibling distance=10mm},
]
\node {IF}
    child {node {IF\_ELEM}
      child {node {...}  edge from parent node[label,left] {cond}}
      child {node {STMT\_LIST}  
        child {node [fill=taintedsrc] {ASSIGN} 
            child {node [fill=taintedtgt] {\shortstack[l]{VAR\\name:$v$}} edge from parent node[label,left] {var}}
            child {node [fill=taintedsrc] {\shortstack[l]{VAR\\name:$v_t$}} edge from parent node[label,right] {expr}}
        }
      edge from parent node[label,right] {stmts}
      }
    };
\end{tikzpicture}
\end{adjustbox}
   \\\begin{minipage}{0.3\columnwidth}
\begin{minted}{php}
 if (...) {
   $url = 'http://'.|\tikzmark{s2}{\colorbox{taintedsrc}{\$\_GET[\textquotesingle{}url\textquotesingle{}]}}|;
 } else {
   $url = |\tikzmark{s1}{\colorbox{taintedsrc}{\$\_GET[\textquotesingle{}full\textquotesingle{}]}}|;
 }
|\tikzmark{t}{\colorbox{taintedtgt}{\$url};}|
\end{minted}
    \begin{tikzpicture}[overlay,remember picture]
    \path (s1.west) edge[taintedarrow, -{Latex}] (t.east);
    \path ([xshift=6mm]s2.south) edge[ bend left=20, taintedarrow, -{Latex}] (t.east);
    \end{tikzpicture}
    \end{minipage} & \begin{minipage}{0.15\columnwidth}
\begin{minted}{php}
if (...) {
  $url = |\tikzmark{s1}{\colorbox{taintedsrc}{\$\_GET[\textquotesingle{}full\textquotesingle{}]}}|;
} 
|\tikzmark{t}{\colorbox{taintedtgt}{\$url};}|
\end{minted}
    \begin{tikzpicture}[overlay,remember picture]
    \path (s1.south) edge[taintedarrow, -{Latex}] (t.north);
    \end{tikzpicture}
    \end{minipage}
   & &\\
\end{tblr}
\end{minipage}
\begin{minipage}{0.54\linewidth}
    \begin{tblr}[]{
    hline{1,X,Z} = {solid},
    colspec={|X[4]X[5]|},
    rows={m,rowsep=0pt,font=\tiny},
    cell{1,3,5,7}{1} = {m,bg=gray!20,fg=black},
    cell{1,3,5,7}{2} = {r=2}{m},
    colsep=3pt
}
   {\ballnumber{9} \textbf{Array Initialization:} \\
   $\{v:\tau, max(a_i)=m\}, a:=[...,v] \models \{a:[...,m+1:\tau],v:\tau]\}$
    }
   & \begin{adjustbox}{width=0.45\textwidth,keepaspectratio}
\begin{minipage}{0.3\columnwidth}
\vspace{0pt}
    \begin{tikzpicture}[
    every node/.style={draw=none, rectangle, align=left,font=\tiny\ttfamily},
    level distance=6mm,
    edge from parent/.style={draw},
    sibling distance=10mm
]
\node {ASSIGN}
    child {node {\shortstack[l]{VAR\\name:$a$}} edge from parent node[label,left] {var}}
    child {node {ARRAY}
        child {node {ARRAY\_ELEM}
            child[fill=taintedtgt] {node {null}
            edge from parent node[label,left] {key}}
            child {node[fill=taintedsrc] {\shortstack[l]{VAR\\name:$v$}} edge from parent node[label,right] {value}}
        } edge from parent node[label,right] {expr}
    };
\end{tikzpicture}
\end{minipage}
\begin{minipage}{0.18\columnwidth}
\vspace{0pt}
\begin{tikzpicture}[
    every node/.style={draw=none, rectangle, align=left,font=\tiny\ttfamily},
    level distance=6mm,
    edge from parent/.style={draw},
    sibling distance=10mm
]
\node[fill=taintedtgt] {DIM}
    child {node[fill=taintedtgt] {\shortstack[l]{VAR\\name:$a$}}edge from parent node[label,left] {expr}}
    child {node[fill=taintedtgt] {\shortstack[l]{$m+1$}}edge from parent node[label,right] {dim}};
\end{tikzpicture}
\end{minipage}
\end{adjustbox}
   \\\begin{minipage}{0.5\columnwidth}
\begin{minted}{php}
|\tikzmark{t}{\$a}| = [1=>...,|\tikzmark{s1}{\colorbox{taintedsrc}{\$url}}|];|\tikzmark{t1}{\colorbox{taintedtgt}{\$a[2]}}|;
\end{minted}
        \begin{tikzpicture}[overlay,remember picture]
            \path (s1.north) edge[ bend left=10, taintedarrow, -{Latex}] (t1.north);
        \end{tikzpicture}
        \vspace{5pt}
     \end{minipage}
    &\\
    \SetHline{2-2}{dotted}
    {$\{v:\tau\}, a:=[c=>v, ...] \models \{a:[c:\tau,...],v:\tau\}$}
   & \begin{adjustbox}{width=0.48\textwidth,keepaspectratio}
\begin{minipage}{0.3\columnwidth}
\vspace{0pt}
    \begin{tikzpicture}[
    every node/.style={draw=none, rectangle, align=left,font=\tiny\ttfamily},
    level distance=6mm,
    edge from parent/.style={draw},
    sibling distance=10mm
]
\node {ASSIGN}
    child {node {\shortstack[l]{VAR\\name:$a$}} edge from parent node[label,left] {var}}
    child {node {ARRAY}
        child {node {ARRAY\_ELEM}
            child {node(cases) {\shortstack[l]{ $c$}} edge from parent node[label,left] {key}}
            child {node [fill=taintedsrc] {\shortstack[l]{VAR\\name:$v$}} edge from parent node[label,right] {value}}
        } edge from parent node[label,right] {expr}
    };
\end{tikzpicture}
\end{minipage}
\begin{minipage}{0.18\columnwidth}
\vspace{0pt}
\begin{tikzpicture}[
    every node/.style={draw=none, rectangle, align=left,font=\tiny\ttfamily},
    level distance=6mm,
    edge from parent/.style={draw},
    sibling distance=10mm
]
\node [fill=taintedtgt] {DIM}
    child {node[fill=taintedtgt] {\shortstack[l]{VAR\\name:$a$}} edge from parent node[label,left] {expr}}
    child {node[fill=taintedtgt](cases) {\shortstack[l]{$c$}} edge from parent node[label,right] {dim}};
\end{tikzpicture}
\end{minipage}
\end{adjustbox}
   \\\begin{minipage}{0.5\columnwidth}
\begin{minted}{php}
|\tikzmark{t2}{\$a}| = ["k"=>|\tikzmark{s2}{\colorbox{taintedsrc}{\$url}}|];|\tikzmark{t3}{\colorbox{taintedtgt}{\$a[\textquotesingle{}k\textquotesingle{}]}}|;
\end{minted}
        \begin{tikzpicture}[overlay,remember picture]
            \path (s2.north) edge[ bend left=10, taintedarrow, -{Latex}] (t3.north);
        \end{tikzpicture}
        \vspace{5pt}
     \end{minipage}
    &\\
    \SetHline{2-2}{dotted}
    {$\{v:\tau\}, a:=[v_k=>v, ...] \models \{a:[v_k:\tau,...],v:\tau\}$}
   & \begin{adjustbox}{width=0.45\textwidth,keepaspectratio}
\begin{minipage}{0.3\columnwidth}
\begin{tikzpicture}[
    every node/.style={draw=none, rectangle, align=left,font=\tiny\ttfamily},
    level distance=6mm,
    edge from parent/.style={draw},
    sibling distance=10mm
]
\node {ASSIGN}
    child {node {\shortstack[l]{VAR\\name:$a$}} edge from parent node[label,left] {var}}
    child {node {ARRAY}
        child {node {ARRAY\_ELEM}
            child {node {\shortstack[l]{VAR\\name:$v_k$}} edge from parent node[label,left] {key}}
            child {node [fill=taintedsrc] {\shortstack[l]{VAR\\name:$v$}}edge from parent node[label,right] {value}}
        } edge from parent node[label,right] {expr}
    };
\end{tikzpicture}
\end{minipage}%
\begin{minipage}{0.2\columnwidth}
\vspace{0pt}
\begin{tikzpicture}[
    every node/.style={draw=none, rectangle, align=left,font=\tiny\ttfamily},
    level distance=6mm,
    edge from parent/.style={draw},
    sibling distance=12mm
]
\node [fill=taintedtgt] {DIM}
    child {node[fill=taintedtgt] {\shortstack[l]{VAR\\name:$a$}} edge from parent node[label,left] {expr}}
    child {node[fill=taintedtgt](cases) {\shortstack[l]{VAR\\name:$v_k$}} edge from parent node[label,right] {dim}};
\end{tikzpicture}
\end{minipage}
\end{adjustbox}
   \\\begin{minipage}{0.5\columnwidth}
\begin{minted}{php}
|\tikzmark{t6}{\$a}| = [$key=>|\tikzmark{s4}{\colorbox{taintedsrc}{\$url}}|];|\tikzmark{t}{\colorbox{taintedtgt}{\$a[\$key]}}|;
\end{minted}
        \begin{tikzpicture}[overlay,remember picture]
            \path (s4.north) edge[ bend left=10, taintedarrow, -{Latex}] (t.north);
        \end{tikzpicture}
        \vspace{5pt}
     \end{minipage}
    &\\
    {\ballnumbertabsmall{10} \textbf{Element Assignment:}\\
   $\{v:\tau,max(a_i)=m\}, a[]:=v \models \{a:[...,m+1:\tau], v:\tau\}$
    }
   & \begin{adjustbox}{width=0.25\textwidth,keepaspectratio}
\begin{minipage}{0.3\columnwidth}
\vspace{0pt}
\begin{tikzpicture}[
    every node/.style={draw=none, rectangle, align=left,font=\tiny\ttfamily},
    level distance=6mm,
    edge from parent/.style={draw},
    sibling distance=12mm,
]
\node {ASSIGN}
    child {node {DIM}
        child {node [fill=taintedtgt] {\shortstack[l]{VAR\\name:$a$}} edge from parent node[label,left] {expr}}
        child {node (cases) {\shortstack[l]{null}} edge from parent node[label,right] {dim}} edge from parent node[label,left] {var}
    } 
    child {node [fill=taintedsrc] {\shortstack[l]{VAR\\name:$v$}} edge from parent node[label,right] {expr}
    };
\end{tikzpicture}
\end{minipage}
\end{adjustbox}
\hspace{0.05\textwidth}
\begin{adjustbox}{width=0.15\textwidth,keepaspectratio}
\begin{minipage}{0.18\columnwidth}
\vspace{0pt}
\begin{tikzpicture}[
    every node/.style={draw=none, rectangle, align=left,font=\tiny\ttfamily},
    level distance=6mm,
    edge from parent/.style={draw},
    sibling distance=10mm
]
\node[fill=taintedtgt] {DIM}
    child {node[fill=taintedtgt] {\shortstack[l]{VAR\\name:$a$}}edge from parent node[label,left] {expr}}
    child {node[fill=taintedtgt] {\shortstack[l]{$m+1$}}edge from parent node[label,right] {dim}};
\end{tikzpicture}
\end{minipage}
\end{adjustbox}
   \\\begin{minipage}{0.5\columnwidth}
\begin{minted}{php}

|\tikzmark{t1}{\$a}|[]=|\tikzmark{s1}{\colorbox{taintedsrc}{\$url}}|;|\tikzmark{t2}{\colorbox{taintedtgt}{\$a[m+1]}}|;
\end{minted}
        \begin{tikzpicture}[overlay,remember picture]
            \path (s1.south) edge[bend right=10, taintedarrow, -{Latex}] (t2.south);
        \end{tikzpicture}
        \vspace{13pt}
     \end{minipage}
   \\
\end{tblr}
\end{minipage}
\begin{minipage}{0.415\linewidth}
\begin{tblr}[]{
    hline{1,Z} = {solid},
    colspec={|X[2]X[2]|},
    rows={m,rowsep=0pt,font=\tiny},
    cell{1,3,5}{1} = {m,bg=gray!20,fg=black},
    cell{1,3,5}{2} = {r=2}{m},
    colsep=3pt
}
   {$\{v:\tau,a:[...]\}, a[c]:=v \models \{a:[...,c:\tau], v:\tau\}$}
   & \begin{tikzpicture}[
    every node/.style={draw=none, rectangle, align=left,font=\tiny\ttfamily},
    level distance=6mm,
    edge from parent/.style={draw},
    sibling distance=12mm,
]
\node {ASSIGN}
    child {node [fill=taintedtgt] {DIM}
        child {node [fill=taintedtgt] {\shortstack[l]{VAR\\name:$a$}} edge from parent node[label,left] {expr}}
        child {node [fill=taintedtgt] (cases) {\shortstack[l]{$c$} } edge from parent node[label,right] {dim}} edge from parent node[label,left] {var}
    } 
    child {node [fill=taintedsrc] {\shortstack[l]{VAR\\name:$v$}} edge from parent node[label,right] {expr}
    };
\end{tikzpicture}
   \\\begin{minipage}{0.5\columnwidth}
\begin{minted}{php}
|\tikzmark{t2}{\colorbox{taintedtgt}{\$a[1]}}|=|\tikzmark{s2}{\colorbox{taintedsrc}{\$url}}|; 
\end{minted}
        \begin{tikzpicture}[overlay,remember picture]
            \path (s2.south) edge[bend left=10, taintedarrow, -{Latex}] (t2.south);
        \end{tikzpicture}
        \vspace{4pt}
     \end{minipage}
   & \\\SetHline{2-2}{dotted}
   {$\{v:\tau,a:[...]\}, a[v_k]:=v \models \{a:[...,v_k:\tau], v:\tau\}$}
   & \begin{tikzpicture}[
    every node/.style={draw=none, rectangle, align=left,font=\tiny\ttfamily},
    level distance=6mm,
    edge from parent/.style={draw},
    sibling distance=12mm,
]
\node {ASSIGN}
    child {node[fill=taintedtgt] {DIM}
        child {node [fill=taintedtgt] {\shortstack[l]{VAR\\name:$a$}} edge from parent node[label,left] {expr}}
        child {node [fill=taintedtgt] (cases) {\shortstack[l]{VAR\\name:$v_k$}} edge from parent node[label,right] {dim}} edge from parent node[label,left] {var}
    } 
    child {node [fill=taintedsrc] {\shortstack[l]{VAR\\name:$v$}} edge from parent node[label,right] {expr}
    };
\end{tikzpicture}
   \\\begin{minipage}{0.5\columnwidth}
\begin{minted}{php}
|\tikzmark{t6}{\colorbox{taintedtgt}{\$a[\$key]}}|=|\tikzmark{s4}{\colorbox{taintedsrc}{\$url}}|;
\end{minted}
        \begin{tikzpicture}[overlay,remember picture]
            \path (s4.south) edge[bend left=10, taintedarrow, -{Latex}] (t6.south);
        \end{tikzpicture}
     \end{minipage}
   & \\       
\end{tblr}    
       \begin{tblr}[]{
    hline{1,Z} = {solid},
    colspec={|X[5]X[5]|},
    rows={m,rowsep=0pt,font=\tiny},
    cell{1,3,5}{1} = {m,bg=gray!20,fg=black},
    cell{1,3,5}{2} = {r=2}{m},
}
   {\ballnumbertabsmall{11} \textbf{Element Retrieval:}\\
   $\{a:[...,c:\tau]\}, v:=a[c] \models \{v:\tau,a:[...,c:\tau]\}$
    }
  & \begin{tikzpicture}[
    every node/.style={draw=none, rectangle, align=left,font=\tiny\ttfamily},
    level distance=6mm,
    edge from parent/.style={draw},
    sibling distance=13mm,
]
\node {ASSIGN}
    child {node[fill=taintedtgt] {\shortstack[l]{VAR\\name:$v$}} edge from parent node[label,left] {var}}
    child {node[fill=taintedsrc] {DIM}
        child {node[fill=taintedsrc] {\shortstack[l]{VAR\\name:$a$}} edge from parent node[label,left] {expr}}
        child {node[fill=taintedsrc] (cases) {\shortstack[l]{$c$}}edge from parent node[label,right] {dim}} edge from parent node[label,right] {expr}
    };
\end{tikzpicture}
  \\\begin{minipage}{0.5\columnwidth}
\begin{minted}{php}
|\tikzmark{t}{\colorbox{taintedtgt}{\$url}} =\tikzmark{s}{\colorbox{taintedsrc}{\$urls[0]}}|;
\end{minted}
        \begin{tikzpicture}[overlay,remember picture]
            \path (s.north) edge[ bend right=10, taintedarrow, -{Latex}] (t.north);
        \end{tikzpicture}
    \end{minipage}
   & \\
   \SetHline{2-2}{dotted}
   $\{a:[k:\tau,...],k\equiv v_k\}, v:=a[v_k] \models \{v:\tau,a:[k:\tau,...]=\}$
  & \begin{adjustbox}{width=0.45\textwidth,keepaspectratio}
\begin{tikzpicture}[
    every node/.style={draw=none, rectangle, align=left,font=\tiny\ttfamily},
    level distance=6mm,
    edge from parent/.style={draw},
    sibling distance=13mm,
]
\node {ASSIGN}
    child {node[fill=taintedtgt] {\shortstack[l]{VAR\\name:$v$}} edge from parent node[label,left] {var}}
    child {node[fill=taintedsrc] {DIM}
        child {node[fill=taintedsrc] {\shortstack[l]{VAR\\name:$a$}} edge from parent node[label,left] {expr}}
        child {node[fill=taintedsrc] {\shortstack[l]{VAR\\name:$v_k$}} edge from parent node[label,right] {dim}} edge from parent node[label,right] {expr}
    };
\end{tikzpicture}
\end{adjustbox}
  \\\begin{minipage}{0.5\columnwidth}
\begin{minted}{php}
$key1=$key;
|\tikzmark{t2}{\colorbox{taintedtgt}{\$url}} =\tikzmark{s3}{\colorbox{taintedsrc}{\$urls[\$key1]}}|;
\end{minted}
        \begin{tikzpicture}[overlay,remember picture]
            \path (s3.south) edge[ bend left =10, taintedarrow, -{Latex}] (t2.south);
        \end{tikzpicture}
    \end{minipage}
   & \\
\end{tblr}
\end{minipage}
\end{table*}

\textbf{Taint Propagation: Generalized Rules.}
When a statement is evaluated, the states of the corresponding variables are updated. Table~\ref{tab:propagation-var} (rules \ballnumber{1}-\ballnumber{7}) shows all the generalized taint propagation rules in our language abstraction used by \textsc{Artemis}.
Among all rules, those for variable assignment, type casting, and unary operations %
are consistent with those used in state-of-the-art approaches~\cite{1-phan, rips, tchecker, psalm, phpjoern} and require no modification.  %

Upon encountering branching, as seen in \texttt{if-else} structures, \textsc{Artemis} employs the phi assignment rule to merge the states originating from different branches. 
For composite conditions, e.g., \texttt{if(cond1 \&\& cond2)} and \texttt{switch-case} blocks, \textsc{Artemis} flattens them into a series of nested conditional statements with atomic conditions, e.g., \texttt{if(cond1)\{if(cond2)\}}.

During method calls, states transition from actual arguments to formal arguments, navigate through the statements within the method body, and eventually extend outward from the method call to the returned or yielded variable at the caller's site.
In the case of API calls, such as \mintinline{php}{$u = trim($_GET['q']}, where the implementation (e.g., \mintinline{php}{trim()}) is inaccessible to \textsc{artemis}, the propagation flow simplifies---\textsc{artemis} copies and merges the states from actual arguments, e.g., \mintinline{php}{$_GET['q']}, to the receiving variable, e.g., \mintinline{php}{$u}.

\newcommand{\ballnumbersmall}[1]{\tikz[baseline=(myanchor.base)] \node[circle,fill=.,inner sep=0.3pt,align=center] (myanchor) {\color{-.}\footnotesize #1};}

\textbf{Taint Propagation: Array-Specific Rules.}
We design propagation rules (\ballnumber{8}-\ballnumbersmall{11} in Table~\ref{tab:propagation-var}) for array-related operations specifically because prior approaches either tend to over-taint a whole array with one tainted element or overlook taint propagation on the arrays in the \texttt{foreach} loop. 

During array initialization, \textsc{Artemis} stores the state of all array elements in either $\texttt{arr}_s$ or $\texttt{arr}_r$ depending on the corresponding keys. 
When the key has a statically known value $c$, the state of the element is stored in $\texttt{arr}_s$ under the key $c$. 
When the value of the key can only be inferred at runtime, the element's state is stored in $\texttt{arr}_r$ under a symbolic key denoted by $v_k$. 
In cases where no explicit key is specified, we interpret the key as an integer which increments by one from the largest previously used numeric key. 
If all previous keys are statically known, i.e., $\texttt{arr}_r$ is empty, a concrete integer $k_c$ is computed based on existing numeric keys in $\texttt{arr}_s$, denoted by $k_c = 1+ \max\limits_{k \in \texttt{arr}_s} k$. The state of the corresponding element is stored in $\texttt{arr}_s$ under $k_c$. 
Otherwise, a new symbolic value $\max(a_i)+1$ is created as the key and the state of the corresponding element is stored in $\texttt{arr}_r$. 
This aligns with the definition of arrays in PHP.
In all cases, \texttt{self} is maintained to track whether all array elements have the same tainted state.

\begin{figure}[t]
\begin{subfigure}[t]{0.49\textwidth}
    \centering
\begin{minted}[fontsize=\tiny,escapeinside=||]{php}
$config = array(...
  'updateUrl'  => 'http://...', 
  );
...
$config[|\tikzmark{t1}{\colorbox{taintedtgt}{\textquotesingle{}download\textquotesingle{}}}|] = |\tikzmark{s1}{\colorbox{taintedsrc}{\$\_GET[\textquotesingle{}download\textquotesingle{}]}}|;
...
file_get_contents($config['updateUrl'], ...);
\end{minted}
\begin{tikzpicture}[overlay,remember picture]
\path (s1.north) edge[ bend right=10,-{Latex}, dataflow] (t1.north);
\end{tikzpicture}
    \caption{\rev{Array with both tainted and safe elements.}
    }
    \label{fig:array-fp}
\end{subfigure}
\begin{subfigure}[t]{0.49\textwidth}
    \centering
\begin{minted}[fontsize=\tiny,linenos,
startinline, 
breaklines=true, 
numbersep=1em,
escapeinside=||
]{php}
|\tikzmark{t1}{\colorbox{taintedtgt}{\$urls}}| = |\tikzmark{s1}{\colorbox{taintedsrc}{\$\_POST}}|[...];
...
foreach (|\tikzmark{s2}{\colorbox{taintedsrc}{\$urls}}| as |\tikzmark{t2}{\colorbox{taintedtgt}{\$imgUrl}}|) {
  ...
  readfile(|\tikzmark{t3}{\colorbox{taintedtgt}{\$imgUrl}}|, ...);
}
\end{minted}
\begin{tikzpicture}[overlay,remember picture]
  \path (s1.north) edge[ bend right=5,-{Latex}, dataflow] (t1.north);
  \path (s2.north) edge[ bend left=10,-{Latex}, dataflow] (t2.north);
  \path (t2.south) edge[ -{Latex}, dataflow] (t3.north);
\end{tikzpicture}
\caption{\rev{CVE-2022-40357 with tainted \texttt{foreach} loop.}
}
\label{fig:array-foreach}
\end{subfigure}
    \caption{\rev{Examples of array-specific taint propagation rules.  {\protect\proparrow} represents the taint data flow from \colorbox{taintedsrc}{source} to \colorbox{taintedtgt}{target}.}}
    \vspace{-5pt}
    \label{fig:array-rules-example}
\end{figure}

During array element assignment, \textsc{Artemis} propagates the state from the assigning variable to the corresponding array element. 
If the key is a statically known value $c$, the state of the element is stored in $\texttt{arr}_s$ under $c$. 
If the value of the key can only be inferred at runtime, a symbolic key $v_k$ using the name of the key variable is used to store the state of the corresponding element in $\texttt{arr}_r$. 
In cases where no explicit key is specified and the exact number of elements in the array is challenging to determine due to loops, a symbolic key $\max(a)$ is employed to store the state of the corresponding element in $\texttt{arr}_r$. \texttt{self} is also updated to track whether all array elements have the same tainted state. 
This design is based on the observation that when this syntax is used, developers care less about the exact value of the key, rather, they just want to append an element to the array. Additionally, when used inside loops, all elements have the same taint. Therefore, this approximation models the propagation of this syntax well. %

During array element retrieval (with a given key $k$), \textsc{Artemis} propagates the state from the corresponding array element to the receiving variable. 
\textsc{Artemis} first checks the array's \texttt{self} field. If \texttt{self} is either $\tau$ or $\mu$, indicating that all elements share the same state, the receiving variable adopts this state.
\blue{Otherwise, this array is partially tainted, we need to check the $\texttt{arr}_s$ and $\texttt{arr}_r$ fields.
If $k$ is statically known, we search for $k$ in $\texttt{arr}_s$, and the state of the receiving variable is updated to match the state of the element retrieved from $\texttt{arr}_s$.
If $k$ can only be inferred at runtime, 
we search in $\texttt{arr}_r$ for an exact match with $k$ and all of $k$'s aliases, where an exact match means $k$ has the same variable name and type with the array key.} The state is copied to the receiving variable only when a match is found. \rev{For example, in Figure~\ref{fig:array-fp}, at line \#1-3, the \mintinline{php}{$config} array is initialized with the key \texttt{updateUrl} and a constant string. Therefore, the value is untainted, i.e, $\texttt{self}=\mu, \texttt{arr}_s=[\texttt{updateUrl}=\mu]$. At line \#5, the array element with key \texttt{download} is assigned a tainted value, leading to \mintinline{php}{$config} being partially tainted, i.e, $\texttt{self}=\accentset{\halfapprox}{\tau}, \texttt{arr}_s=[\texttt{updateUrl}=\mu, \texttt{download}=\tau]$. At line \#7, since \texttt{self} is $\accentset{\halfapprox}{\tau}$ and \texttt{updateUrl} exists in $\texttt{arr}_s$, the result taint is correctly retrieved as $\mu$, i.e., not tainted. 
Existing tools~\cite{rips} mark the whole \mintinline{php}{$config} array as tainted after line \#5, therefore, the access to the element at line \#7 is also marked as tainted, leading to a false positive.
}

Upon encountering \texttt{foreach} loops, when all elements in the iterated array $a$ are tainted, i.e.,  $a:\tau$, or when the array is partially tainted (i.e., $a:\accentset{\halfapprox}{\tau}$) but there are no branching paths in the loop body, the created value variable will be considered tainted. Although this is a conservative approximation, it does not lead to false negatives in practice because in cases where a vulnerability path involves a \texttt{foreach} loop, the array being iterated is typically derived from a source variable or initialized within a loop, resulting in all elements having the same taint. \rev{For example, in Figure~\ref{fig:array-foreach}, at line \#1, \mintinline{php}{$urls} is retrieved from \mintinline{php}{$_POST}, which is a built-in source, i.e., $\texttt{self}_{\texttt{POST}}=\tau$. Therefore, \mintinline{php}{$urls} is considered tainted, i.e., $\texttt{self}_{\texttt{urls}}=\tau$. Inside the \texttt{foreach} loop at line \#3, \mintinline{php}{$imgUrl} is created by \mintinline{php}{$urls}, thus it is also tainted since $\texttt{self}_{\texttt{urls}}=\tau$. We model the \texttt{foreach} loop to avoid under-tainting, while existing tools~\cite{phpjoern,tchecker} fail to maintain taint propagation for variables created inside \texttt{foreach}, resulting in false negatives.
}

\textbf{Taint Continuity: Implicit Dataflow Reconstruction Rule. %
} 
In addition to applying the aforementioned taint propagation rules, we must also trace taint propagation in cases where data flows are implicitly present due to the dynamic features of PHP.
\rev{We manually go through the PHP documentation and discover that} the built-in \texttt{extract}~\cite{extract} function creates variables implicitly by extracting them from an array. \rev{Specifically, a variable is created implicitly for every key/value pair in the first array parameter.} The created variable names are derived from the array keys with a prefix specified in the third parameter. When an \texttt{extract} function is encountered, both the array and the prefix are recorded. If a variable is subsequently accessed without preceding data flow and the prefix matches, a synthetic assignment from the array is generated to explicitly establish the data flow. 
For example, between two consecutive statements \mintinline{php}{extract($_GET,...,'req');} and \mintinline{php}{return $req_url;}, \textsc{Artemis} interprets an intermediate hidden statement \mintinline{php}{$req_url=$_GET['url'];} to uncover the implicit data flow generated by the \texttt{extract} function.

\textbf{Taint
Clearance: Safety String Assurance Rule.}
To exploit an SSRF vulnerability, the tainted string must either be an arbitrary file URL fully controlled by an attacker or a request-sending URL where the attacker have full control over where the IP the request is sent. 
However, when tainted strings are concatenated with other strings to form a new string, the resulting string does not necessarily become a file URL or request-sending URL. 
Simply tainting the result if any of its components are tainted leads to over-tainting. 
\textsc{Artemis} applies safety string assurance rules to prevent such over-tainting cases where SSRF exploitation is no longer possible. In Section~\ref{subsec:fp-comp}, we demonstrate that these safety string assurance rules are crucial in reducing over-tainting and false positives. 

\blue{We take string concatenation in the form of \mintinline{php}{$v=$v1.$v2} without loss of generality to illustrate our rules. Other string manipulations, such as interpolation and formatting, can be converted to equivalent string concatenation form when handling safety strings. 
For example, string interpolation such as \mintinline{php}{$v="http://$domain"} is equivalent to \mintinline{php}{$v="http://".$domain}. 
After the concatenation step, for $T_{f}$ (taint for file URLs), we consider a variable $v$ to be untainted, represented as $T_{f_{v}}=(\mu, [], [])$, if $v1$ and $v2$ are not both tainted. This is because if any part of the string is not attacker-controlled, the concatenated file path cannot be arbitrary. 
For $T_{r}$ (taint for request-sending URLs), the taint status of $v1$ is first checked. If $v1$ is tainted as part of a request URL (i.e., $T_{f_{v1}}$ is set), then $v$ is also considered tainted. If $v1$ is not tainted, its value is evaluated. If $v1$ is a constant or literal string, we compare the value of $v1$ against valid URL schemes and also use a URL parser to parse $v1$. 
If $v1$ represents a valid scheme (e.g., \texttt{http} or \texttt{tcp}) or appending $v2$ could form a new host (e.g., \mintinline{php}{$v = "http://a" . $v2} and \mintinline{php}{$v2=".evil.com"}), the taint depends only on $v2$, because now $v2$ determines the IP that the request is sent to. Otherwise, $T_{r}$ remains untainted. 
This approach ensures that the tainted input affects the domain segment of the URL, controlling the IP address to which the request is sent as defined by the standard~\cite{rfc-url}. In other cases, the result follows conventional binary operation rules, where the result is tainted if any part of the string is tainted.
}

\subsection{False Positive Pruning}
\label{subsec:path-constraint}

\blue{The rule-based taint analysis described in Section~\ref{sec:taintanalysis} is path insensitive, which can lead to false positives by reporting infeasible SSRF propagation paths. This occurs when the code conditions always cannot be satisfied or reject URLs.}
For instance, in Figure~\ref{fig:fp-patterns-2}, we identify the tainted path from the source \mintinline{php}{downloadFile} at line \#8 to the sink \mintinline{php}{fopen} at line \#5. 
However, there is a \rev{conditional check} at line \#2 along the path. When the attacker manipulates \rev{\mintinline{php}{$d['filename']}} to be a URL, the user-defined function \mintinline{php}{isFilenameValid} always returns \texttt{false} because the URL contains a forward slash (/). It means that SSRF exploitation is impossible along the path, which should be flagged as a false positive. \rev{We refer to the conditional checks on the taint propagation paths as path conditions.}

\rev{To prune infeasible propagation paths, we perform a lightweight static analysis with a focus on SSRF-specific string conditions}. 
We start by extracting a consolidated list of conditions for each \rev{tainted path}. Next, we prune paths that contain conditions that are always unsatisfied or reject URLs before they reach the sensitive sink. \rev{By modeling and checking only conditions that are relevant in SSRF, we prune path-related false positives without expensive symbolic execution.} \blue{Section~\ref{subsec:fp-comp} shows that analyzing path conditions is important in reducing false positives.}

\begin{figure*}[t]
    \centering
    \setlength\fboxsep{1pt}
    \begin{subfigure}{0.45\textwidth}
        \centering
        \begin{minted}[fontsize=\tiny,escapeinside=@@]{php}
private function downloadFile($d) {
  if (@\colorbox{lightblue}{!\$this->isFilenameValid(\$d[\textquotesingle{}filename\textquotesingle{}])}@)
    throw ...;
  else {
    @\colorbox{darkorange!20}{fopen(\$d[\textquotesingle{}filename\textquotesingle{}], ...)}@; @\textcolor{darkorange}{\textbf{\slash{}\slash{}sink}}@
  }
}
$this->downloadFile($_REQUEST); 
        \end{minted}
    \end{subfigure}
    \begin{subfigure}{0.45\textwidth}
        \centering
        \begin{minted}[fontsize=\tiny,escapeinside=@@]{php}
public function isFilenameValid($f) {
  // ...
  foreach (['/', '\0'] as $char)
    if (@\colorbox{lightblue}{strpos(\$f, \$char) !== false}@)
      return false;
  // ...
  @\colorbox{darkorange!20}{return true}@; @\textcolor{darkorange}{\textbf{\slash{}\slash{}Expected return stmt}}@
}
        \end{minted}
    \end{subfigure}

\vspace{2mm}
    
\begin{subfigure}{0.45\textwidth}
    \centering
    \begin{tikzpicture}[node distance=1.5cm, auto,every node/.style={font=\tiny}]
  \node[rectangle, draw] (cond1) {\shortstack[l]{\textbf{Block\#1(@2)}\vspace{-1mm}\\$v_1$:=\texttt{\$d['k']}\vspace{-1mm}\\$v_2$:=\colorbox{lightblue}{\texttt{isFilenameValid()}}\\$v_3$:=\colorbox{lightblue}{!$v_2$}\\\texttt{JumpIf}( $v_3$)}};
  
  \node[rectangle, draw] (if) [left = 5mm of cond1] {\shortstack[l]{\textbf{Block\#2(@3)}\vspace{-1mm}\\\texttt{throw ...}}};
  
  \node[rectangle, draw] (else) [below = 3mm of cond1] {\shortstack[l]{\textbf{Block\#3(@5)}\\$v_4$:=\texttt{\$d['k']}\\\textcolor{darkorange}{\texttt{fopen(\$v4)}}\\\texttt{Jump}}};
  
  \node[rectangle, draw] (ret) [right = 5mm of else] {\shortstack[l]{\textbf{Block\#4(@7)}\vspace{-1mm}\\\texttt{return}}};

   \draw[arrows = -{Latex}] ($(cond1.west)-(0mm,0mm)$) -- node[xshift=0mm,yshift=0mm] {if} ($(if.east)-(-0mm,0mm)$);
   \draw[arrows = -{Latex}] (cond1.south) -- node[xshift=0mm,yshift=0mm] {else} ($(else.north)-(0mm,0mm)$);
    \draw[arrows = -{Latex}] (else.east) -- node[xshift=-1.5mm,yshift=-1mm]{} ($(ret.west)-(-0mm,0mm)$);
    
    \draw[red, dashed, arrows = -{Latex}] ($(else.north)-(4mm,0mm)$) -- ($(cond1.south)-(4mm,0mm)$);

\end{tikzpicture}
\caption{CFG of \texttt{downloadFile}}
\end{subfigure}
\begin{subfigure}{0.45\textwidth}
    \centering
    \begin{tikzpicture}[node distance=1.5cm, auto,every node/.style={font=\tiny}]
  \node[rectangle, draw] (forheader) {\shortstack[l]{\textbf{Block\#1(@3)}\vspace{-1mm}\\$v_1$:=\texttt{['/', '\textbackslash0']}\vspace{-1mm}\\\texttt{IterReset(\$v1)}\\\texttt{Jump}}};

  \node[rectangle, draw] (forcond)[right = 4mm of forheader] {\shortstack[l]{\textbf{Block\#2(@3)}\vspace{-1mm}\\$v_2$:=\texttt{IterValid}($v_1$)\vspace{-1mm}\\\texttt{JumpIf}($v_2$)}};
  
  \node[rectangle, draw] (ifcond) [below = 8mm of forcond] {\shortstack[l]{\textbf{Block\#3(@4)}\vspace{-1mm}\\$v_3$:=\texttt{IterValue}($v_2$)\vspace{-1mm}\\$v_4$:=\colorbox{lightblue}{\texttt{strpos}($f$, $v_3$)}\vspace{-1mm}\\$v_5$:=\colorbox{lightblue}{$v_4$!=\texttt{false}}\vspace{-1mm}\\\texttt{JumpIf}($v_5$)}};
  
  \node[rectangle, draw] (if) [left = 5mm of ifcond] {\shortstack[l]{\textbf{Block\#4(@5)}\vspace{-1mm}\\\texttt{return(false)}}};
  
  \node[rectangle, draw] (forelse) [right = 5mm of forcond] {\shortstack[l]{\textbf{Block\#5(@7)}\vspace{-1mm}\\\textcolor{darkorange}{\texttt{return(true)}}}};

   \draw[arrows = -{Latex}] ($(forheader.east)-(0mm,0mm)$) -- node[xshift=0mm,yshift=0mm] {} ($(forcond.west)-(-0mm,0mm)$);
   \draw[arrows = -{Latex}] ($(forcond.south)-(2mm,0mm)$) -- node[xshift=-5mm,yshift=0mm] {if} ($(ifcond.north)-(2mm,0mm)$);
   \draw[arrows = -{Latex}] (forcond.east) -- node[xshift=0mm,yshift=0mm] {else} ($(forelse.west)-(0mm,0mm)$);
   \draw[arrows = -{Latex}] ($(ifcond.north)+(2mm,0mm)$) -- node[xshift=6mm,yshift=0mm] {else} ($(forcond.south)+(2mm,0mm)$);
    \draw[arrows = -{Latex}] ($(ifcond.west)-(0mm,0mm)$) -- node[xshift=0mm,yshift=0mm]{if} ($(if.east)-(-0mm,0mm)$);

    \draw[red, dashed, arrows = -{Latex}] ($(forelse.west)-(0mm,1mm)$) -- ($(forcond.east)-(0mm,1mm)$);
    \draw[red, dashed, arrows = -{Latex}] ($(forcond.south)+(1mm,0mm)$) -- ($(ifcond.north)+(1mm,0mm)$);
    \draw[red, dashed, arrows = -{Latex}] ($(ifcond.north)-(1mm,0mm)$) -- ($(forcond.south)-(1mm,0mm)$);
\end{tikzpicture}
\caption{CFG of \texttt{isFilenameValid}}
\end{subfigure}
    \caption{An example of false positive caused by \colorbox{lightblue}{path condition}, with the source code block and the corresponding CFG in SSA form. 
{\protect\redarrowdashed} represents the (backward) control dependency edges. {\protect\blockarrow} represents the block linkage in a CFG.
}
\vspace{-5pt}
    \label{fig:fp-patterns-2}
\end{figure*}

\textbf{Path Condition Extraction.} 
Given a taint propagation path from source to sink (Section~\ref{sec:taintanalysis}), \textsc{Artemis} identifies all function invocations along the path, extracts path conditions from each function's control flow graph (CFG), and merges these conditions into a consolidated path condition list. 
For each function in a taint path, \textsc{Artemis} identifies the sink block containing the sink function invocation (leaf node in the call graph) or the end block containing a return statement (intermediate node in the call graph). 
Starting from the identified sink or end block, we conduct a backward dependency analysis by iteratively traversing preceding blocks until we reach either the entry block of the CFG or a previously visited block, which indicates the presence of a loop.
\textsc{Artemis} extracts condition labels (i.e., true/false) derived from dependence edges. 
\textsc{Artemis} connects the traversal flow using the constructed call graphs (Section~\ref{sec:cg}) whenever it encounters function calls. It merges the extracted conditions from each function in the call graph by applying the corresponding logical operators (AND/OR) according to the control flow dependencies.

We use the example in Figure~\ref{fig:fp-patterns-2} to illustrate how \textsc{Artemis} extracts interprocedural path conditions within a tainted path. 
Given a taint propagation path where the tainted source \mintinline{php}{$_REQUEST} is passed as an argument to the \mintinline{php}{downloadFile()} function at line \#8 and subsequently flows to the sink function \mintinline{php}{fopen()} at line \#5, \textsc{Artemis} identifies Block \#3 within the \mintinline{php}{downloadFile()} function as the tainted block where the sink function is invoked.
Starting from Block \#3, Artemis tracks back to Block \#1 with an edge labeled as \texttt{false}, indicating that: 1) a path condition is defined in Block \#1, 2) branching occurs at Block \#1, and 3) the path from Block \#1 to Block \#3 is taken when the condition is false. 
\textsc{Artemis} then retrieves the source code line number (i.e., 2) from Block \#1 to extract the condition's source code as \mintinline{php}{!$this->isFilenameValid($d['filename'])}.
\textsc{Artemis} considers the path label to normalize the path condition from Block \#1 to Block \#3 as \mintinline{php}{isFilenameValid()} returning true.
\textsc{Artemis} then locates the definition of \mintinline{php}{isFilenameValid()} function from the constructed call graphs and identifies Block \#5 as the target block, which is the only block that returns true.
Starting from Block \#5, Artemis tracks back to Block \#2 with an edge labeled as \texttt{false}. \textsc{Artemis} then retrieves the source code line number (i.e., 3) from Block \#2 to extract the condition's source code as \mintinline{php}{foreach (['/', '\0'] as $char)}, indicating that: 1) the path from Block \#2 to Block \#5 marks the end of a loop, and 2) any condition within the loop that would cause early termination must be negated; otherwise, the path from Block \#2 to Block \#5 would be infeasible.
From Block \#2, \textsc{Artemis} backtracks every path until it reaches Block \#2 again and identifies that branching and condition definition occur at Block \#3.   
\textsc{Artemis} then retrieves the source code line number (i.e., 4) from Block \#3 to extract the condition's source code as \mintinline{php}{strpos($f, $char) !== false}.
With negation, \textsc{Artemis} normalizes the path condition from Block \#2 to Block \#5 in \mintinline{php}{isFilenameValid()} function as \mintinline{php}{strpos($f, $char) == false}.
It is also the path condition from the source \mintinline{php}{$_REQUEST} to the sink \mintinline{php}{fopen()} in the \mintinline{php}{downloadFile()} function.

\begin{table}[t]
    \centering
    \caption{String checking functions and core translation rules. \texttt{s} represents the string from user input, \texttt{c} and \texttt{p} represents constants used in condition.}
    \label{tab:constraint}
    \begin{tblr} {
    hline{1,2,Z}={1pt},
    colspec = {X[2,l]X[1,l]X[3,l]},
    rows={m,rowsep=0pt,font=\scriptsize},
    row{1} = {font=\scriptsize\bfseries},
    row{even} = {bg=gray!10},
    }
Function  & Type & Translation \\
\texttt{in\_array(s,a)} & Allowlist & / \\
\texttt{array\_key\_exists(s,a)} & Allowlist & / \\
\texttt{strpos(s,c)} & URL-blocking & \texttt{(str.indexof s c 0)} \\
\texttt{stripos(s,c)} & URL-blocking &  \texttt{(str.indexof (str.to\_lower s) (str.to\_lower c) 0)} \\
\texttt{strstr(s,c)} & URL-blocking  & \texttt{(str.contains s c)} \\
\texttt{stristr(s,c)} & URL-blocking  & \texttt{(str.contains (str.to\_lower s) (str.to\_lower c))} \\
\texttt{preg\_match(p,s)} & URL-blocking  & \texttt{(str.in\_re s (re.from\_ecma2020 p))} \\ 
\texttt{preg\_match\_all(p,s)} & URL-blocking  & \texttt{(str.in\_re s (re.from\_ecma2020 p))} \\
    \end{tblr}
\end{table}

\textbf{Always \blue{Unsatisfied} Conditions.}
\textsc{Artemis} prunes unreachable paths after identifying that their conditions are always unsatisfied, including 1) parameter and argument mismatch; and 2) conflicting logic. 
When a condition depends on a function parameter but the passed argument at the call site does not match, it results in an unsatisfiable condition. \rev{We denote the condition on formal argument involving constants as $C(v_f, S_1)$ and the called method is $m$ with constant parameter $S_2$, then if: }
\[
\color{formula}
C(v_f,S_1) \land \text{call}(m, S_2 \to v_f) \equiv \bot 
\]
\rev{, the path is pruned.}
For example, if a function checks whether a parameter equals a specific value (e.g., \mintinline{php}{if($param === "A")}), but the passed argument is \mintinline{php}{"B"}, the condition is always false.
Conflicting logic arises when contradictory conditions exist, such as in the cases of \mintinline{php}{if ($x == 10)} and \mintinline{php}{if ($x != 10)}, which are mutually exclusive. \rev{Formally, for conditions $C_1(v)$ and $C_2(v)$ that checks variable $v$ with statically known conditions, if:}
\[
\color{formula}
C_1(v) \land C_2(v) \equiv \bot
\]
\rev{then the path is pruned.}

\textbf{URL Rejection Conditions.} \textsc{Artemis} prunes secure paths where attacker-controlled URLs cannot reach the sink due to path conditions that perform string checks. 
Specifically, two types of path conditions need to be identified: 1) allowlist checks, where tainted user input is restricted to a predefined list of values; and 2) string checks that block URLs from passing because URL-required characters are disallowed, such as \texttt{/} or \texttt{.}.
To accurately identify the two types of conditions, first, \textsc{Artemis} identifies potential conditions by locating related functions shown in Table~\ref{tab:constraint}. Allowlist checks are recognized directly by their function names, while URL-blocking string checks are modeled based on their logic. Conditions involving 1) external variables (e.g., class properties or globals) or 2) non-constant variables not directly derived from user input (e.g., from function calls) are excluded. Then, each string check is translated into a string constraint, with the core rules for this process outlined in Table~\ref{tab:constraint}. Lastly, \textsc{Artemis} uses the \textsc{OSTRICH} SMT solver~\cite{ostrich} to verify whether \rev{valid URLs in the form of \texttt{protocol://domain.tld}} can satisfy the constraints.
For example, in Figure~\ref{fig:fp-patterns-2}, the condition on line \#4 in \mintinline{php}{isFilenameValid} uses \mintinline{php}{strpos}, making it a candidate for URL rejection. The conditions are modeled as string constraints \texttt{(assert (= (str.indexof w char 0) (- 1)))} \rev{where \texttt{char} can be \texttt{/} and \texttt{\textbackslash0}. Constraint for valid URL \texttt{w} is modeled by regular expression. Then, we use the SMT solver to determine that the the union of the constraints are unsatisfiable for a valid URL. Therefore, we mark the path as a false positive.}

\section{Evaluation}
\label{sec:evaluation}
In this section, we present our experimental evaluation. \rev{We first evaluate the detection capability of \textsc{Artemis} against five generic static PHP vulnerability detection tools based on taint analysis. Next, we evaluate the generalizability of the source and sink identification module, the performance of the call graph construction module, and finally, the detection speed and scalability of \textsc{Artemis}.}
We have implemented a prototype of \textsc{Artemis}. 
The source/sink identification module is developed using the GPT-4o-2024-08-06 model~\cite{gpt4} with default configurations, accessed via API. The maximum output token length is set to 1024.
The call graph construction and rule-based taint analysis module are built on top of \textsc{Phan}~\cite{1-phan, 1-phan-plugin}, which utilizes the php-ast extension~\cite{phpast} to extract abstract syntax trees (ASTs). 
The false positive pruning module is built on top of the Joern framework~\cite{joern}.
All experiments are run on a system with an Intel i7-10700 CPU with 8 cores, 32GB RAM, and running 64-bit Ubuntu 22.04 with kernel version 5.15.0. The applications are set up using the PHP version recommended in their respective installation documentation.

\subsection{Methodology}
\subsubsection{Target Applications and Vulnerability Collection}
\label{subsec:apps}
We select target applications based on two criteria. First, we collect vulnerable applications from the Common Vulnerabilities and Exposures (CVE) database~\cite{cve} to evaluate whether \textsc{Artemis} is able to detect known SSRFs. Using keywords such as ``server-side request forgery'' and ``SSR'', we filter applications that meet the following conditions: 1) the application is written in PHP language and open-source; and 2) the CVE reports a true SSRF vulnerability in the application code. From this list, we prioritize frequently occurring applications and download both the vulnerable and latest versions for evaluation. We collect 55 applications with reported SSRFs.
Second, we gather popular open-source PHP applications from GitHub (via the Awesome-Selfhosted project~\cite{1-selfhost}) and the WordPress plugin repository~\cite{wp-plugins} to assess \textsc{Artemis}'s ability to discover new SSRFs. We sample a total of 195 applications. 
We use the latest available versions at the time of the experiment. 
In total, we collected 250 PHP applications using the two criteria. %
The applications have varying complexity, with line of code (LoC) ranging from 780 to 872506, with an average of 173049.

The collected SSRF CVEs cover a variety of root causes, including 1) missing URL validation, where user-provided URLs are used without validation; 2) missing URL segment validation, where user input modifies parts of the URL's domain without validation; 3) incomplete input validation, where basic checks are bypassed with attacker-controlled domains; and 4) flawed input validation, where allowlists are used but flawed, leading to bypasses.
SSRFs usually have severe impacts on running applications. The collected CVEs cause impacts including 1) access control bypass; 2) sensitive data leakage; 3) denial of service (DoS); 4) privilege escalation; and 5) arbitrary remote code execution. For example, the SSRF in Figure~\ref{fig:array-foreach} results in data leakage as the response of the crafted request is returned to the attacker. In contrast, the SSRF in Figure~\ref{fig:motivating} allows attackers to circumvent access controls to communicate with internal network hosts.   

\begin{table}[t]
    \centering
    \caption{Feature comparison of \textsc{Rips}, \textsc{Phpjoern}, \textsc{TChecker}, \textbf{Psalm}, \textsc{Phan}, and \textsc{Artemis}.%
    }
    \label{tool-comp}
    \begin{tblr}{
        hline{1,2,Z}={1pt},
        colspec = {X[-1,c]X[1,c]X[1,c]X[1,c]X[1,c]X[1,c]X[1,c]},
        rowsep=-2pt,
        rows={m, rowsep=0pt, font=\scriptsize},
        row{1} = {font=\footnotesize\bfseries},
        row{even} = {bg=gray!10},
        cell{even}{2,4,6} = {bg=gray!10!babyblueeyes!40},
        cell{odd}{2,4,6} = {bg=babyblueeyes!10},
        colsep=0pt,
    }
    Feature                     & \textsc{Rips} & \textsc{Phpjoern} & \textsc{TChecker} & \textsc{Psalm} & \textsc{Phan} & \textsc{Artemis} \\
    \textbf{Sources/Sinks}               & PHP built-in  & PHP built-in      & PHP built-in      & PHP built-in  & PHP built-in  & PHP built-in and third-party \\
    \textbf{Explicit Calls}              & Unique function name matching & Unique function/method name matching & Type inference & Type inference & Type inference & Type inference \\
    \textbf{Implicit Calls}              & \xmark       & \xmark           & \xmark           & \xmark       & \xmark       &  Relaxed implicit call graph construction \\
    \textbf{Taint Propagation}     & Generalized rules, sanitizer functions & Generalized rules, sanitizer functions & Generalized rules, sanitizer functions & Generalized rules, sanitizer functions & Generalized rules, sanitizer functions & Generalized rules, refined array rules, implicit dataflow rules, and safety string analysis \\
    \textbf{False Positive Pruning}    & \xmark & \xmark        & \xmark & \xmark & \xmark & Path condition analysis \\
    \end{tblr}
    \vspace{-5pt}
\end{table}

\subsubsection{Alternative Approaches}
\label{subsec-alternate}
We compare \textsc{Artemis} with five generic static PHP vulnerability detection tools based on taint analysis, i.e., \textsc{Rips}~\cite{rips},  \textsc{TChecker}~\cite{tchecker}, \blue{\textsc{phpjoern}~\cite{phpjoern}, \textsc{psalm}~\cite{psalm} and \textsc{Phan}~\cite{1-phan-plugin}}. We summarize their analysis features and compare them with \textsc{Artemis} in Table~\ref{tool-comp}.
To tune the five generic tools for SSRF detection, we configure them with the same PHP built-in sources and sinks as \textsc{Artemis} to ensure a fair comparison.
For call graph construction, \textsc{Rips} matches only function names, while \textsc{Phpjoern} matches unique function and method names. \textsc{TChecker}, \textsc{Psalm}, and \textsc{Phan} employ type inference for explicit calls, but none of the five tools handle implicit calls. \textsc{TChecker} starts call graph construction and taint analysis from the top-level function of each PHP file, while the other tools start their analysis from each defined function.
In taint propagation, the five tools use generalized taint rules (rules \ballnumber{1}-\ballnumber{7} in Table~\ref{tab:propagation-var}) same as \textsc{Artemis} with manually defined sanitizers for non-SSRF vulnerabilities, such as SQL injection and XSS. \textsc{Rips} ignores object-oriented features, therefore if the right-hand side includes objects, the taint is not propagated for rules \ballnumber{1}-\ballnumber{4} in Table~\ref{tab:propagation-var} %
None of the five tools prunes false positives with path condition analysis as \textsc{Artemis} does.

To assess the impact of the third-party sources and sinks, we extend all alternative approaches by incorporating both built-in and third-party sources and sinks identified by \textsc{Artemis}, resulting in five modified tools: \textsc{Rips$^{*}$}, \textsc{TChecker$^{*}$}, \textsc{phpjoern$^{*}$}, \textsc{psalm$^{*}$}, and \textsc{Phan$^{*}$}.

To assess the impact of the false positive pruning module in \textsc{Artemis}, we integrate it into \textsc{TChecker}, \textsc{Psalm} and \textsc{Phan} which have type inference support. We refer to these modified versions as \textsc{TChecker}$^{\dag}$, \textsc{Psalm}$^{\dag}$, and \textsc{Phan}$^{\dag}$. %

\subsubsection{Ablated Versions.}
To evaluate the contribution of each component in \textsc{Artemis}, we conduct an ablation study. We systematically remove one module from taint analysis at a time to assess the module's impact on detection performance. We compare \textsc{Artemis} against four ablated versions.
 
\textbf{$\textsc{Artemis}^{a}$}. We remove the \textit{third-party source/sink identification} module from \textsc{Artemis}. Specifically, we only use the PHP built-in functions as the source and sink functions. 

\textbf{$\textsc{Artemis}^{c}$}. 
We remove the \textit{statically inferred call graph construction} module from \textsc{Artemis}. Specifically, we only use the explicit call graph construction in \textsc{Phan}.

\textbf{$\textsc{Artemis}^{t}$}. We remove the \textit{rule-based taint analysis} module from \textsc{Artemis}. Specifically, we use the taint propagation rules in \textsc{Phan}. 

\textbf{$\textsc{Artemis}^{p}$}. We remove the \textit{false positive pruning} module from \textsc{Artemis}. The results from the \textit{rule-based taint analysis} module are directly reported as candidate SSRFs.

\subsubsection{Exploit Generation}\label{subsec:exploit}
Creating exploits manually from taint analysis reports is a time-consuming process requiring domain expertise.
We automate the exploit generation process by leveraging LLM with a multi-turn conversation to generate exploits. The conversation prompt template is shown in Appendix~\ref{subsec:exploit-prompt}.

Due to context limitations in LLMs~\cite{context-window}, we equip them with tools for dynamic context retrieval like code search, extraction, execution, and database queries~\cite{1-prompt-engineering}. When additional context is required, the LLM generates tool calls that convert to PHP function calls that gather information and return it to the LLM. \rev{In our implementation, we choose GPT-4o as the backbone LLM because 1) it has a relatively large context window; and 2) it has best built-in support for tool calling.} 

\blue{The exploit generation is complex, therefore we break it down into three subtasks:}

\blue{\textbf{Payload Generation.} The LLM identifies the user input type (e.g., GET/POST data, cookies) and generates the payload needed to trigger SSRF. For example, it may generate a POST request with the key \texttt{url} containing \texttt{blogspot} in value for our motivating example in Figure~\ref{fig:motivating}.}

\blue{\textbf{Route Formation.} The LLM determines the URL route to reach the vulnerable code by analyzing routing code or framework knowledge. For instance, in Figure~\ref{fig:related-fn}, the LLM identifies the correct route to the vulnerable function via domain knowledge about the CakePHP~\cite{1-cakephp} framework it uses.}

\blue{\textbf{Value Inference.} The LLM infers additional required field values, such as valid user IDs, which are necessary for the request to be accepted. For example, in Figure~\ref{fig:related-fn}, the LLM needs to fetch a valid product associated with a manufacturer for the request to be accepted.}

\blue{To enhance accuracy, we automate exploit testing on a live server, logging responses and execution traces. If an exploit fails, the feedback is used to refine it iteratively with the LLM. Unresolved cases after three attempts are marked as undetermined and reviewed manually. During manual review, we revise the payload, route, and fields generated by the LLM. We then test the corrected exploit on a server to confirm its validity as a true or false positive.
}

\subsubsection{Result Validation}
If a path is automatically exploited by the LLM, it is marked as a true positive (TP). Otherwise, we manually verify if it is a TP or a false positive (FP) and write exploitation payloads for TPs. For reports from alternative methods, we check for overlap with TP reports from \textsc{Artemis}. Overlapping reports are marked as TPs, while non-overlapping ones undergo manual code review to identify false positives. 

Note that multiple true positive paths can share the same patch, so they are reported as a single vulnerability, a common practice in CVE reporting. For example, in CVE-2018-1000138, the function \mintinline{php}{getFromWeb} is used to fetch web resources with three different sources, creating three true positive paths. However, as the patch is applied in the \mintinline{php}{getFromWeb} function, the CVE report combines all three paths under a single CVE.

\subsection{SSRF Detection Results}
\label{subsec:compare}

\newcommand{\amark}{\ding{51}$^{*}$}%
\begin{table*}[tbp]
\caption{The comparision of detected number of TP paths by \textsc{Artemis}, \textsc{Rips}, \textsc{Phpjoern}, \textsc{TChecker}, \textsc{Psalm}, and \textsc{Phan} on 106 SSRF vulnerabilities (71 known and \colorbox{green!5}{35 new}). 
\xmark ~means no TP path is detected for the corresponding vulnerability. 
}
\label{tab:new-vul}
\begin{adjustbox}{width=\textwidth,totalheight=0.75\textheight,keepaspectratio}
\begin{minipage}[t]{0.49\linewidth}
\vspace{0pt}
\NewColumnType{I}[1][]{>{\ifnum \value{rownum}=1 \relax\else \number\numexpr\value{rownum}-1\relax\fi}Q[co=1,#1]}
\begin{tblr}[]{
    hline{1,2,Z} = {1pt},
    colspec={I[0.5,c]X[6,c]X[-1,c]X[-1,c]X[-1,c]X[-1,c]X[-1,c]X[-1,c]},
    rows={m,font=\tiny},
    row{1} = {m,font=\scriptsize\bfseries},
    rowsep=-2pt,
    colsep=5pt,
    row{even} = {bg=gray!10},
}
\textbf{\#} & \textbf{CVE/Vuln ID}  & \rot[20]{\textbf{\textsc{Artemis}}}  & \rot[20]{\textbf{\textsc{Rips}}} & \rot[20]{\textbf{\textsc{Phpjoern}}}  & \rot[20]{\textbf{\textsc{TChecker}}} & \rot[20]{\textbf{\textsc{psalm}}} &   \rot[20]{\textbf{\textsc{Phan}}} \\ 
  & CVE-2015-7816    & 1    & 1       &\xmark    &\xmark    &1        & 1      \\
  & CVE-2016-10926   & 1    & 1       & 1        & 1        &1        & 1      \\
  & CVE-2016-10927   & 1    & 1       & 1        & 1        &1        & 1      \\
  & CVE-2016-7964    & 1    &\xmark   &\xmark    &\xmark    &1        & 1      \\
  & CVE-2016-9417    & 2    &\xmark   &\xmark    &\xmark    &2        & 2      \\
  & CVE-2017-1000419 & 1    &\xmark   &\xmark    &\xmark    &\xmark   &\xmark  \\
  & CVE-2017-10973   & 1    & 1       &\xmark    & 1        &1        & 1      \\
  & CVE-2017-14323   & 1    & 1       & 1        & 1        &1        & 1      \\
  & CVE-2017-16870   & 1    &\xmark   & 1        &\xmark    &1        & 1      \\
  & CVE-2017-7566    & 1    &\xmark   &\xmark    &\xmark    &\xmark   &\xmark  \\
  & CVE-2017-9307    & 3    & 3       & 3        & 3        &3        & 3      \\
  & CVE-2018-1000138 & 3    & 3       & 3        & 3        &3        & 3      \\
  & CVE-2018-11031   & 1    &\xmark   &\xmark    &\xmark    &\xmark   &\xmark  \\
  & CVE-2018-14514   & 6    &\xmark   &\xmark    &\xmark    &\xmark   &\xmark  \\
  & CVE-2018-14728   & 2    & 2       & 2        & 2        &2        & 2      \\
  & CVE-2018-15495   & 2    & 2       & 2        & 2        &2        & 2      \\
  & CVE-2018-16444   & 1    &\xmark   &\xmark    & 1        &1        & 1      \\
  & CVE-2018-18867   & 2    & 2       & 2        & 2        &2        & 2      \\
  & CVE-2018-6029    & 3    &\xmark   &\xmark    &\xmark    &\xmark   &\xmark  \\
  & CVE-2018-9302    & 4    & 4       & 4        & 4        &4        & 4      \\
  & CVE-2019-11565   & 3    &\xmark   &\xmark    &\xmark    &\xmark   &\xmark  \\
  & CVE-2019-11574   & 4    &\xmark   &\xmark    &\xmark    &4        & 4      \\
  & CVE-2019-11767   & 1    &\xmark   &\xmark    &\xmark    &\xmark   &\xmark  \\
  & CVE-2019-12161   & 7    &\xmark   &\xmark    &\xmark    &\xmark   &\xmark  \\
  & CVE-2019-15033   & 1    &\xmark   &\xmark    &\xmark    &\xmark   &\xmark  \\
  & CVE-2019-15494   & 1    &\xmark   &\xmark    &\xmark    &\xmark   &\xmark  \\
  & CVE-2020-10212   & 3    & 3       & 3        & 3        &3        & 3      \\
  & CVE-2020-10791   & 1    &\xmark   &\xmark    &\xmark    &\xmark   &\xmark  \\
  & CVE-2020-14044   & 1    &\xmark   & 1        &\xmark    &1        & 1      \\
  & CVE-2020-20341   & 2    &\xmark   &\xmark    &\xmark    &\xmark   &\xmark  \\
  & CVE-2020-20582   & 4    &\xmark   &\xmark    &\xmark    &4        & 4      \\
  & CVE-2020-21788   & 2    &\xmark   &\xmark    &\xmark    &\xmark   &\xmark  \\
  & CVE-2020-23534   & 1    &\xmark   &\xmark    &\xmark    &\xmark   &\xmark  \\
  & CVE-2020-24063   & 1    &\xmark   & 1        & 1        &1        & 1      \\
  & CVE-2020-25466   & 2    & 2       &\xmark    &\xmark    &\xmark   &\xmark  \\
  & CVE-2020-28043   & 1    &\xmark   &\xmark    &\xmark    &\xmark   &\xmark  \\
  & CVE-2020-28976   & 1    & 1       & 1        & 1        &1        & 1      \\
  & CVE-2020-28977   & 1    &\xmark   & 1        & 1        &1        & 1      \\
  & CVE-2020-28978   & 1    &\xmark   & 1        & 1        &1        & 1      \\
  & CVE-2020-35313   & 3    & 2       & 2        &\xmark    &2        & 2      \\
  & CVE-2020-35970   & 1    &\xmark   &\xmark    &\xmark    &1        & 1      \\
  & CVE-2021-24150   & 2    &\xmark   &\xmark    &\xmark    &\xmark   &\xmark  \\
  & CVE-2021-24371   & 2    &\xmark   &\xmark    &\xmark    &\xmark   &\xmark  \\
  & CVE-2021-27329   & 2    &\xmark   &\xmark    &\xmark    &\xmark   &\xmark  \\
  & CVE-2021-28060   & 1    &\xmark   &\xmark    &\xmark    &1        & 1      \\
  & CVE-2021-4075    & 3    &\xmark   &\xmark    &\xmark    &\xmark   & 3      \\
  & CVE-2022-0768    & 3    &\xmark   &\xmark    &\xmark    &\xmark   &\xmark  \\
  & CVE-2022-1037    & 1    &\xmark   &\xmark    &\xmark    &\xmark   &\xmark  \\
  & CVE-2022-1191    & 1    & 1       & 1        & 1        &1        & 1      \\ 
  & CVE-2022-1213    & 1    & 1       & 1        & 1        &1        & 1      \\
  & CVE-2022-1239    & 1    &\xmark   &\xmark    &\xmark    &1        & 1      \\
  & CVE-2022-31386   & 4    & 3       & 4        & 4        &4        & 4      \\
  & CVE-2022-31830   & 1    & 1       & 1        &\xmark    &1        & 1      \\ 
\end{tblr}
\end{minipage} \hfill
\begin{minipage}[t]{0.5\linewidth}
\NewColumnType{J}[1][]{>{\ifnum \value{rownum}=1\relax\else \ifnum\value{rownum}=55\relax\else \number\numexpr\value{rownum}-1+53\relax\fi\relax\fi}Q[co=1,#1]}
\vspace{0pt}
\begin{tblr}[]{
    hline{1,2,Z} = {1pt},
    colspec={J[0.5,c]X[5.5,c]X[-1,c]X[-1,c]X[-1,c]X[-1,c]X[-1,c]X[-1,c]},
    rows={m,font=\tiny},
    row{1} = {m,font=\scriptsize\bfseries},
    row{20-54} = {m,bg=green!5,fg=black},
    row{55} = {font=\scriptsize\bfseries},
    rowsep=-2pt,
    colsep=5pt,
    row{even[2-20]} = {bg=gray!10},
    row{odd[20-54]} = {m,bg=green!5,fg=black},
    row{even[20-54]} = {m,bg=gray!10!green!15,fg=black},
}
\textbf{\#} & \textbf{CVE/Vuln ID}  & \rot[20]{\textbf{\textsc{Artemis}}}  & \rot[20]{\textbf{\textsc{Rips}}} & \rot[20]{\textbf{\textsc{Phpjoern}}}  & \rot[20]{\textbf{\textsc{TChecker}}} & \rot[20]{\textbf{\textsc{psalm}}} &   \rot[20]{\textbf{\textsc{Phan}}} \\ 
  & CVE-2022-38292 & 1    &    \xmark    &\xmark    &\xmark    &\xmark  &\xmark  \\
  & CVE-2022-40357 & 1    &    \xmark    &\xmark    &\xmark    &\xmark  &\xmark  \\  
  & CVE-2022-41477 & 1    &     1        & 1        & 1        & 1      & 1      \\  
  & CVE-2022-41497 & 3    &     2        & 3        & 3        & 3      & 3      \\  
  & CVE-2022-46998 & 1    &    \xmark    &\xmark    &\xmark    & 1      & 1      \\  
  & CVE-2022-47872 & 4    &    \xmark    &\xmark    &\xmark    & 4      & 4      \\  
  & CVE-2023-1938  & 1    &    \xmark    &\xmark    &\xmark    &\xmark  &\xmark  \\  
  & CVE-2023-1977  & 1    &    \xmark    &\xmark    &\xmark    &\xmark  &\xmark  \\  
  & CVE-2023-2927  & 6    &    \xmark    &\xmark    &\xmark    &\xmark  & 6      \\  
  & CVE-2023-34959 & 1    &    \xmark    &\xmark    &\xmark    & 1      & 1      \\
  & CVE-2023-3744  & 1    &     1        & 1        & 1        & 1      & 1      \\  
  & CVE-2023-39108 & 2    &    \xmark    & 2        &\xmark    & 2      & 2      \\  
  & CVE-2023-39109 & 2    &    \xmark    & 2        &\xmark    & 2      & 2      \\  
  & CVE-2023-39110 & 1    &    \xmark    & 1        &\xmark    & 1      & 1      \\  
  & CVE-2023-40969 & 1    &    \xmark    &\xmark    &\xmark    &\xmark  &\xmark  \\  
  & CVE-2023-41054 & 1    &    \xmark    &\xmark    &\xmark    & 1      & 1      \\  
  & CVE-2023-41055 & 1    &    \xmark    &\xmark    &\xmark    & 1      & 1      \\  
  & CVE-2023-4651  & 1    &     1        &\xmark    &\xmark    & 1      & 1      \\
  & BMLT-1         & 4    &    \xmark    &\xmark    &\xmark    &\xmark  &\xmark  \\
  & BN-1           & 1    &    \xmark    &\xmark    &\xmark    &\xmark  &\xmark  \\  
  & CHL-1          & 2    &    \xmark    &\xmark    &\xmark    &\xmark  &\xmark  \\ 
  & COL-1          & 1    &    \xmark    &\xmark    &\xmark    &\xmark  &\xmark  \\  
  & CR-1           & 10   &    \xmark    &\xmark    &\xmark    &\xmark  &\xmark  \\  
  & CVE-2023-38515 & 1    &     1        &\xmark    &\xmark    & 1      & 1      \\  
  & CVE-2023-46725 & 1    &    \xmark    &\xmark    &\xmark    &\xmark  &\xmark  \\  
  & CVE-2023-46730 & 1    &    \xmark    &\xmark    &\xmark    &\xmark  &\xmark  \\  
  & CVE-2023-46736 & 1    &    \xmark    &\xmark    &\xmark    &\xmark  &\xmark  \\  
  & CVE-2023-48005 & 2    &     2        & 2        & 2        & 2      & 2      \\  
  & CVE-2023-48006 & 1    &    \xmark    &\xmark    &\xmark    &\xmark  &\xmark  \\  
  & CVE-2023-4878  & 1    &     1        & 1        & 1        & 1      & 1      \\  
  & CVE-2023-49159 & 2    &    \xmark    &\xmark    &\xmark    &\xmark  &\xmark  \\  
  & CVE-2023-49746 & 2    &    \xmark    &\xmark    &\xmark    &\xmark  &\xmark  \\  
  & CVE-2023-50374 & 1    &    \xmark    &\xmark    &\xmark    &\xmark  &\xmark  \\  
  & CVE-2023-50621 & 4    &    \xmark    &\xmark    &\xmark    &\xmark  &\xmark  \\  
  & CVE-2023-50622 & 1    &    \xmark    &\xmark    &\xmark    & 1      & 1      \\  
  & CVE-2023-51676 & 4    &    \xmark    &\xmark    &\xmark    &\xmark  &\xmark  \\  
  & CVE-2023-52233 & 1    &    \xmark    &\xmark    &\xmark    &\xmark  &\xmark  \\  
  & CVE-2023-5798  & 1    &    \xmark    &\xmark    &\xmark    &\xmark  &\xmark  \\  
  & CVE-2023-5877  & 1    &     1        & 1        & 1        & 1      & 1      \\  
  & CVE-2024-22134 & 3    &    \xmark    &\xmark    &\xmark    &\xmark  &\xmark  \\  
  & CVE-2024-32430 & 1    &    \xmark    &\xmark    &\xmark    &\xmark  &\xmark  \\
  & CVE-2024-33629 & 2    &    \xmark    &\xmark    &\xmark    &\xmark  &\xmark  \\
  & CVE-2024-35633 & 1    &    \xmark    &\xmark    &\xmark    &\xmark  &\xmark  \\
  & CVE-2024-35635 & 1    &    \xmark    &\xmark    &\xmark    &\xmark  &\xmark  \\
  & CVE-2024-35637 & 1    &    \xmark    &\xmark    &\xmark    &\xmark  &\xmark  \\
  & CVE-2024-37098 & 2    &    \xmark    &\xmark    &\xmark    &\xmark  &\xmark  \\
  & CVE-2024-38791 & 1    &    \xmark    &\xmark    &\xmark    &\xmark  &\xmark  \\
  & IC-1           & 4    &    \xmark    &\xmark    &\xmark    &\xmark  &\xmark  \\
  & MAC-1          & 6    &    \xmark    &\xmark    &\xmark    &\xmark  &\xmark  \\
  & NONE-1         & 3    &    \xmark    &\xmark    &\xmark    &\xmark  &\xmark  \\
  & FR-1           & 2    &    \xmark    &\xmark    &\xmark    &\xmark  &\xmark  \\
  & PL-1           & 3    &    \xmark    &\xmark    &\xmark    &\xmark  &\xmark  \\
  & SI-1           & 2    &    \xmark    &\xmark    &\xmark    &\xmark  &\xmark  \\
  \SetCell[c=2]{c} \textbf{Total Detected Path\#}  &&207 &45 &51 & 43  & 79 &88 \\
\end{tblr}
\end{minipage}
\end{adjustbox}
\vspace{-5pt}
\end{table*}

\subsubsection{True Positives}
\label{subsec:tp-comp} 
Table~\ref{tab:new-vul} and Table~\ref{tab:fp-compare} present the detection accuracy results for \textsc{Artemis}, \textsc{Rips}, \textsc{Phpjoern}, \textsc{TChecker}, \textsc{Psalm}, \textsc{Phan}, and ablated versions of \textsc{Artemis}. Overall, \textsc{Artemis} significantly outperforms other static analysis tools, detecting up to \textbf{5} times more SSRFs. In total, \textsc{Artemis} identifies \textbf{207} true positive paths, corresponding to \textbf{106} true positive SSRFs. Among the 106 SSRFs, \textbf{35} are newly discovered by \textsc{Artemis}. \rev{\textsc{Artemis} generates exploits for 194 of 222 detected paths, covering 99 of 106 SSRFs. Among the 28 remaining paths, 15 are confirmed as false positives and 13 as vulnerable. }
We show the detected vulnerabilities' affected application with versions, root causes and system impacts in Table~\ref{tab:cve-app} of Appendix~\ref{subsec:vul-detect-details}. %
\begin{wrapfigure}{R}{0.5\textwidth}
\definecolor{Red}{HTML}{E74C3C}
\definecolor{Blue}{HTML}{2980B9}
\definecolor{Green}{HTML}{27AE60}
\definecolor{Purple}{HTML}{8E44AD}
\definecolor{Orange}{HTML}{E67E22}
    \centering
\begin{minted}[fontsize=\tiny,escapeinside=||]{php}
class Request {
 public function |\colorbox{Green!50}{getData}|(...) {} |\textcolor{Green}{\textbf{\slash{}\slash{}PHPJoern ignores}}|
}                                  |\textcolor{Green}{\textbf{\slash{}\slash{}calls to getData}}|
                                   |\textcolor{Green}{\textbf{\slash{}\slash{}due to duplicate}}|
class PdfWriterService {           |\textcolor{Green}{\textbf{\slash{}\slash{}function names in }}|
 public function |\colorbox{Green!50}{getData}|(...) {} |\textcolor{Green}{\textbf{\slash{}\slash{}different classes}}| 
}

class ApiController { |\textcolor{Blue}{\textbf{\slash{}\slash{}TChecker skips updateProducts due}}|
 |\colorbox{Blue!50}{public function updateProducts()}||\textcolor{Blue}{\textbf{\slash{}\slash{}to missing call path}}|
 {       |\textcolor{Purple}{\textbf{\slash{}\slash{}Unknown type to Phan and Psalm}}| 
  $this->|\colorbox{Purple!50}{Product}|=$this->getTable()->get('Products'); 
  $productsData=$this->getRequest()->|\colorbox{Green!50}{getData}|('data.data');
  $products=[]; |\textcolor{Blue}{\textbf{\slash{}\slash{}foreach is not modeled by TChecker}}|
  |\colorbox{Blue!50}{foreach}|($productsData as $product){
   $manufacturerIsOwner=$this->Product->find(...)->count();
   if (!$manufacturerIsOwner) {
    throw ...;
   }
   $products[]=[$product['image']];
  }      |\textcolor{Purple}{\textbf{\slash{}\slash{}Unknown type to Phan and Psalm}}|
  $this->|\colorbox{Purple!50}{Product}|->|\colorbox{Red!50}{changeImage}|($products); 
 }                  |\textcolor{Red}{\textbf{\slash{}\slash{}Method call ignored by Rips}}|
}

class ProductsTable {
 // $products propagates to sink
 public function changeImage($products) {...}
}
\end{minted}
    \caption{A newly found CVE-2023-46725 has been confirmed by the developer. 
    \textsc{rips}, \textsc{phpjoern}, \textsc{tchecker}, \textsc{psalm}, and \textsc{phan} all fail to detect the vulnerability.}
    \label{fig:related-fn}
\end{wrapfigure}

\textbf{Comparison with \textsc{Rips}es.}
\textsc{Rips} produces 45 true positive paths (27 SSRFs) with only built-in sources and sinks. When third-party sources and sinks are added, \textsc{Rips}$^{*}$ produces 58 true positive paths (4 more SSRFs). The detection rate is 78.3\% and 72.0\% lower than \textsc{Artemis}'s. The primary factor is that \textsc{Rips} does not support object-oriented features, overlooking vulnerabilities in class methods. For example, in Figure~\ref{fig:related-fn}, \textsc{Rips} fails to detect the taint propagation from the method call on line \#22 to the method on line \#28, resulting in a false negative.

\textbf{Comparison with \textsc{Phpjoern}s.}
\textsc{Phpjoern} detects 51 true positive paths (30 SSRFs) with only built-in sources and sinks, which is 75.4\% fewer than \textsc{Artemis}. After incorporating third-party sources and sinks, \textsc{Phpjoern}$^{*}$ identifies 80 true positive paths (49 SSRFs), still 61.4\% fewer than \textsc{Artemis}. The primary reason is \textsc{Phpjoern}'s incomplete call graph construction. \textsc{Phpjoern} struggles with object-oriented programming where methods in different classes may have the same name. For example, in Figure~\ref{fig:related-fn}, two methods named \mintinline{php}{getData} exist in different classes, but \textsc{Phpjoern} cannot distinguish them, leading to early termination of taint paths on line \#13 and therefore false negatives.

\textbf{Comparison with \textsc{TChecker}s.}
\textsc{TChecker} and \textsc{TChecker$^{\dag}$} identify 43 true positive paths (25 SSRFs) with only built-in sources and sinks, and \textsc{TChecker}$^{*}$ identifies 52 true positive paths (28 SSRFs) when third-party sources and sinks are included. The detection rate is 79.2\% and 74.9\% lower than \textsc{Artemis}'s. 
\textsc{TChecker}'s design of starting analysis from the top-level function of each PHP file leads to a comparative low detection rate of SSRFs. 
Specifically, many true SSRF paths do not have explicit caller-callee relationship with the top-level functions, causing \textsc{TChecker}'s mis-detection. 
Furthermore, \textsc{TChecker}'s propagation rules are incomplete. \textsc{TChecker} does not model tainted arrays iterated through \texttt{foreach} loops. \blue{For example, in Figure~\ref{fig:related-fn}, method \mintinline{php}{updateProducts} is skipped from the analysis of \textsc{Tchecker}. Additionally, the taint is propagated in a \texttt{foreach} loop at line \#15, which is also ignored by \textsc{TChecker}.}

\textbf{Comparison with \textsc{psalm}s.}
\textsc{Psalm} and \textsc{Psalm$^{\dag}$} detect 79 true positive paths (48 SSRFs) with only built-in sources and sinks, which is 61.8\% fewer than \textsc{Artemis}. With third-party sources and sinks, \textsc{Psalm}$^{*}$ detects 132 true positive paths (65 SSRFs), still 36.2\% fewer. The main reason is that \textsc{Psalm} fails to handle implicit call graph connection, particularly in cases where type inference fails. \textsc{Psalm} relies on type annotations in comments to determine variable types. If the type of a variable cannot be inferred, method calls on that variable are ignored, cutting off the taint propagation path. For example, in Figure~\ref{fig:related-fn}, \textsc{Psalm} fails to infer the type of \mintinline{php}{$this->Product}, ignoring the \mintinline{php}{changeImage} method call, leading to a false negative.

\textbf{Comparison with \textsc{Phan}s.}
\textsc{Phan} and \textsc{Phan$^{\dag}$} produce 88 true positive paths (50 SSRFs) with only built-in sources and sinks, and \textsc{Phan}$^{*}$ produces 172 true positive paths (85 SSRFs) when third-party sources and sinks are added. Although surpassing other tools, \textsc{Phan}'s detection rate falls short of \textsc{Artemis} by 57.5\% and 16.9\%, respectively. Like \textsc{Psalm}, \textsc{Phan} fails to create implicit call graph connections and cannot infer the type of \mintinline{php}{$this->Product} in Figure~\ref{fig:related-fn}, missing the implicit call target on line \#22 and thus failing to propagate the taint. Additionally, \textsc{Phan} does not model implicit data flow relationships, such as those established by the \texttt{extract} function, leading to further false negatives.

As shown in the preceding comparisons, while existing approaches boost their true positive rates by including \textit{third-party sources and sinks}, they still suffer low detection coverage (16.9\% to 74.9\% fewer than \textsc{Artemis}) due to their lack of support for implicit call targets and implicit data flows.

\textbf{Comparison with Ablated \textsc{Artemis}es.}
Removing the \textit{third-party source and sink identification} module from \textsc{Artemis} results in only 97 true positive paths (53 SSRFs), 53.1\% fewer than the full version. The low number of true positive paths highlights the widespread use of third-party sources and sinks in modern PHP applications and the critical importance of this module. 

Removing \textit{statically inferred call graph construction} results in 181 true positive paths (89 SSRFs), 12.6\% fewer than \textsc{Artemis}. The 26 false negatives are due to lack of support for implicit call targets: 4 from missing magic methods, 18 from known method names with variable class names, 3 from known class names with variable method names, and 2 from both class and method names being unknown. Although implicit call targets are not frequently used, ignoring them causes false negatives. One case (CVE-2023-40969) involves both magic methods and known class names with variable method names.

By replacing the \textit{rule-based taint analysis} module with generic propagation rules from \textsc{Phan}, 198 TPs are produced, which is 4.3\% fewer than \textsc{Artemis} due to implicit data flow. %

Removing the \textit{false positive pruning} module does not impact detection coverage, because this module is specifically designed to reduce false positives deterministically, and conditions that cannot be statically verified are ignored.

\begin{table}[!t]
    \centering
     \caption{Detection results of \textsc{Artemis}, \textsc{TChecker}, \textsc{Phan}, \textsc{Rips}, \textsc{Phpjoern}, \textsc{Psalm} and four ablated versions of \textsc{Artemis}. The superscripts $a$, $c$, $s$, $p$ represent \textsc{Artemis} without \textbf{third-party source/sink identification}, \textbf{call graph construction}, \textbf{safety string analysis} and \textbf{false positive pruning} respectively. A superscript * represents the tool using both built-in and third-party sources and sinks identified by \textsc{Artemis}. \rev{A superscript $\dag$ represents the tools with false positive pruning module from \textsc{Artemis}.} 
     }
    \label{tab:fp-compare}
    \begin{adjustbox}{width=0.8\textwidth,keepaspectratio}
    \begin{tblr}{
    hline{1,2,Z}={1pt},
    colspec = {X[2,l]X[2,r]X[-1,l]X[2,r]X[-1,l]X[2,r]X[-1,l]X[2,c]},
    cell{1}{2,4,6} = {c=2}{c},
    rows={m,rowsep=0pt,font=\footnotesize},
    row{1} = {font=\footnotesize\bfseries},
    rowsep=-2pt,
    row{even} = {bg=gray!10},
    cell{even}{2,3,6,7} = {bg=gray!10!babyblueeyes!40},
    cell{odd}{2,3,6,7} = {bg=babyblueeyes!10},
    column{3,5,7} = {fg=hotmagenta!70},
    cell{18}{3,5} = {fg=greenpigment},
    cell{16,17}{7} = {fg=greenpigment},
}
Approaches             & TP CVE Nums & & TP Path Nums          & & FP Path Nums            & & Precision \\
\textsc{Artemis}          & 106  &          & 207 &         & 15  &            &93.2\% \\
\textsc{Rips}             & 27   & -74.5\%  & 45  & -78.3\% & 139 & +8.27x     &24.5\% \\
\textsc{Rips$^{*}$}       & 31   & -70.8\%  & 58  & -72.0\% & 149 & +8.93x     &28.0\% \\
\textsc{Phpjoern}         & 30   & -71.7\%  & 51  & -75.4\% & 100 & +5.67x     &33.8\% \\
\textsc{Phpjoern$^{*}$}   & 49   & -53.8\%  & 80  & -61.4\% & 113 & +6.53x     &41.5\% \\
\textsc{TChecker}         & 25   & -76.4\%  & 43  & -79.2\% & 57  & +2.80x     &43.0\% \\
\textsc{TChecker$^{*}$}   & 28   & -73.6\%  & 52  & -74.9\% & 63  & +3.20x     &45.2\% \\
\textsc{TChecker$^{\dag}$}& 25   & -76.4\%  & 43  & -79.2\% & 36  & +1.40x     &54.4\% \\
\textsc{Psalm}            & 48   & -54.7\%  & 79  & -61.8\% & 138 & +8.20x     &36.4\% \\
\textsc{Psalm$^{*}$}      & 65   & -38.7\%  & 132 & -36.2\% & 149 & +8.93x     &47.0\% \\
\textsc{Psalm$^{\dag}$}   & 48   & -54.7\%  & 79  & -61.8\% & 92  & +5.13x     &46.2\% \\
\textsc{Phan}             & 50   & -52.8\%  & 88  & -57.5\% & 156 & +9.40x     &36.1\% \\
\textsc{Phan$^{*}$}       & 85   & -19.8\%  & 172 & -16.9\% & 166 & +10.07x    &50.9\% \\
\textsc{Phan$^{\dag}$}    & 50   & -52.8\%  & 88  & -57.5\% & 124 & +7.27x     &41.5\% \\
\textsc{Artemis$^{a}$}    & 53   & -50.0\%  & 97  & -53.1\% & 11  & -26.7\%    &89.8\% \\
\textsc{Artemis$^{c}$}    & 89   & -16.0\%  & 181 & -12.6\% & 15  & +0\%       &92.3\% \\
\textsc{Artemis$^{t}$}    & 102  & -3.8\%   & 198  & -4.3\% & 153 & +9.20x     &56.4\% \\
\textsc{Artemis$^{p}$}    & 106  & +0\%     & 207 & +0\%    & 35  & +1.33x   &85.5\% \\
\end{tblr}
\end{adjustbox}
\end{table}
 
\subsubsection{False Positives}
\label{subsec:fp-comp}
Table~\ref{tab:fp-compare} presents the detection precision of \textsc{Artemis}, \textsc{Rips}, \textsc{Phpjoern}, \textsc{TChecker}, \textsc{Psalm}, \textsc{Phan}, and the ablated versions of \textsc{Artemis}. \textsc{Artemis} produces 15 false positives, achieving a precision rate of 93.2\%.

\textbf{Comparison with \textsc{Ripses}.}
\textsc{Rips} and \textsc{Rips}$^{*}$ generate 149 false positives with third-party sources and sinks, and 139 without, having the lowest precision rates of 28.0\% and 24.5\%. This is mainly due to \textsc{Rips}'s over-tainting, assuming tainted parameters make return values tainted. Additionally, \textsc{Rips} lacks support for safety string analysis and considers any string constructed with a tainted variable as tainted, regardless of whether the tainted input controls the critical parts of the string. These over-tainting issues lead to a significant number of false positives.

\textbf{Comparison with \textsc{Phpjoerns}}
\textsc{Phpjoern} and \textsc{Phpjoern}$^{*}$ generate 100 false positives with third-party sources and sinks and 113 without, with a precision rate of 41.5\% and 33.8\%, respectively. \textsc{Phpjoern} has high false positive rates mainly because the lack of safety string analysis and false positive pruning. 

\textbf{Comparison with \textsc{TCheckers}.}
\textsc{TChecker} and \textsc{TChecker}$^{*}$ produce 63 false positives with third-party sources and sinks and 57 without, resulting in a precision rate of 45.2\% and 43.0\%, respectively. The relatively low number of false positives is due to many functions (that are not explicitly invoked by the top-level functions) not being analyzed by \textsc{TChecker}, which also results in fewer true positive paths. Consequently, \textsc{TChecker} remains less precise compared to \textsc{Artemis}. Compared to \textsc{TChecker}, \textsc{TChecker$^{\dag}$} reduces false positives to 36, a decrease of 36.8\%.

\textbf{Comparison with \textsc{Psalms}.}
\textsc{Psalm} and \textsc{Psalm}$^{*}$ produce 149 false positives with third-party sources and sinks and 138 without, achieving a precision rate of 47.0\% and 36.4\%. \textsc{Psalm}$^{\dag}$ produces 92 false positives, a 33. 3\% decrease compared to \textsc{Psalm}.

\textbf{Comparison with \textsc{Phans}.}
\textsc{Phan} and \textsc{Phan}$^{*}$ have the most false positives, 166 with and 156 without, resulting in precision rates of 50.9\% and 36.1\%. With false positive pruning module, \textsc{Phan}$^{\dag}$ produces 124 false positives, resulting in a 20.5\% decrease.

\textsc{TChecker}s, \textsc{Psalm}s, and \textsc{Phan}s have high false positive rates due to their generic propagation rules. %
Without \textit{safety string analysis}, they over-taint URL strings when tainted input does not affect critical URL parts. 
The absence of \textit{path condition analysis} also leads to reporting infeasible paths, further increasing their false positives. 
We have observed that integrating the \textit{path condition analysis} module into existing approaches significantly reduces false positives, for example, \textsc{TChecker$^{\dag}$} reduces the false positives reported by \textsc{TChecker} by 36.8\%.
However, without the augmented taint propagation and clearance rules in \textsc{Artemis}, existing approaches relying solely on post-processing for pruning false positives are imprecise.

\textbf{Comparison with Ablated \textsc{Artemis}es.}
Removing the \textit{third-party source and sink identification} module results in 11 false positives, slightly reducing the number of false positives, but at the cost of 53.1\% fewer true positive paths, highlighting the importance of this module.

Replacing the \textit{rule-based taint analysis} module with propagation rules from \textsc{Phan} results in 153 false positives, 9.2 times more than the full version of \textsc{Artemis}. This sharp increase is due to the lack of safety string assurance rules, which prevents proper handling of cases where user-controlled input cannot control critical parts of request-sending URLs, such as query parameters or parts of file-accessing URLs.

Removing the \textit{false positive pruning} module leads to 35 false positives, an increase of 133.3\% compared to the full version of \textsc{Artemis}. The increase in false positives suggests the importance of handling path conditions in SSRF detection.

\textbf{Discussion.} It is important to note that \textsc{Artemis} still has 15 false positives, which is attributed to three key factors. 
First, four false positives arise from always unsatisfied conditions involving array variables, which \textsc{Artemis} cannot accurately recognize. Accurately tracking each element in PHP arrays remains an open challenge~\cite{semantic} due to the lack of formal semantics for PHP and the dynamic nature of arrays. 
\blue{Second, six false positives are caused by user-defined functions that remove characters from tainted input strings on one branch. If these removed characters include URL components, such as a slash (/), SSRF becomes impossible. However, since \textsc{Artemis} merges taint information from different branches, input taking the branch that does not remove characters is considered tainted and the return value is tainted according to the phi rule, leading to false positives. 
To effectively prune these cases, static symbolic execution would be required to determine which branch actually propagates the taint, but this remains an open challenge due to the dynamic nature of PHP and is out of the scope of current work~\cite{semantic, symphp}.
Third, five false positives are due to intended features. In some applications, developers design functions that can send arbitrary requests for debugging purposes. These debugging functions are typically protected by file-based authentication, meaning that prior access to the server’s file system is required. 
Pruning these cases requires domain knowledge about the intended use of these functions, which is difficult to automate.} 

\addtolength{\textfloatsep}{-0.2in}
\begin{figure}[t]
\centering
        \includegraphics[width=\linewidth]{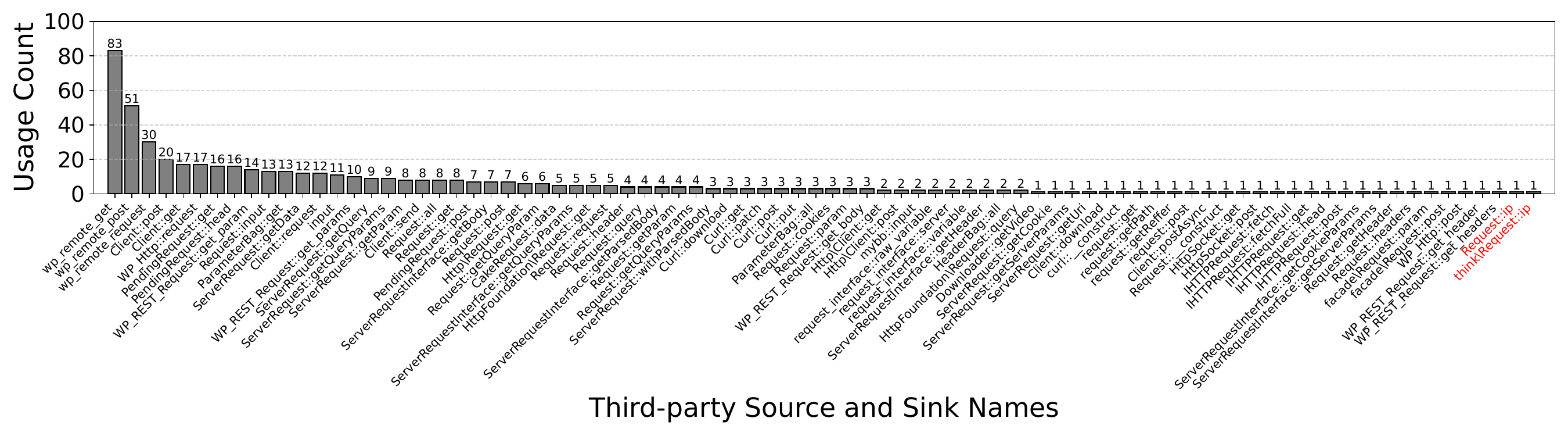}
        \caption{\rev{Usage frequency of third-party sources and sinks. The two third-party sources missed by GPT-4o are marked in red.}} %
        \label{fig:freq-fig}
        \vspace{-5pt}
\end{figure}

\rev{\subsection{ Source and Sink Identification Results}}
\rev{
\subsubsection{ Generalization Analysis}
\label{subsec:llm}
We evaluate the generalizability of our LLM-based source and sink identification by extending our analysis to multiple state-of-the-art models, including GPT-4o, LLaMA-3.1~\cite{llama3}, and Claude-3.5-Sonnet~\cite{claude3-5}, without human refinement. 
}

\begin{wraptable}{r}{0.35\textwidth}
\centering
\caption{\rev{Precision, recall, f1-score and cost of GPT-4o, LLaMA-3.1, and Claude-3.5.}}
    \label{tab:llm-comp}
    \begin{tblr}{
    hline{1,2,Z}={1pt},
    colspec = {X[1,c]X[1,c]X[1,c]X[1,c]X[-1,c]},
    rows={m,rowsep=0pt,font=\tiny},
    row{1} = {font=\bf\tiny},
    rowsep=-2pt,
    row{even} = {bg=gray!10},
    cell{even}{2,4} = {bg=gray!10!babyblueeyes!40},
    cell{odd}{2,4} = {bg=babyblueeyes!10},
    colsep=1pt,
    }
    LLM        & Precision & Recall & F1 & Cost (\$) \\
    GPT-4o     & 91.1\%    & 97.6\% & \textbf{94.3\%} & 21         \\
    Claude-3.5 & \textbf{92.0\%}    & 96.4\% & 94.2\%  & 40         \\
    LLaMA-3.1  & 87.3\%    & \textbf{98.8\%} & 92.7\% & \textbf{19}         
    \end{tblr}
\end{wraptable}%

First, we manually create the ground truth by examining PHPDocs and code examples, classifying all 7,406 third-party APIs from 250 applications as sources, sinks, or neither. 
This \textit{manual} process yields 44 sources and 40 sinks as our ground truth.
We then measure precision and recall by comparing the identification outputs of state-of-the-art models with our ground truth.
We present the precision, recall, and cost of GPT-4o, Claude-3.5, and LLaMA-3.1 in Table~\ref{tab:llm-comp}. We select GPT-4o as our main assistant model primarily because (1) it has a balanced precision and recall with the highest F1 score; (2) it is relatively cost-effective;
and (3) its standard API format enables easy extraction of raw model outputs in JSON. Although GPT-4o misses two sources which are associated with the MipCMS application, they do not lead to any SSRFs. %
GPT-4o misses the two sources due to insufficient context in PHPDocs. %

\begin{wrapfigure}{r}{0.35\textwidth}
\centering
        \includegraphics[width=\linewidth]{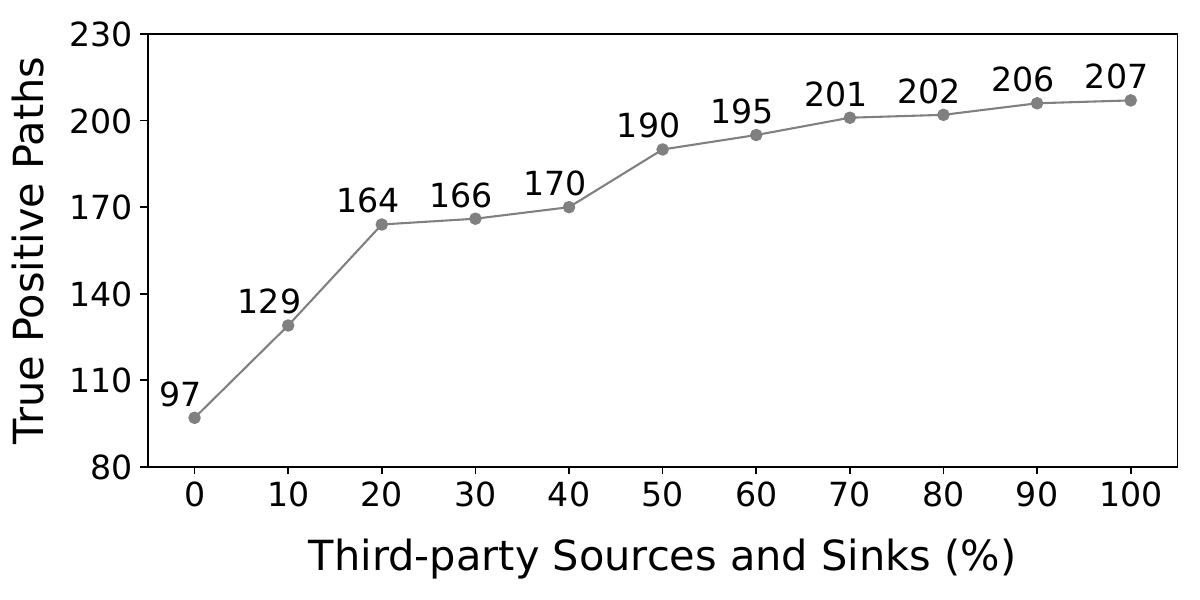}
        \caption{\rev{Number of true positive paths detected with 0\% to 100\% of frequently-used third-party sources and sinks.} %
        }
        \label{fig:tp-sub-fig}
        \vspace{-10pt}
\end{wrapfigure}

\rev{
\subsubsection{ Sensitivity Analysis}
To evaluate how scalable \textsc{Artemis} is in identifying third-party sources and sinks for use in detecting SSRFs in other PHP applications, we conduct a sensitivity analysis on all 84 third-party sources and sinks. First, we calculate how many applications each third-party source and sink appears in, as shown in Figure~\ref{fig:freq-fig}. We conducted the experiment 11 times, progressively incorporating 0\%, 10\%, ..., up to 100\% of the most frequently used sources and sinks in each iteration to identify SSRFs across all applications. \footnote{\rev{The two sources missed by GPT-4o are added in the last 10\% iteration and do not impact our result.}}
}

\rev{As shown in Figure~\ref{fig:tp-sub-fig}, the number of detected true positive paths increases logarithmically from 97 to 207 as the inclusion of third-party sources and sinks rises from 0\% to 100\%, converging sublinearly. With an inclusion of 0\%, 97 true positive paths involving only PHP built-in sources and sinks are detected. We also observe that with 50\% of inclusion, we achieve 84.5\% coverage. This aligns with the frequency distribution in Figure~\ref{fig:freq-fig}, where infrequently used sources and sinks appear only once or twice, having minimal impact on the overall detection rate. Therefore, we believe that our identified set of third-party sources and sinks can be applied to future application SSRF detection, offering a solid coverage rate. }

\begin{table}[!t]
    \centering
     \caption{\rev{Precision, recall, and construction time of call graphs in \textsc{Artemis}, \textsc{Rips}, \textsc{Phpjoern}, \textsc{TChecker}, \textsc{Psalm}, and \textsc{Phan} on 10 sample applications.}
     }
    \label{tab:cg-result}
    \begin{tblr}{
    hline{1,3,Z}={1pt},
    colspec = {X[-1,c]X[1,c]X[1,c]X[1,c]X[1,c]X[1,c]X[1,c]},
    rows={m,rowsep=0pt,font=\scriptsize,},
    row{1} = {font=\scriptsize\bfseries},
    row{2} = {font=\tiny\bfseries},
    rowsep=-2pt,
    cell{3-Z}{1-Z} = {font=\tiny},
    row{even[3-Z]} = {bg=gray!10},
    cell{even[3-Z]}{2,4,6} = {bg=gray!10!babyblueeyes!40},
    cell{1,2}{2,4,6} = {bg=babyblueeyes!10},
    cell{odd[3-Z]}{2,4,6} = {bg=babyblueeyes!10},
    cell{1}{1} = {r=2}{c},
    colsep=1pt,
}
App       & \textsc{Artemis}           & \textsc{Rips}              & \textsc{Phpjoern}         & \textsc{TChecker}         & \textsc{Psalm}            & \textsc{Phan}         \\

          & P / R / Time (ms)  & P / R / Time (ms)  & P / R / Time (ms) & P / R / Time (ms) & P / R / Time (ms) & P / R / Time (ms)    \\
          
Cockpit   & 70.9\% / \textbf{99.0\%} / 19.8   & 98.1\% / 10.9\% / 1.1  & \textbf{99.8\%} / 11.9\% / 11.7  & 95.9\% / 38.9\% / 233.2  & 98.7\% / 64.4\% / 17.8  & 96.3\% / 41.6\%  / 7.9    \\

i-librarian & 96.4\% / \textbf{100\%} / 48.5  & \textbf{100\%} / 76.1\% / 4.4  & \textbf{100\%} / 80.1\% / 26.7  &  99.4\% / 92.9\% / 47.3  &  99.0\% / 94.1\% / 169.1  &  99.1\% / 93.2\% / 35.5               \\

leadin    & \textbf{100\%} / \textbf{100\%} / 9.5  &  \textbf{100\%} / 72.6\% / 1.9 & \textbf{100\%} / 85.1\% / 7.1 & \textbf{100\%} / \textbf{100\%} / 16.9 & \textbf{100\%} / \textbf{100\%} / 43.1 & \textbf{100\%} / \textbf{100\%} / 9.3 \\

LibreY    & \textbf{100\%} / \textbf{100\%} / 4.2 & \textbf{100\%} / 89.4\% / 0.6 & \textbf{100\%} / 98.3\% / 6.9 & \textbf{100\%} / \textbf{100\%} / 6.1 & \textbf{100\%} / \textbf{100\%} /  13.7    & \textbf{100\%} / 95.0\%  / 3.4    \\

LinkAce   & 66.9\% / \textbf{98.3\%} / 168.1 & \textbf{98.2\%} / 5.5\% / 1.4      & 98.1\% / 5.4\%  / 22.5     & 96.3\% / 86.0\% / 451.8     & 96.7\% / 87.3\% / 77.1    & 96.5\% / 86.8\%  / 110.2    \\

NoneCms   & 67.4\% / \textbf{98.5\%} / 17.3  & 99.5\% / 50.5\%  / 1.5    & \textbf{99.7\%} / 51.0\% / 12.2     & 98.0\% / 71.2\% / 951.1     & 98.3\% / 75.5\%  /  33.8  & 98.2\% / 73.5\%   / 9.2   \\

rconfig   & 97.4\% / \textbf{100\%} / 37.1 & \textbf{100\%} / 79.8\% / 7.5 & \textbf{100\%} / 82.7\% / 23.4 & 99.8\% / 92.9\% / 536.7 & 99.8\% / 97.4\% / 301.3 & \textbf{100\%} / 95.5\% / 30.7 \\

WeBid     & 99.1\% / \textbf{99.5\%} / 67.9  & \textbf{99.8\%} / 50.6\% / 16.5 & \textbf{99.8\%} / 52.1\% / 28.3      & 99.5\% / 96.1\% / 282.5    & 99.1\% / 96.6\% / 94.2   & 99.6\% / 96.4\% /    57.1 \\

wp-fastest-cache & 98.3\% / \textbf{100\%} / 44.5 & \textbf{100\%} / 83.7\% / 1.6 & \textbf{100\%} / 87.7\% / 12.4 & 98.9\% / 95.8\% / 18.4 & 99.4\% / 96.5\% / 127.3 & \textbf{100\%} / 94.6\% / 31.3 \\

yzmcms    & 78.7\% / \textbf{98.8\%} / 124.3 & 99.6\% / 32.7\% / 8.1 & \textbf{99.8\%} / 33.9\% / 29.7 & 99.1\% / 39.9\% / 329.7 & 99.2\% / 41.8\% / 215.9    & 99.4\% / 48.5\% / 104.9  \\

\end{tblr}
\end{table}

\rev{\subsection{ Call Graph Construction Results}}
\rev{In this section, we conduct a small-scale experiment to evaluate the effectiveness of our call graph construction approach in Section~\ref{sec:cg}. We compare \textsc{Artemis} with alternative methods, including the unique name matching approach used in \textsc{Rips} and \textsc{PhpJoern}, and the type inference-based approach used in \textsc{TChecker}, \textsc{Psalm}, and \textsc{Phan}. 
We select ten moderately sized applications from our benchmark and manually extract and verify their caller-callee pairs as the ground truth.
} 

\rev{
Table~\ref{tab:cg-result} presents the precision, recall, and speed of different call graph construction approaches.
Among all tools, \textsc{Artemis} consistently achieves the highest recall, ranging from 98.3\% to 100\%, indicating superior call graph coverage.
However, existing tools exhibit poor recall, significantly lower than \textsc{Artemis} in some cases. Among them, the name matching-based call graphs (\textsc{Rips} and \textsc{PhpJoern}) show the lowest recall, with a minimum of just 5.4\%. The reason \textsc{Artemis} cannot reach 100\% recall is due to incorrect type inference, which is an open problem~\cite{type-inference, type-phan}.} 
\rev{
On the other hand, \textsc{Artemis} has the lowest precision due to our over-approximation strategy to identify as many implicit calls as possible. This relaxation is then tightened by our safety string rules in Section~\ref{sec:taintanalysis} and false positive pruning in Section~\ref{subsec:path-constraint} to prevent excessive false positives, as many of them are either untainted or never reach the sink.
Therefore, the final SSRF detection in \textsc{Artemis} maintains high precision despite the lower precision of its call graph module. As shown in Table~\ref{tab:fp-compare}, when including implicit call graph construction, \textsc{Artemis} does not produce any additional false positives.
}

\rev{\textsc{Rips} and \textsc{Phpjoern} are the fastest because they use simple unique name matching. \textsc{TChecker}, \textsc{Psalm}, \textsc{Phan}, and \textsc{Artemis} take longer due to type inference. Although \textsc{Artemis} is about 30\% slower than \textsc{Phan} (on which \textsc{Artemis} is based) for implicit call targets, it still completes in less than 1 second for all cases, making the extra construction time worthwhile for better SSRF coverage.}

\subsection{Detection Time}
\rev{\textsc{Artemis} executes in 10.4 to 328.1 seconds, averaging 69.5 seconds. \textsc{Rips} completes analysis in 0.2 to 2045.2 seconds, averaging 34.3 seconds. \textsc{TChecker} analyzes projects in 0.4 to 652.1 seconds, with an average of 33.7 seconds. \textsc{Phpjoern} takes 0.5 to 210.7 seconds, averaging 22.4 seconds. \textsc{Psalm} requires 0.7 to 921.3 seconds, with an average of 28.1 seconds. \textsc{Phan} finishes analysis in 0.6 to 157.4 seconds, averaging 20.9 seconds. 
Compared to other tools, \textsc{Artemis} on average requires twice as much time to analyze a project primarily because the call graph generated by \textsc{Artemis} contains implicit calls ignored by other tools. Each connected implicit call site uncovers additional hidden call chains, thereby increasing the workload for taint propagation tracking and false positive pruning.
} 

\rev{
\textsc{Rips}, \textsc{TChecker}, and \textsc{Phpjoern} cannot complete the analysis on some projects because they are designed for PHP 7.0 and lack forward compatibility. 
For example, in \textsc{Rips}, \textsc{Phpjoern}, and \textsc{TChecker}, the iterative and recursive analysis of PHP 7.4 arrow syntax in lambdas~\cite{arrow-lambda} leads to repeated evaluations of the same functions or string elements, with intermediate results stored in memory during each iteration or recursion. This continuous accumulation of data without termination eventually results in an out-of-memory (OOM) condition.
The detailed detection time for each application with all detection tools can be found in Table~\ref{tab:time-app} in Appendix.} 

\subsection{Limitations}
\label{subsec:limit}

\begin{wrapfigure}{r}{0.38\textwidth}
    \centering
\begin{minted}[linenos,
startinline, 
breaklines=true, 
fontsize=\tiny,
escapeinside=||
]{php}
$f = esc_sql($_POST['file']);
$q = "INSERT INTO $t VALUES('$f',...)";
$wpdb->query($q);
...
$r = $wpdb->get_row("SELECT...FROM $t...");
readfile($r->file);
\end{minted}
\caption{A second-order SSRF (CVE-2020-24141) missed by \textsc{Artemis}.}
\label{fig:fn-case-study}
\vspace{-5pt}
\end{wrapfigure}

\textbf{False Negatives.}
\textsc{Artemis} cannot detect second-order SSRF where user input is first stored in a database and later retrieved. For example, in Figure~\ref{fig:fn-case-study}, user input is saved in the database on line \#3, read from the database on line \#5, and finally used in a sink function on line \#6. To detect such vulnerabilities, we need to model database APIs, infer database schemas, and identify the affected columns during data transfer, which is outside the scope of our current work. Past studies~\cite{splendor, second-order} focusing on application-specific database operations in second-order vulnerabilities can help improve \textsc{Artemis} to detect such SSRF vulnerabilities.

\textbf{Language Features.}
So far, \textsc{Artemis} does not support all dynamic PHP features, such as argument unpacking, which involves spreading array elements into argument lists. Integrating this feature into \textsc{Artemis} requires inferring all array elements when constructing implicit call graphs and updating our taint propagation rules accordingly. To make matters worse, nested arrays, where an array element is another array, further complicate this feature. Although we have not observed these features being exploited in existing vulnerabilities, they could lead to undetected vulnerabilities.

\section{Conclusion}
\label{sec:conclusion}
In this paper, we introduce \textsc{Artemis}, a static taint analysis tool %
to effectively identify SSRF vulnerabilities. \textsc{Artemis} achieves an enhanced detection coverage and accuracy by integrating LLM-based source and sink identification, explicit and implicit call graph construction, augmented rule-based taint analysis, and false positive pruning based on path conditions. 
\textsc{Artemis} operates entirely automatically without any application-specific domain knowledge. 
We have implemented a prototype of \textsc{Artemis} and tested it with 250 open-source applications. The results reveal that \textsc{Artemis} successfully identifies 207 true vulnerable paths (106 true SSRFs) and 15 false positives, significantly outperforming existing tools. 
Of the 106 SSRFs, 35 are newly discovered. 
We have reported the 35 new SSRFs to the developers, and 24 SSRFs have been confirmed by the developers and assigned CVE IDs.

\begin{acks}
We would like to thank the anonymous reviewers for their
insightful feedback and valuable comments. This work was supported by the Shanghai Sailing Program 22YF1428600. Yutian Tang's work was supported in part by the National Natural Science Foundation of China under grant 62202306.
\end{acks}

\section*{Data-Availability Statement}
\textsc{Artemis}'s source code and detailed reports of newly detected SSRFs are available at \url{https://zenodo.org/records/13353039}.

\appendix %
\section{Appendix}

\subsection{Prompt for Source and Sink Identification}
\label{subsec:identify-prompt}
We use the following prompt for third-party source and sink identification (Section~\ref{subsec:source-sink}):

\begin{tcblisting}{colback=usercolor, sharp corners, boxrule=0pt, top=0pt, bottom=0pt, left=4pt, right=4pt, boxsep=0pt, frame hidden, listing only, beforeafter skip=0pt,
listing options={language=prompt}}
User: You will be given a PHP function/method/property from PHP web applications along with its doc comment. Your task is to determine whether it is a potential taint source according to its name and doc comment. Here taint source is defined to be something that can return data from an incoming HTTP request. Think step by step.

For example:
```
Name: \app\migration\exception::getParameters
DocComment: Get the exception parameters @return array
```
This is not a source because it gets information from exception.
```
Name: \app\request_interface::get_param
DocComment: Returns the request data in an array.
```
This is a source because it returns data from incoming request (user input)

User: Name: signature
      DocComment: Full doc comment
\end{tcblisting}
\begin{tcblisting}{colback=assistantcolor, sharp corners, boxrule=0pt, boxsep=0pt, top=0pt, bottom=0pt, left=4pt, right=4pt, frame hidden, listing only, beforeafter skip=0pt,
listing options={language=prompt}}
Assistant: [Analyze Process and Result]
\end{tcblisting}
\begin{tcblisting}{colback=usercolor, sharp corners, boxrule=0pt, top=0pt, bottom=0pt, left=4pt, right=4pt, boxsep=0pt, beforeafter skip=0pt, frame hidden, listing only,
listing options={language=prompt}}
User: Summarize your result in JSON format.
\end{tcblisting}
\begin{tcblisting}{colback=assistantcolor, sharp corners, boxrule=0pt, top=0pt, bottom=0pt, left=4pt, right=4pt, boxsep=0pt, frame hidden, beforeafter skip=0pt, listing only,
listing options={language=prompt}}
Assistant:
{
  "result": true/false
}
\end{tcblisting}

\subsection{Language Abstraction}
\label{subsec:language-abstraction}
Our abstraction selectively models statements that are related to our taint analysis. 

Arrays play a fundamental role in PHP because of their versatile and distinctive syntax. In PHP, arrays are essentially ordered maps,
where keys can be strings, integers, a mix of both, or denoted by a (statically unknown) variable~\cite{phparray}. Therefore, we specifically abstract arrays' initialization, element assignment, and element retrieval.

We model the \texttt{foreach} loop statement because it contains an iterable variable assigned to a key-value pair or a value. In other types of loop statements, including the while loop, for loop, and do-while loop, the loop header structure contains normal expressions whose effect is covered by other rules.  %
We should note that the loop body statements are modeled separately from the loop head structure itself.
Similarly, blocks of statements within branches of if-else structures, switch structures, and ternary operators are also modeled separately. 
Following branching, we introduce a phi operator to consolidate variable assignments from different branches of control flow, ensuring a consistent representation of variable assignments. %

We model method calls for both passing arguments at the call site and returning values to the caller. 
Argument passing takes place when each formal argument corresponds to an actual argument. %
Returning value assignment occurs when the returned value from a method call is assigned to a variable at the caller's site.
Yielding value assignment occurs when the yielded value from a generator function is assigned to a variable at the caller's site.

Object instantiation is symbolized by the \texttt{new} keyword, 
where each formal argument corresponds to an actual argument, resulting in the creation of a new object variable.

\subsection{Prompt for Exploit Generation}
\label{subsec:exploit-prompt}
We use the following prompt for exploit generation
(Section~\ref{subsec:exploit}):

\begin{tcblisting}{colback=usercolor, sharp corners, boxrule=0pt, top=0pt, bottom=0pt, left=4pt, right=4pt, boxsep=0pt, frame hidden, listing only, beforeafter skip=0pt,
listing options={language=prompt}}
System: You are an expert on PHP web security. 
You will be given PHP code snippets. 
Your task is to analyze the potential server-side request forgery (SSRF) vulnerability in the code snippet. 
When there is an SSRF, construct a potential exploit input. 
After that, retrieve relevant information with appropriate tool to figure out the API (route) and the parameter values in the actual request to trigger SSRF.
If the code indicate the use of a well-known framework, such as WordPress or Laravel, make full use of your domain knowledge about them in your analysis.
In our definition, SSRF happens when 1) input can change the request destination so that attacker can induce the server-side application to make HTTP requests to an internal resource not accessible by attacker or 2) the input uses `file://` protocol and can read arbitrary file or directory such as `/etc/passwd` or `/secret-dir`. 
Think step by step and provide concise results. Do not provide advice.

ToolDef: searchKeywords: {...}, executePhpCode: {...}, ...

User: Here's the code snippets of functions (and their position in files) containing a path from source (user input) to sink (a function that may send a request or read files). 
Statements involved in the taint path are marked with comment `// Tainted` above the line. 
Pay attention to conditions required to reach statements that may trigger SSRF. 
Analyze the code, especially the taint path and construct the input that may trigger SSRF. 
[Additional detail]
Try construct a minimal input where only necessary fields are populated.
```
[Code]
```
\end{tcblisting}
\begin{tcblisting}{colback=assistantcolor, sharp corners, boxrule=0pt, boxsep=0pt, top=0pt, bottom=0pt, left=4pt, right=4pt, frame hidden, listing only, beforeafter skip=0pt,
listing options={language=prompt}}
ToolCall: [Use tool ...]
ToolCallResult: [Result]
[More tool calls]
Assistant: [Infer core payload]
\end{tcblisting}
\begin{tcblisting}{colback=usercolor, sharp corners, boxrule=0pt, top=0pt, bottom=0pt, left=4pt, right=4pt, boxsep=0pt, beforeafter skip=0pt, frame hidden, listing only,
listing options={language=prompt}}
User: Now try to infer the API (route) to trigger this SSRF. This route would trigger the previously provided code.
Here the route is the part after the base address (e.g., `/api/to/route` in `http://example.com/base/api/to/route`).
Use appropriate tools to get information that you need.
To identify the route, use the following process:
[Steps]
After you have determined the route, briefly analyze how it reaches the previously vulnerable code and double check your result.
First present your plan, then execute it accordingly.
Do not guess the route.
\end{tcblisting}
\begin{tcblisting}{colback=assistantcolor, sharp corners, boxrule=0pt, top=0pt, bottom=0pt, left=4pt, right=4pt, boxsep=0pt, frame hidden, beforeafter skip=0pt, listing only,
listing options={language=prompt}}
ToolCall: [Use tool ...]
ToolCallResult: [Result]
[More tool calls]
Assistant: [Infer route]
\end{tcblisting}
\begin{tcblisting}{colback=usercolor, sharp corners, boxrule=0pt, top=0pt, bottom=0pt, left=4pt, right=4pt, boxsep=0pt, beforeafter skip=0pt, frame hidden, listing only,
listing options={language=prompt}}
User: Now check the generated request, aside from the URL-related field, try to obtain a valid value for the other fields if they are required for the request to be sent.
Be sure to analyze additional constraints along the route path.
Determine the values using the following process:
[Steps]
For each field, generate a concise plan. Then execute the plan accordingly.
If all fields already have a valid value, return `Completed`.
\end{tcblisting}
\begin{tcblisting}{colback=assistantcolor, sharp corners, boxrule=0pt, top=0pt, bottom=0pt, left=4pt, right=4pt, boxsep=0pt, frame hidden, beforeafter skip=0pt, listing only,
listing options={language=prompt}}
ToolCall: [Use tool ...]
ToolCallResult: [Result]
[More tool calls]
Assistant: [Infer related values]
\end{tcblisting}
\begin{tcblisting}{colback=usercolor, sharp corners, boxrule=0pt, top=0pt, bottom=0pt, left=4pt, right=4pt, boxsep=0pt, beforeafter skip=0pt, frame hidden, listing only,
listing options={language=prompt}}
User: Now conclude your result in JSON format as specified below, be sure to use concrete values.
Note that for GET request, all query parameters should also be included in the `Path` field.
Request body should be included in the `Data` field. 
When it is POST form data, it should be in the form of key-value pairs encoded as valid JSON.
When both `http://` and `file://` are possible, always generate `http://` only. `file://` is only used when `http://` is not possible.

Example Response schema:
[Schema]
\end{tcblisting}
\begin{tcblisting}{colback=assistantcolor, sharp corners, boxrule=0pt, top=0pt, bottom=0pt, left=4pt, right=4pt, boxsep=0pt, frame hidden, beforeafter skip=0pt, listing only,
listing options={language=prompt}}
Assistant: {
  "Method" : "Request method",
  "Path": "/api/to/route?with=params",
  "Data": {
    "field1": "value1",        
   },
   "Cookie": {
     "key1": "value1"
   },
   "Header": {
     "Key": "Value"
   }
}
\end{tcblisting}
\begin{tcblisting}{colback=usercolor, sharp corners, boxrule=0pt, top=0pt, bottom=0pt, left=4pt, right=4pt, boxsep=0pt, beforeafter skip=0pt, frame hidden, listing only,
listing options={language=prompt}}
[Send the request to server and validate whether SSRF is triggered]
[Following is executed when SSRF not triggered]
User: The tainted lines/functions are not present in the trace. The route you generated is not correct./The SSRF is not triggered. The input you generated for URL/file path is not correct. 
Please check the error message and try to fix the route.
The following functions are not triggered in the runtime trace:
```
[Missing functions]
```
The request does not trigger SSRF. The response is:
```
[Response]
```
\end{tcblisting}
\begin{tcblisting}{colback=assistantcolor, sharp corners, boxrule=0pt, top=0pt, bottom=0pt, left=4pt, right=4pt, boxsep=0pt, frame hidden, beforeafter skip=0pt, listing only,
listing options={language=prompt}}
ToolCall: [Use tool ...]
ToolCallResult: [Result]
[More tool calls]
Assistant: [Analysis to correct exploit]
\end{tcblisting}
\begin{tcblisting}{colback=usercolor, sharp corners, boxrule=0pt, top=0pt, bottom=0pt, left=4pt, right=4pt, boxsep=0pt, beforeafter skip=0pt, frame hidden, listing only,
listing options={language=prompt}}
User: Now conclude your result in JSON format as specified below, be sure to use concrete values.
Note that for GET request, all query parameters should also be included in the `Path` field.
Request body should be included in the `Data` field. 
When it is POST form data, it should be in the form of key-value pairs encoded as valid JSON.
When both `http://` and `file://` are possible, always generate `http://` only. `file://` is only used when `http://` is not possible.

Example Response schema:
[Schema]
\end{tcblisting}
\begin{tcblisting}{colback=assistantcolor, sharp corners, boxrule=0pt, top=0pt, bottom=0pt, left=4pt, right=4pt, boxsep=0pt, frame hidden, beforeafter skip=0pt, listing only,
listing options={language=prompt}}
Assistant: [Request data in JSON]
[Iterate until no improvement for two turns or succeeds]
\end{tcblisting}

\subsection{More Results of the Detected Vulnerabilities}
\label{subsec:vul-detect-details}
Table~\ref{tab:cve-app} presents the detected vulnerabilities' affected application versions, root causes, impacts, and \textsc{Artemis}'s detection time. 

\begin{table*}
\caption{Root cause and impact of the detected SSRFs with corresponding application names. For root causes, D, P, I, F represents missing URL validation, missing URL segment validation, incomplete validation, and flawed validation respectively. For impacts, B, L, D, P represents bypassing access control, sensitive information leakage, DoS, privilege escalation, and remote code execution respectively. %
}
\label{tab:cve-app}
\begin{adjustbox}{totalheight=0.88\textheight,keepaspectratio}
\begin{minipage}[t]{0.49\linewidth}
\vspace{0pt}
\NewColumnType{K}[1][]{>{\ifnum \value{rownum}=1 \relax\else \number\numexpr\value{rownum}-1\relax\fi}Q[co=1,#1]}
\begin{tblr}[]{
    hline{1,2,Z} = {1pt},
    colspec={K[-1,c]X[3,l]X[4,l]X[1,c]X[1,c]},
    rows={m,rowsep=0pt,font=\scriptsize},
    row{1} = {font=\tiny},
    cell{16}{3} = {r=2}{m},
    cell{38}{3} = {r=3}{m},
    rowsep=-1pt,
    colsep=2pt,
    cell{even[2-16]}{3} = {bg=gray!10},
    cell{odd[17-Z]}{3} = {bg=gray!10},
}
\textbf{\#} & \textbf{CVE/Vuln ID} & \textbf{Application} & \textbf{Cause} & \textbf{Impact}  \\ 
& CVE-2015-7816        & Matomo v2.14.3                     & D          & R       \\
& CVE-2016-10926       & nelio ab testing  v4.5.9           & D          & L;B     \\  
& CVE-2016-10927       & nelio ab testing  v4.5.11          & D          & L;B     \\  
& CVE-2016-7964        & dokuwiki v2016-06-26a              & D          & B       \\  
& CVE-2016-9417        & mybb v1.8.7                        & D          & L;B     \\    
& CVE-2017-1000419     & phpbb v3.2.0                       & D          & L;B     \\    
& CVE-2017-10973       & FineCMS v2017-07-06                & D          & R;P     \\  
& CVE-2017-14323       & Onethink v1.1                      & D          & L;B     \\  
& CVE-2017-16870       & UpdraftPlus v1.13.12               & D          & L;D     \\  
& CVE-2017-7566        & mybb v1.8.11                       & I          & L;B     \\    
& CVE-2017-9307        & Allen Disk v1.6                    & D          & L;B     \\  
& CVE-2018-1000138     & i-librarian v4.8                   & D          & L;B     \\  
& CVE-2018-11031       & phprap v1.0.4                      & D          & L;B     \\  
& CVE-2018-14514       & icms v7.0.9                        & D          & L;B     \\    
& CVE-2018-14728       & Responsive FileManager v9.13.1     & D          & L;B     \\  
& CVE-2018-15495       &                                    & D          & L;B     \\  
& CVE-2018-16444       & SeaCMS v6.61                       & D          & B;D     \\  
& CVE-2018-18867       & Responsive FileManager v9.13.4     & I          & L;B     \\  
& CVE-2018-6029        & NoneCms v1.3.0                     & F          & L;B     \\  
& CVE-2018-9302        & Cockpit v0.5.5                     & D          & R       \\
& CVE-2019-11565       & Print My Blog v1.6.6               & D          & B       \\
& CVE-2019-11574       & SMF v2.1                           & D          & B       \\
& CVE-2019-11767       & phpbb v3.2.6                       & I          & L;B     \\    
& CVE-2019-12161       & webpagetest v19.04                 & D          & L;B     \\  
& CVE-2019-15033       & pydio core v6.0.8                  & D          & B       \\  
& CVE-2019-15494       & openITCOCKPIT v3.7.1               & D          & B       \\  
& CVE-2020-10212       & Responsive FileManager v9.14.0     & I          & L;B     \\  
& CVE-2020-10791       & openITCOCKPIT v3.7.2               & D          & B       \\  
& CVE-2020-14044       & Codiad v1.7.8                      & D          & R       \\
& CVE-2020-20341       & yzmcms v5.3.0                      & I          & L;D     \\  
& CVE-2020-20582       & MipCMS v5.0.1                      & D          & L;B     \\  
& CVE-2020-21788       & CRMEB v3.0                         & F          & B       \\
& CVE-2020-23534       & gopeak v2.1.5                      & D          & R       \\
& CVE-2020-24063       & Canto plugin v1.3.0                & D          & L;B     \\  
& CVE-2020-25466       & CRMEB v3.0                         & D          & L;R     \\  
& CVE-2020-28043       & MISP v2.4.133                      & D          & B       \\  
& CVE-2020-28976       & Canto plugin v1.3.0                & D          & L;B     \\  
& CVE-2020-28977       &                                    & D          & L;B     \\  
& CVE-2020-28978       &                                    & D          & L;B     \\  
& CVE-2020-35313       & WonderCMS v3.1.3                   & D          & L;R     \\  
& CVE-2020-35970       & yzmcms v5.3.0                      & D          & L;B     \\  
& CVE-2021-24150       & likebtn-like-button v2.6.31        & F          & L;B     \\  
& CVE-2021-24371       & rsvpmaker v8.7.2                   & D          & B       \\
& CVE-2021-27329       & friendica v2021.01                 & D          & B       \\  
& CVE-2021-28060       & Group Office v6.4.196              & D          & L;B     \\    
& CVE-2021-4075        & snipe-it v5.3.3                    & D          & B       \\
& CVE-2022-0768        & alltube v3.0.1                     & D          & L;B     \\    
& CVE-2022-1037        & EXMAGE v1.0.7                      & D          & B       \\
& CVE-2022-1191        & livehelperchat v3.96               & D          & B;D     \\  
& CVE-2022-1213        & livehelperchat v3.97               & I          & B;D     \\  
& CVE-2022-1239        & leadin v8.8.15                     & D          & L;B     \\  
& CVE-2022-31386       & Nbnbk 532bfdc                      & D          & R       \\
& CVE-2022-31830       & Kity Minder v1.3.5                 & D          & B;D     \\  
& CVE-2022-38292       & SLiMS v9.4.2                       & D          & B       \\
& CVE-2022-40357       & Z-BlogPHP v1.7.2                   & D          & B       \\
\end{tblr}
\end{minipage} \hfill
\begin{minipage}[t]{0.49\linewidth}
\NewColumnType{L}[1][]{>{\ifnum \value{rownum}=1\relax\else \number\numexpr\value{rownum}-1+55\relax\fi}Q[co=1,#1]}
\vspace{0pt}
\begin{tblr}[]{
    hline{1,2,Z} = {1pt},
    colspec={L[-1,c]X[3,l]X[4,l]X[1,c]X[1,c]},
    rows={m,rowsep=0pt,font=\scriptsize},
    row{1} = {font=\tiny},
    cell{11}{3} = {r=3}{m},
    cell{15}{3} = {r=2}{m},
    rowsep=-1pt,
    colsep=2pt,
    cell{even[2-14]}{3} = {bg=gray!10},
    cell{odd[15-Z]}{3} = {bg=gray!10},
}
\textbf{\#} & \textbf{CVE/Vuln ID} & \textbf{Application} & \textbf{Cause} & \textbf{Impact}  \\ 
& CVE-2022-41477                & WeBid    v1.2.2                           & D       & L;B     \\  
& CVE-2022-41497                & ClipperCMS v1.3.3                         & D       & B       \\ 
& CVE-2022-46998                & taocms v3.0.2                             & D       & B       \\
& CVE-2022-47872                & maccms10 v2021.1000.2000                  & D       & B       \\
& CVE-2023-1938                 & wp-fastest-cache v1.1.3                   & D       & B       \\
& CVE-2023-1977                 & Booking Manager v2.0.29                   & D       & B       \\
& CVE-2023-2927                 & JIZHICMS v2.4.5                           & D       & B       \\
& CVE-2023-34959                & Chamilo v1.11.18                          & D       & B       \\ 
& CVE-2023-3744                 & SLiMS v9.6.0                              & D       & L;B     \\  
& CVE-2023-39108                & rconfig v3.9.4                            & D       & B       \\
& CVE-2023-39109                &                                           & D       & B       \\
& CVE-2023-39110                &                                           & D       & B       \\
& CVE-2023-40969                & SLiMS v9.6.1                              & P       & B;P     \\  
& CVE-2023-41054                & LibreY commit 8f9b98                      & F       & L;D     \\  
& CVE-2023-41055                &                                           & P       & D       \\
& CVE-2023-4651                 & icms2 v2.16.0                             & D       & L;B     \\  
& BMLT-1                        & BMLT Meeting Map v2.5.0                   & P       & B       \\
& BN-1                          & benotes v2.8.2                            & D       & B       \\
& CHL-1                         & Chyrp   Lite v2023.02                     & D       & B       \\
& COL-1                         & CommentLuv v3.0.4                         & D       & B       \\
& CR-1                          & CRMEB v5.4                                & F       & L;B     \\  
& CVE-2023-38515                & Church   Admin v3.6.57                    & D       & B       \\
& CVE-2023-46725                & Foodcoopshop   v3.6.0                     & D       & L;B     \\  
& CVE-2023-46730                & Group   Office v6.8.14                    & D       & L;B     \\   
& CVE-2023-46736                & EspoCRM   v8.0.4                          & D       & L;B     \\   
& CVE-2023-48005                & PictShare commit 5c3ee9e                  & D       & L;B     \\  
& CVE-2023-48006                & LinkAce   v1.12.2                         & D       & B       \\ 
& CVE-2023-4878                 & icms2   v2.16.1                           & I       & L;B     \\  
& CVE-2023-49159                & CommentLuv v3.0.4                         & D       & L;B     \\  
& CVE-2023-49746                & SpeedyCache  v1.1.2                       & D       & B       \\
& CVE-2023-50374                & CMP   v4.1.10                             & P       & B       \\
& CVE-2023-50621                & Pixelfed   v0.11.9                        & D       & B       \\ 
& CVE-2023-50622                & IFM   v4.0.0                              & D       & L;B     \\  
& CVE-2023-51676                & Happy   Addons for Elementor v3.9.11      & P       & B       \\
& CVE-2023-52233                & POST   SMTP Mailer v2.8.5                 & D       & B       \\
& CVE-2023-5798                 & Assistant   v1.3.5                        & D       & L;B     \\  
& CVE-2023-5877                 & affiliate-toolkit   v3.2.0                & D       & L;B     \\  
& CVE-2024-22134                & CF7 Extension For Mailchimp v0.5.70       & P       & B       \\
& CVE-2024-32430                & ActiveCampaign v8.1.13                    & P       & B       \\
& CVE-2024-33629                & Auto   Featured Image v3.9.18             & D       & L;B     \\  
& CVE-2024-35633                & blocksy-companion v2.0.42                 & P       & B       \\
& CVE-2024-35635                & ninja-tables v5.0.7                       & D       & B       \\
& CVE-2024-35637                & Church   Admin v3.6.57                    & P       & B       \\
& CVE-2024-37098                & BlossomThemes   Email Newsletter v2.2.5   & P       & B       \\
& CVE-2024-38791                & ai-engine v2.2.94                         & D       & B       \\
& IC-1                          & iCMS v8.0.0                               & D       & L;B     \\   
& MAC-1                         & maccms10 v2021.1000.2000                  & D       & B       \\
& NONE-1                        & NoneCms v1.3.0                            & D       & B       \\
& FR-1                          & friendica v2021.01                        & D       & B       \\ 
& PL-1                          & Pixelfed   v0.11.9                        & D       & B       \\ 
& SI-1                          & snipe-it v5.3.3                           & D       & B       \\  
\end{tblr}
\end{minipage}
\end{adjustbox}
\end{table*}

\begin{table*}
\caption{\rev{Execution time of different tools. The superscripts $A$, $R$, $T$, $J$, $Ps$, $Ph$ denote execution time for \textsc{Artemis}, \textsc{Rips}, \textsc{TChecker}, \textsc{PHPJoern}, \textbf{Psalm} and \textsc{Phan}, unit is seconds. / represents execution failed due to out of memory.}
}
\label{tab:time-app}
\begin{adjustbox}{totalheight=0.92\textheight,keepaspectratio}
\begin{minipage}[t]{0.49\linewidth}
\vspace{0pt}
\begin{tblr}[]{
    hline{1,2,Z} = {1pt},
    colspec={X[5,l]X[-1,l]X[-1,c]X[-1,c]X[-1,c]X[-1,c]X[-1,c]X[-1,c]},
    rows={m,rowsep=0pt,font=\tiny},
    rowsep=-1pt,
    colsep=2pt,
    row{1} = {font=\tiny},
    row{even[2]} = {bg=gray!10},
}
\textbf{Application} & \textbf{LoC} & {\textbf{T$^{A}$}} & {\textbf{T$^{R}$}} & {\textbf{T$^{T}$}} & {\textbf{T$^{J}$}} & {\textbf{T$^{Ps}$}} & {\textbf{T$^{Ph}$}}  \\ 
ActiveCampaign v8.1.13                  &  2.3k   & 33.1  & 0.4     & 0.7    &    0.9   &  7.0   &  2.8  \\ 
affiliate-toolkit   v3.2.0              &  49.0k  & 59.1  & 3.4     & 16.2   &    14.7  &  9.0   &  26   \\ 
ai-engine v2.2.94                       &  8.7k   & 19.1  & 5.4     & 8.3    &    7.5   &  11.3  &  12.4 \\  
Allen Disk v1.6                         &  4.8k   & 14.6  & 0.7     & 1      &    1.4   &  0.7   &  1.2  \\ 
alltube v3.0.1                          &  34.5k  & 142.5 & 10.4    & 25.1   &    19.2  &  7.2   &  15.6 \\   
Assistant   v1.3.5                      &  6.1k   & 27.1  & 0.3     & 0.9    &    1.2   &  6.8   &  7.1  \\ 
Auto   Featured Image v3.9.18           &  17.0k  & 37.4  & 0.5     & 1.6    &    1.5   &  10.4  &  11.4 \\   
benotes v2.8.2                          &  5.5k   & 75.8  & 1.7     & 8.3    &    7.6   &  1.4   &  3.4  \\ 
blocksy-companion v2.0.42               &  18.0k  & 62.7  & 6.5     & 12.8   &    11.3  &  4.5   &  8.8  \\  
BlossomThemes Email Newsletter v2.2.5   &  6.6k   & 28.9  & 1.8     & 5.8    &    4.9   &  6.9   &  11.4 \\  
BMLT Meeting Map v2.5.0                 &  0.7k   & 32.9  & 1.3     & 1.4    &    1.8   &  6.4   &  2.6  \\ 
Booking Manager v2.0.29                 &  12.0k  & 31.7  & 1.1     & 1.3    &    2.4   &  2.0   &  6.2  \\  
Canto plugin v1.3.0                     &  1.5k   & 35.9  & 0.2     & 0.7    &    1.9   &  1.8   &  8.5  \\ 
CF7 Extension For Mailchimp v0.5.70     &  2.3k   & 34.4  & 1.7     & 1.6    &    1.2   &  6.5   &  13.6 \\  
Chamilo v1.11.18                        &  774.9k & 183.2 & 2045.2  & /      &    98.5  &  921.3 &  123.3\\     
Church   Admin v3.6.57                  &  60.3k  & 58.7  & 7.2     & 7.8    &    6.1   &  13.2  &  23.3 \\   
Chyrp   Lite v2023.02                   &  39.5k  & 37    & 3.6     & 1.7    &    1.2   &  12.4  &  7.9  \\  
ClipperCMS v1.3.3                       &  98.4k  & 116.8 & 9.1     & 16.9   &    14.3  &  15.1  &  46.3 \\   
CMP   v4.1.10                           &  4.1k   & 34.2  & 0.7     & 0.9    &    1.5   &  9.2   &  4.5  \\ 
Cockpit v0.5.5                          &  24.3k  & 21.1  & 0.8     & 2.3    &    3.5   &  0.9   &  7.2  \\  
Codiad v1.7.8                           &  90.0k  & 22.4  & 0.8     & 1.3    &    2.1   &  6.3   &  1.5  \\  
CommentLuv v3.0.4                       &  15.2k  & 39.4  & 3.6     & 1.7    &    1.2   &  8.7   &  7.9  \\  
CRMEB v3.0                              &  34.3k  & 80.1  & 19.4    & /      &    59.1  &  3.6   &  17.6 \\   
CRMEB v5.4                              &  34.5k  & 94.4  & 18.9    & /      &    61.4  &  5.9   &  18.3 \\   
dokuwiki v2016-06-26a                   &  15.2k  & 227.6 & 37.1    & 23.1   &    20.4  &  11.9  &  75.1 \\   
EspoCRM   v8.0.4                        &  200.2k & 101.6 & 3.4     & 213.2  &    66.1  &  41.1  &  45.8 \\    
EXMAGE v1.0.7                           &  1.1k   & 24.5  & 0.5     & 0.5    &    1.7   &  0.7   &  5.6  \\ 
FineCMS v2017-07-06                     &  18.9k  & 31.8  & 0.5     & 2      &    2.3   &  2.0   &  1.6  \\  
Foodcoopshop   v3.6.0                   &  63.7k  & 97.8  & /       & /      &    63.1  &  23.5  &  27.1 \\   
friendica v2021.01                      &  39.5k  & 130.4 & 2.7     & 13.2   &    12.7  &  11.3  &  22.4 \\   
gopeak v2.1.5                           &  20.7k  & 40.8  & 12.1    & 26.7   &    20.1  &  7.5   &  10.9 \\   
Group Office v6.8.14                  &  24.3k  & 225.5 & 16.1    & 58.7   &    55.8  &  19.8  &  47.4 \\   
Group Office v6.4.196                   &  24.4k  & 217.8 & 15.4    & 58.1   &    55.2  &  18.4  &  46.8 \\   
Happy  Addons v3.9.11    &  77.0k  & 39    & 0.9     & 3.6    &    3.1   &  10.3  &  13.6 \\   
iCMS v7.0.9                             &  53.4k  & 138.1 & 5.9     & 11.4   &    10.5  &  6.9   &  28.1 \\   
iCMS v8.0.0                             &  53.8k  & 135.6 & 6.5     & 11.1   &    10.2  &  9.9   &  29.3 \\   
icms2   v2.16.1                         &  129.9k & 47.3  & 4.7     & 21.8   &    28.4  &  10.2  &  13.1 \\    
icms2 v2.16.0                           &  129.9k & 45.6  & 4.8     & 21.5   &    28.4  &  10.1  &  13.2 \\    
IFM   v4.0.0                            &  1.3k   & 13.6  & 0.9     & 0.5    &    0.6   &  1.2   &  0.6  \\ 
i-librarian v4.8                        &  26.2k  & 23.6  & 10.1    & 3.7    &    3.1   &  17.6  &  3.5  \\  
JIZHICMS v2.4.5                         &  43.6k  & 45.5  & 0.3     & 1.2    &    2.8   &  4.4   &  12   \\ 
Kity Minder v1.3.5                      &  16.7k  & 22.6  & 1.5     & 2.7    &    2.1   &  2.4   &  2.7  \\  
leadin v8.8.15                          &  2.7k   & 25    & 0.4     & 0.7    &    0.5   &  6.4   &  5.1  \\ 
LibreY commit 8f9b98                    &  1.8k   & 10.4  & 0.3     & 1.5    &    1.2   &  0.7   &  1.4  \\ 
likebtn v2.6.31             &  12.5k  & 46.9  & 1.1     & 1.9    &    1.8   &  15.0  &  9.5  \\  
\end{tblr}
\end{minipage} \hfill
\begin{minipage}[t]{0.49\linewidth}
\vspace{0pt}
\begin{tblr}[]{
    hline{1,2,Z} = {1pt},
    colspec={X[5,l]X[-1,l]X[-1,c]X[-1,c]X[-1,c]X[-1,c]X[-1,c]X[-1,c]},
    rows={m,rowsep=0pt,font=\tiny},
    rowsep=-1pt,
    colsep=2pt,
    row{1} = {font=\tiny},
    row{even[2]} = {bg=gray!10},
}
\textbf{Application} & \textbf{LoC} & {\textbf{T$^{A}$}} & {\textbf{T$^{R}$}} & {\textbf{T$^{T}$}} & {\textbf{T$^{J}$}} & {\textbf{T$^{Ps}$}} & {\textbf{T$^{Ph}$}}  \\ 
LinkAce   v1.12.2                 &  18.3k   & 131.6  & 21.3  &  /     &  /    & 23.1  &  26.1   \\  
livehelperchat v3.96              &  274.5k  & 94.1   & 13.9  &  126.6 &  105.8  & 3.0   &  24.6   \\   
livehelperchat v3.97              &  274.5k  & 95.7   & 14.5  &  125.9 &  105.2  & 3.1   &  24.3   \\   
maccms10 v2021.1000.2000          &  79.9k   & 41.2   & 2.8   &  11.5  &  12.1   & 12.9  &  10.5   \\  
Matomo v2.14.3                    &  600.5k  & 91.7   & 8.8   &  36.7  &  21.4   & 2.1   &  19.2   \\   
MipCMS v5.0.1                     &  204.5k  & 24.8   & 9.6   &  39.5  &  35.4   & 20.4  &  5.1    \\  
MISP v2.4.133                     &  368.6k  & 328.1  & 15.3  &  /     &  50.4   & 28.4  &  157.4  \\    
mybb v1.8.11                      &  128.8k  & 125.9  & 23.5  &  48.8  &  42.1   & 15.4  &  51.4   \\   
mybb v1.8.7                       &  126.8k  & 118.2  & 21.7  &  41.4  &  39.2   & 14.3  &  48.9   \\   
Nbnbk 532bfdc                     &  27.1k   & 93.4   & 8.3   &  10.4  &  12.1   & 3.5   &  6.6    \\ 
nelio ab testing  v4.5.11         &  25.2k   & 29.1   & 12.5  &  2.7   &  2.5    & 2.5   &  9.8    \\ 
nelio ab testing  v4.5.9          &  25.2k   & 28.7   & 12.5  &  2.4   &  2.6    & 2.5   &  9.6    \\ 
ninja-tables v5.0.7               &  9.5k    & 68.2   & 1.7   &  11.5  &  10.4   & 3.1   &  20.1   \\ 
NoneCms v1.3.0                    &  44.3k   & 62.1   & 6.7   &  6.5   &  5.8    & 10.8  &  3.8    \\ 
Onethink v1.1                     &  26.8k   & 62     & 17.3  &  3.5   &  4.6    & 5.4   &  5.3    \\ 
openITCOCKPIT v3.7.1              &  756.1k  & 140.3  & /     &  648.3 &  210.7  & 58.5  &  74.7   \\   
openITCOCKPIT v3.7.2              &  756.1k  & 143.2  & /     &  652.1 &  209.2  & 57.9  &  76.2   \\   
phpbb v3.2.0                      &  241.8k  & 201.4  & 5.2   &  46.4  &  38.2   & 18.6  &  35.8   \\   
phpbb v3.2.6                      &  242.1k  & 205.3  & 5.6   &  52.1  &  41.9   & 18.9  &  36.1   \\   
phprap v1.0.4                     &  29.8k   & 21.4   & 3.8   &  8.8   &  6.3    & 1.3   &  1.6    \\ 
PictShare commit 5c3ee9e          &  2.7k    & 10.6   & 0.4   &  0.7   &  1.3    & 2.3   &  1.6    \\
Pixelfed   v0.11.9                &  810.3k  & 126.7  & /     &  /     &  34.1   & 20.7  &  40.6   \\   
POST   SMTP Mailer v2.8.5         &  116.6k  & 61.2   & 7.2   &  20.5  &  18.3   & 17.1  &  28.5   \\   
Print My Blog v1.6.6              &  1.0k    & 33.6   & 0.2   &  0.7   &  1.6    & 6.5   &  2.6    \\
pydio core v6.0.8                 &  128.4k  & 135.9  & 12.1  &  45.6  &  31.1   & 18.3  &  32.6   \\   
rconfig v3.9.4                    &  48.3k   & 26.5   & 9.4   &  5.4   &  6.2    & 10.1  &  20.1   \\  
Responsive FileManager v9.13.1    &  13.5k   & 21.7   & 0.8   &  1.5   &  1.9    & 4.8   &  1.7    \\ 
Responsive FileManager v9.13.4    &  13.5k   & 23.4   & 1.1   &  1.4   &  2.1    & 4.9   &  1.8    \\ 
Responsive FileManager v9.14.0    &  13.6k   & 22.2   & 0.9   &  1.5   &  1.9    & 4.9   &  1.8    \\ 
rsvpmaker v8.7.2                  &  52.9k   & 44.8   & 6.1   &  6.4   &  5.8    & 13.9  &  31.9   \\  
SeaCMS v6.61                      &  37.2k   & 27.3   & 70.6  &  5.2   &  5.4    & 11.7  &  5.8    \\ 
SLiMS v9.4.2                      &  230.6k  & 47.8   & 39.2  &  15.2  &  14.1   & 6.4   &  32.5   \\   
SLiMS v9.6.0                      &  341.6k  & 53.4   & 63.5  &  21.9  &  22.8   & 5.9   &  30.8   \\   
SLiMS v9.6.1                      &  342.6k  & 52.9   & 63.9  &  22.6  &  22.3   & 6.7   &  31     \\ 
SMF v2.1                          &  131.9k  & 39.1   & 15.1  &  28.9  &  25.3   & 9.4   &  28.5   \\   
snipe-it v5.3.3                   &  872.5k  & 90     & /     &  /     &  /    & 29.0  &  37.2   \\   
SpeedyCache  v1.1.2               &  9.8k    & 35.3   & 0.7   &  1.2   &  1.5    & 3.2   &  12.5   \\ 
taocms v3.0.2                     &  3.1k    & 81.1   & 1.7   &  0.8   &  1.2    & 0.8   &  1.4    \\
UpdraftPlus v1.13.12              &  133.2k  & 80.5   & 5.6   &  18.3  &  14.4   & 102.3 &  26.4   \\   
WeBid    v1.2.2                   &  35.8k   & 15.5   & 51.2  &  4.6   &  3.9    & 5.9   &  4.8    \\ 
webpagetest v19.04                &  367.6k  & 92.2   & 43.4  &  77.9  &  75.2   & 146.5 &  31.5   \\   
WonderCMS v3.1.3                  &  1.7k    & 11.5   & 0.2   &  0.4   &  0.9    & 1.4   &  1.2    \\
wp-fastest-cache v1.1.3           &  9.2k    & 25.1   & 0.3   &  1.2   &  1.8    & 12.5  &  12     \\
yzmcms v5.3.0                     &  16.4k   & 33     & 0.9   &  2.4   &  3.1    & 1.9   &  2.9    \\ 
Z-BlogPHP v1.7.2                  &  56.9k   & 48.1   & 44.5  &  6.4   &  6.2    & 12.6  &  7.9    \\ 
\end{tblr}
\end{minipage}
\end{adjustbox}
\end{table*}

\bibliographystyle{ACM-Reference-Format}
\citestyle{acmnumeric}
\bibliography{ref}

\end{document}